\RequirePackage{ifpdf}
\documentclass[hyper,letterpaper]{JHEP3}
\usepackage{amsmath,amssymb,amsfonts,bm,amscd}
\usepackage{cite}
\usepackage{stackrel}
\usepackage{graphicx}
\usepackage{multirow}
\usepackage{verbatim}
\usepackage{appendix}
\usepackage{fancybox}
\usepackage{array}
\usepackage{url}
\usepackage{float}
\usepackage{framed}
\usepackage[all,cmtip]{xy}

\newcommand{\bea}{\begin{eqnarray}}
\newcommand{\eea}{\end{eqnarray}}
\newcommand{\be}{\begin{equation}}
\newcommand{\ee}{\end{equation}}

\def\Tr{{\rm Tr \,}}

\def\tilde{\widetilde}
\def\hat{\widehat}
\def\bar{\overline}

\def\CC{{\mathcal C}}

\def\CF{{\mathcal F}}

\def\CH{{\mathcal H}}
\def\CI{{\mathcal I}}

\def\CL{{\mathcal L}}

\def\CN{{\mathcal N}}
\def\CO{{\mathcal O}}

\def\CQ{{\mathcal Q}}

\def\CT{{\mathcal T}}

\renewcommand{\bar}{\overline}
\renewcommand{\hat}{\widehat}

\renewcommand{\d}{\partial}

\newcommand{\ttR}{{\mathtt R}}
\newcommand{\ttL}{{\mathtt L}}
\newcommand{\jacobi}{\eta}
\newcommand{\SU}{{\rm SU}}
\newcommand{\UU}{{\rm U}}
\newcommand{\SO}{{\rm SO}}
\newcommand{\EE}{{\rm E}}
\newcommand{\frH}{{\mathfrak H}}
\newcommand{\frG}{{\mathfrak G}}
\newcommand{\frh}{{\mathfrak h}}
\newcommand{\frg}{{\mathfrak g}}
\newcommand{\ph}{\phi}
\newcommand{\aph}{{\bar \phi}}
\newcommand{\modsub}{{\tilde \Gamma}}

\title{Exact Solutions of 2d Supersymmetric Gauge Theories}

\author{Abhijit Gadde$^{1}$, Sergei Gukov$^{1,2}$, and Pavel Putrov$^{1}$
\\
$^1$ California Institute of Technology, Pasadena, CA 91125, USA \\
$^2$ Max-Planck-Institut f\"ur Mathematik, Vivatsgasse 7, D-53111 Bonn, Germany}

\abstract{We study dynamics of two-dimensional non-abelian gauge theories with $\CN=(0,2)$ supersymmetry
that include $\CN=(0,2)$ supersymmetric QCD and its generalizations.
In particular, we present the phase diagram of $\CN=(0,2)$ SQCD and determine its massive and low-energy spectrum.
We find that the theory has no mass gap, a nearly constant distribution of massive states, and lots of massless states
that in general flow to an interacting CFT.
For a range of parameters where supersymmetry is not dynamically broken at low energies,
we give a complete description of the low-energy physics in terms of 2d $\CN=(0,2)$ SCFTs using anomaly matching and modular invariance.
Our construction provides a vast landscape of new $\CN=(0,2)$ SCFTs which, for small values of the central charge,
could be used for building novel heterotic models with no moduli and, for large values of the central charge,
could be dual to AdS$_3$ string vacua.}

\preprint{CALT 68-2886}
\begin{document}
\cornersize{1}

%%%%%%%%%%%%%%%%%%%%%%%%%%%%%%%%%%%%%%%%%%%%%%%%%%%%%%%%%%%%%%%%%%%%%

\section{Introduction and summary}

Our goal is to study the quantum physics of two-dimensional SQCD with $\CN=(0,2)$ supersymmetry,
{\it i.e.} a gauge theory with $\UU(N_c)$ or $\SU(N_c)$ gauge group and matter fields (quarks and squarks)
in the fundamental representation. Following the standard terminology of the four-dimensional QCD,
we shall refer to $N_c$ as the number of {\it colors} and the number of matter fields as the number of {\it flavors}.

\begin{figure}[htb] \centering
\includegraphics[width=3.0in]{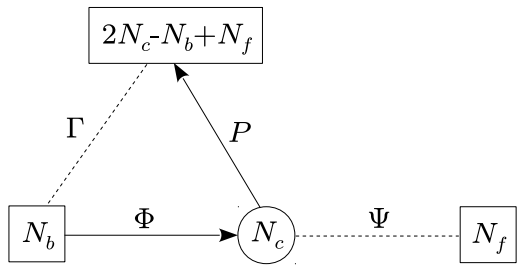}
\caption{\label{fig:qcd-su}
The field content of $(0,2)$ SQCD with $\SU(N_c)$ gauge group,
$N_f$ (resp. $N_b$) Fermi (resp. chiral) multiplets in the fundamental representation,
and $2N_c - N_b + N_f$ chiral multiplets in the anti-fundamental representation.
A similar theory with $\UU(N_c)$ gauge group requires two extra Fermi multiplets in the determinant representation
to cancel the abelian gauge anomaly.}
\end{figure}

In two-dimensional theories with $\CN=(0,2)$ supersymmetry, however, there are two types of matter multiplets,
namely the Fermi multiplets and $(0,2)$ chiral multiplets whose lowest components are fermionic and bosonic, respectively.
Therefore, allowing both types of multiplets, we define $(0,2)$ SQCD as a gauge theory with $N_f$ Fermi quark
multiplets and $N_b$ chiral multiplets, plus the minimal completion of the theory that cancels gauge anomaly and yields a normalizable vacuum for generic values of $N_f$ and $N_b$.
The resulting field content is shown in Figure~\ref{fig:qcd-su}.

It turns out that such theories enjoy a very non-trivial triality symmetry \cite{Gadde:2013lxa},
which can be best described by introducing
\be
N_1 = 2N_c + N_f - N_b \,, \quad
N_2 = N_b \,, \quad
N_3 = N_f
\label{trialityframe}
\ee
and which, among other things, acts by a cyclic permutation of these numbers $N_1 \to N_2 \to N_3 \to N_1$.
In the present paper, however, the triality will play a secondary role and our aim will be to understand
the physics of $(0,2)$ SQCD in a fixed duality frame, without relying on the triality.

\begin{figure}[htb] \centering
\includegraphics[scale=1.2]{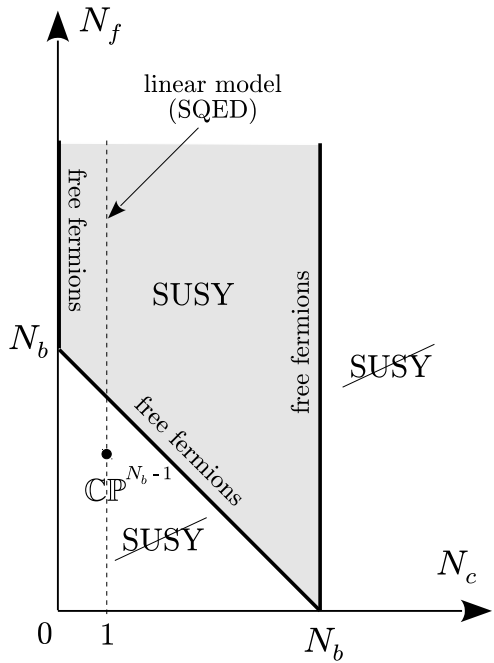}
\caption{\label{fig:SQCD1}
The phase diagram of $\UU(N_c)$ $(0,2)$ SQCD as a function of $N_c$ and $N_f$ (with $N_b$ kept fixed).}
\end{figure}

The low energy behavior of $(0,2)$ SQCD is summarized in the ``phase diagram" in Figure~\ref{fig:SQCD1}.
The supersymmetry is dynamically broken except in the shaded region \cite{Gadde:2013lxa}.
On the other hand, 't Hooft anomaly matching condition suggests that the theory is gapless at all the points in the phase diagram.
Together these statements imply that the low energy physics is described by an $\CN=(0,2)$ superconformal field theory in the shaded region and by a (non-supersymmetric) conformal field theory everywhere else. The flavor symmetries of the theory are promoted to affine Kac-Moody symmetries
(to put this principle in a broader context, see {\it e.g.} \cite{Frishman:1992mr,Kutasov:1994xq,Witten:2005px,Nekrasov:2005wg}).

When the supersymmetry is not dynamically broken, the affine Kac-Moody symmetries are purely left-moving.
The physical Hilbert space $\CH$ consists of states $|\psi\rangle:=|\psi_\ttL\rangle \otimes |\psi_\ttR\rangle$ with $|\psi_\ttL\rangle\in \CH_\ttL$ and $|\psi_\ttR\rangle \in \CH_\ttR$ furnishing a representation (or module) of the left-moving, \emph{i.e.} holomorphic, affine algebra and right-moving, \emph{i.e.} anti-holomorphic, $\CN=2$ superconformal algebra respectively. The physical pairing of left-moving and right-moving modules is dictated by modular invariance. We propose a novel modular invariant pairing between the modules of affine symmetry and modules of $\CN=2$ symmetry, thus solving for the complete spectrum of the low-energy SQCD.

Below we present a brief summary of this result, along with several other salient features of $(0,2)$ SQCD that we find:

\begin{itemize}

\item
In many two-dimensional models with $\SU(N_c)$ gauge group and massless matter (in fundamental or adjoint representation)
there is strong numerical evidence that spectrum of the quantum theory contains no massless states,
see \cite{Dalley:1992yy,Kutasov:1993gq,Demeterfi:1993rs,Antonuccio:1998tm}
(and \cite{Aoki:1995dh,Antonuccio:1998kz,Antonuccio:1998jg} for further discussion).
On the contrary, we find that the spectrum of $(0,2)$ SQCD exhibits accumulation of light states and, in particular,
has many massless states (that flow to an interacting CFT).

\item
The infra-red CFT has an affine symmetry on the left-moving side and the $\CN=2$ superconformal symmetry on the right-moving side. The gauge theories have a crucial property that their left-moving central charge matches with the Sugawara central charge of the affine symmetry. This determines the holomorphic stress tensor of the fixed point to be the Sugawara stress tensor and leads to a rational CFT.

\item
The modular invariant pairing of modules of affine symmetry and $\CN=2$ symmetry that solves the gauge theory may be of merit on its own. For example, these models could be relevant in the context of heterotic string phenomenology. In spirit, they are similar to Gepner points in Calabi-Yau moduli space \cite{Gepner:1987vz} and, in addition, do not have any exactly marginal
deformations.\footnote{Much like Gepner models ``solve'' the GLSM description of Calabi-Yau sigma-models at special points on the conformal manifold,
the conformal theories discussed here solve $(0,2)$ gauge theories. There are several crucial differences, however.
First, unlike gauged linear sigma-models, the $(0,2)$ gauge theories discussed here are non-abelian.
The second crucial difference is that abelian $\CN=(2,2)$ gauge theories often have
exactly marginal deformations that lead to unwanted moduli in four-dimensional physics.}
This could be an attractive aspect for developing ``exactly solvable string phenomenology".

\item
In many two-dimensional models, it was found that low-lying states are surprisingly pure in a sense that mass eigenstates
are almost exactly eigenstates of string length \cite{Dalley:1992yy,Demeterfi:1993rs,Gross:1997mx}.
We find that $(0,2)$ SQCD has qualitatively different distribution of mass eigenstates versus parton number, illustrated in Figure~\ref{fig:dlcq-partons}.

\item
We find continuous massive spectrum which is expected to be a general feature of 2d gauge theories
with adjoint scalars~\cite{Gross:1997mx,Antonuccio:1998jg}.
An interesting feature of the massive spectrum in $\CN=(0,2)$ SQCD is that it is nearly flat.

\end{itemize}

The paper is organized as follows.
In section~\ref{SQCD} we review $(0,2)$ SQCD and discuss a variety of its aspects such as RG flow
to non-linear sigma-model and, further, to a CFT where one sometimes finds a flavor symmetry enhancement.
In section~\ref{CFT}, we focus our attention on theories that \emph{do} preserve supersymmetry in the infra-red
and propose a solution to the complete low-energy spectrum of $(0,2)$ SQCD.
The general case is illustrated in great detail with the simplest non-trivial example of the $N_1=N_1=N_3=2$ theory,
which also happens to exhibit a peculiar enhancement of the flavor symmetry in the infra-red to the exceptional group $\EE_6$,
and whose modular invariant partition function can be conveniently expressed in terms of affine $\EE_6$ characters.
In section~\ref{sec:massive} we study massive states of the quantum theory, not limiting ourselves
to theories which preserve SUSY in the infra-red.
We employ discrete light cone quantization to determine the spectrum of ``$\eta'$ mesons''
and similar color-flavor singlets that dominate in the Veneziano limit.
Appendices contain relevant supplementary material and further details.
In particular, appendix \ref{affinechar} compiles a number of useful facts about affine characters and level-rank duality.

%%%%%%%%%%%%%%%%%%%%%%%%%%%%%%%%%%%%%%%%%%%%%%%%%%%%%%%%%%%%%%%%%%%%%%%%%%%%%%%%%%%%%%%%%%%%%%

\section{Aspects of (0,2) SQCD}
\label{SQCD}

As we already mentioned earlier, sometimes it will be convenient to go from the data $(N_c,N_f,N_b)$
that specifies the field content of the basic $\CN = (0,2)$ SQCD to a more symmetric set of labels $(N_1,N_2,N_3)$
related to the former via \eqref{trialityframe}. In particular, the rank of the gauge group is $N_c=(N_1+N_2-N_3)/2$.
The representation of matter multiplets under the gauge and flavor symmetry group is,
\be
\begin{array}{@{~~}c@{~~}|@{~~}c@{~~~}c@{~~~}c@{~~~}c@{~~}|@{\quad}c@{~~}l@{~~}}
 & \Phi & \Psi & P & \Gamma & & \text{labels} \\
\hline
\SU(N_{c}) & \square & \overline{\square} & \overline{\square} & {\bf 1} & & \alpha,\beta,\gamma \tabularnewline
\SU(N_{1}) & {\bf 1} & {\bf 1} & \square & \overline{\square} & & a,b,c  \tabularnewline
\SU(N_{2}) & \overline{\square} & {\bf 1} & {\bf 1} & \square & & r,s,t \tabularnewline
\SU(N_{3}) & {\bf 1} & \square & {\bf 1} & {\bf 1} & & i,j,k
\end{array}
\label{reps}
\ee
Note, in particular, that the combination $\Tr \Gamma \Phi P$ is a singlet under all of these symmetries; it will play an important role as the superpotential in our models. Specifically, we have
\be\label{J-term}
\CL_J \; = \; \int d \theta^+ \; \Gamma_{a}^s J^{a}_s (\Phi,P)|_{{\bar \theta}^+=0}
\ee
where
\be
J^{a}_s (\Phi,P) \; = \; m \,\Phi^{\alpha}_s P^{a}_{\alpha}
\label{JPPhi}
\ee
Note, in 2d the canonical (mass) dimensions of various fields are
\be
[\phi] = 0 \,, \qquad
[\psi_{\pm}] = \frac{1}{2} \,, \qquad
[A_{\mu}] = [g] = 1 \,, \qquad
[\lambda_- ] = \frac{3}{2} \,, \qquad
[D] = 2
\label{massdimensions}
\ee
In particular, it follows that the coefficient of the $\CN = (0,2)$ superpotential \eqref{JPPhi} has mass dimension 1,
which explains why we denote it by $m$. Moreover, since the gauge coupling $g$ also has mass dimension 1,
the theory has only one continuous {\it dimensionless} parameter $y := \frac{m}{g}$,
such that $y \to \infty$ is the ``weak coupling'' limit, while $y \to 0$ corresponds to the ``strong coupling''.

\begin{figure}[ht]
\centering
\includegraphics[width=3.0in]{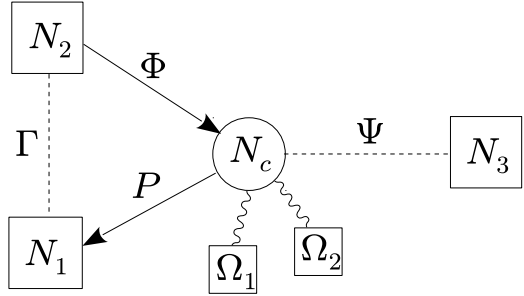}
\caption{\label{sqcd}
The quiver diagram of $\CN=(0,2)$ SQCD with $\UU(N_c)$ gauge group.
The solid lines denote bi-fundamental  chiral multiplets and dotted lines denote bi-fundamental Fermi multiplets.
We have explicitly depicted the two Fermi multiplets in the determinant representation of the gauge group with wiggly lines.}
\end{figure}

When the gauge group of the theory is $\UU(N_c)$ instead of $\SU(N_c)$,
two extra Fermi multiplets in the determinant representation of the gauge group
are required to cancel the anomaly of the abelian part of the gauge group.
We denote them as $\Omega_1$ and $\Omega_2$.
The field content of this theory is shown in Figure~\ref{sqcd}.
The $\UU(N_c)$ theories, henceforth denoted $\CT_{N_1,N_2,N_3}$, will feature prominently in section~\ref{CFT}.

The charge conjugation operator acts on the theory by conjugating all the global symmetry representations. It maps the theory $\CT_{N_1,N_2,N_3}$  to $\CT_{N_2,N_1,N_3}$. In what follows, it will convenient to define
\be
N:=\frac{N_1+N_2+N_3}{2},\qquad \qquad n_i:=N-N_i.
\ee
In this notation, the rank of the gauge group is $N_c=n_3$.

In general, the trace anomaly $k_F$ for flavor symmetry $F$ is given by
\be
k_F={\rm Tr} \, \gamma^3 \, J_F\cdot J_F.
\ee
We can easily calculate
\be
k_{\SU (N_1)}= -\frac{n_1}{2},\qquad k_{\SU (N_2)}= -\frac{n_2}{2}, \qquad k_{\SU (N_3)}=-\frac{n_3}{2}.
\ee
In addition to the non-abelian flavor symmetries, the Fermi multiplets $\Omega_1$ and $\Omega_2$ have an $\SU (2)_\Omega$ symmetry rotating them.
For our purposes, it is convenient to look at its $\UU (1)$ subgroup.
Including this $\UU (1)$, the classical theory has a total of four abelian symmetries $\UU (1)_{(i)}$ $i=1,\ldots, 4$.
Unlike $\SU (N)$ symmetries, the $\UU (1)$ symmetries can have mixed anomalies.
We choose them so that their mixed anomalies vanish. The charges of the matter fields under $\UU (1)_{(i)}$ are listed below:
\begin{center}
\begin{tabular}{c|rrrrrr}
 & $\Phi$ & $\Psi$ & $P$ & $\Gamma$ &$\Omega_1$ & $\Omega_2$ \tabularnewline
\hline
$\UU (1)_{(1)}$ & $0$ & $0$ & $1$ & $-1$ & $-N_1$ & $0$ \tabularnewline
$\UU (1)_{(2)}$ & $-1$ & $0$ & $0$ & $1$ & $-N_2$ & $0$\tabularnewline
$\UU (1)_{(3)}$ & $0$ & $1$ & $0$ & $0$ & $0$ & $N_3$\tabularnewline
$\UU (1)_{(4)}$ & $1$ & $1$ & $-1$ & $0$ & $n_3$ & $-n_3$\tabularnewline
\end{tabular}
\par\end{center}
They have
\be
k_{\UU (1)_{(1)}}= -\frac{NN_1}{2},\qquad k_{\UU (1)_{(2)}}=-\frac{NN_2}{2},\qquad k_{\UU (1)_{(3)}}=-\frac{NN_3}{2}, \qquad k_{\UU (1)_{(4)}}=0.
\ee
The symmetries $\UU (1)_{(i)}$ for $i=1,2,3$ are non-anomalous,
{\it i.e.} they have vanishing mixed anomaly with the gauge symmetry but $\UU (1)_{(4)}$ is anomalous.
The mixed anomaly between the gauge symmetry and $\UU (1)_{(4)}$ is $2N$. It breaks the  symmetry to ${\mathbb Z}_{2N} \subset \UU (1)_{(4)}$.
In fact, as we can see from the table, the ${\mathbb Z}_2\subset {\mathbb Z}_{2N}$
is a subgroup of the abelian part of the gauge group and the remaining ${\mathbb Z}_N$
is a subgroup of $\UU (1)_{(1)}\times \UU (1)_{(2)} \times \UU (1)_{(3)}$.
This means that the classical  $\UU (1)_{(4)}$ symmetry is completely destroyed by quantum anomaly.

The superconformal symmetry relates the right-moving central charge $c_\ttR$ to the R-symmetry anomaly and the left-moving central charge $c_\ttL$ is computed from $c_\ttR$ via gravitational anomaly $k$,
\be
c_\ttR=3{\Tr} \, \gamma^3\, R^2,\qquad c_\ttR-c_\ttL=k={\Tr} \, \gamma^3.
\ee
Here $R$ is the exact superconformal R-symmetry of the theory that extremizes $c_\ttR$.
${\Tr} \gamma^3$ is simply the difference between the number of chiral multiplets and Fermi multiplets.
For $\CT_{N_1,N_2,N_3}$ we get,
\bea\label{gaugec}
c_\ttR&=& \frac{3}{4}\frac{(N_1+N_2-N_3)(N_1-N_2+N_3)(-N_1+N_2+N_3)}{N_1+N_2+N_3},\nonumber\\
c_\ttL&=& c_\ttR-\frac{1}{4}(N_1^2+N_2^2+N_3^2-2N_1N_2-2N_2N_3-2N_3N_1)+2.
\eea
The central charges $c_\ttR$ and $c_\ttL$ as well as flavor symmetry anomalies are invariant under
cyclic permutation of $N_i$'s, as expected from triality.

%%%%%%%%%%%%%%%%%%%%%%%%%%%%%%%%%%%%%%%%%%%%%%%%%%%%%%%%%%%%%%%%%%%%%%%%%%%%%%%

\subsection{RG flow}

The first hints about the balance between massive and massless states  as well as the role of interactions
in our $\CN=(0,2)$ SQCD can be obtained by comparing the central charge of free constituents in Figure~\ref{fig:qcd-su}
to that of an IR fixed point found in \cite{Gadde:2013lxa}.
Specifically, in the triality frame \eqref{trialityframe} we have
\be
c_\ttR^{\text{(IR)}} - c_\ttR^{\text{(UV)}}
= - \frac{3(N_1+N_2-N_3) \left(N_1^2+N_2^2+N_1 N_3+N_2 N_3\right)}{2(N_1+N_2+N_3)}
\label{ccdiffernc}
\ee
First, note that, in accordance with the Zamolodchikov's $c$-theorem \cite{Zamolodchikov:1986gt},
we always have $c_\ttR^{\text{(IR)}} - c_\ttR^{\text{(UV)}} < 0$.
For the theories that preserve supersymmetry in the infra-red,
the plot of $\frac{1}{12N^2}(c_\ttR^{\text{(IR)}}-c_\ttR^{\text{(UV)}})$ is presented in Figure~\ref{fig:cRnnplot}.
\FIGURE[H]{
\includegraphics[width=2.5in]{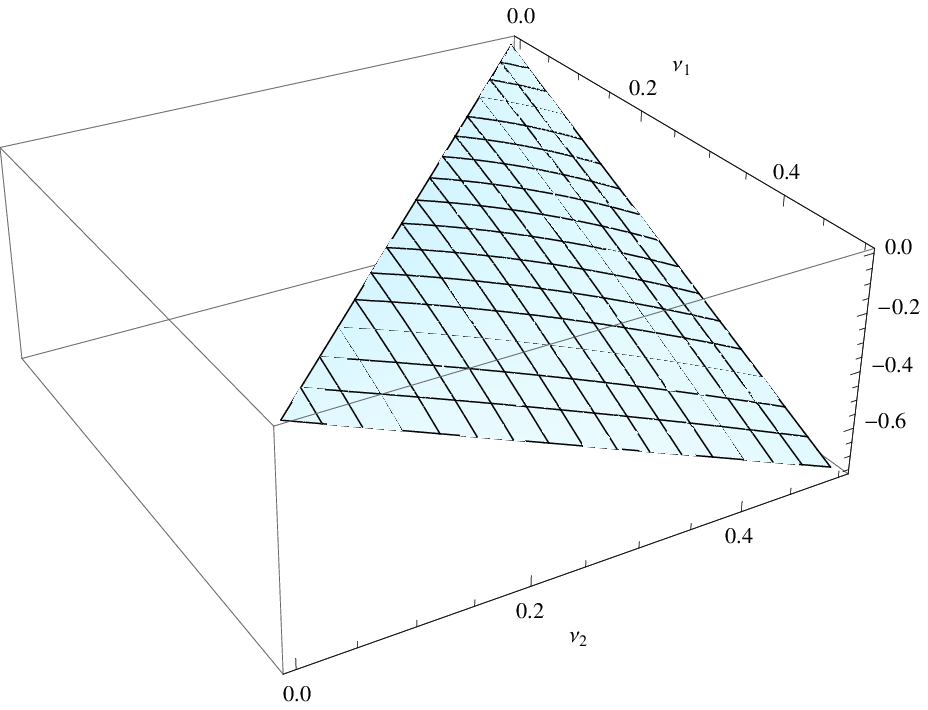}
\caption{\label{fig:cRnnplot}
The plot of $c_\ttR^{\text{(IR)}} - c_\ttR^{\text{(UV)}}$ as a function of $\nu_1$ and $\nu_2$.
}}
In this plot, $\nu_i$ are the ``homogeneous coordinates" on the parameter space
\be
\nu_i := \frac{N_i}{N_1 + N_2 + N_3} \,, \quad i = 1,2,3
\label{nudef}
\ee
that are also convenient for studying the large-$N_c$ limit and will play role later.
As a function of $\nu_1$ and $\nu_2$, the difference \eqref{ccdiffernc}
attains a maximum value of $c_\ttR^{\text{(IR)}} - c_\ttR^{\text{(UV)}} = 0$ at $\nu_1 + \nu_2 = \frac{1}{2}$,
{\it i.e.} at $N_1 + N_2 - N_3 = 0$. At these values of $N_i$ (or, equivalently, $\nu_i$)
the right-moving central charge $c_\ttR$ does not flow too much, indicating that the spectrum
of the quantum theory is mostly massless and is not too much ``distorted'' by the interactions.
This happens for a good (physical) reason because $\frac{N_1 + N_2 - N_3}{2}$
is precisely the rank of the gauge group in the duality frame at hand.
For larger values of $N_c$ the RG flow of $c_\ttR$ is more significant.

The other ``triangle inequalities'', $N_2 + N_3 \ge N_1$ and $N_1 + N_3 \ge N_2$,
in the triality frame \eqref{trialityframe} read $N_c \le N_b$ and $N_c + N_f \ge N_b$, respectively.
In particular, their boundaries --- shown in Figure~\ref{fig:SQCD1} --- correspond to the regimes
where $c_\ttR^{\text{IR}}$ is well approximated by the central charge of free $\Phi \Psi$ and $\Psi P$ mesons, respectively
(with $P$, $\Gamma$ or $\Phi$, $\Gamma$ integrated out in each case).

The most convenient description of $(0,2)$ SQCD depends on the energy scale at which it is studied.
At ultra-low energies it is best to use the language of CFT while at intermediate energies,
it is most convenient to use the description in terms of non-linear sigma-models that we present next.
As the gauge theory flows to a non-linear sigma-model, the gauge coupling and the $J$-term coupling $m$
are washed out, {\it i.e.} they correspond to irrelevant deformations of the sigma-model, {\it cf.} \eqref{massdimensions}.
The FI parameter $t$ becomes the Kahler modulus of the target space.
Even though it is a marginal modulus, it can have a non-vanishing beta function.
When the theory flows to the infra-red fixed point, the RG flow can be seen as,
$$
\CT_{\rm gauge}(g,m,t) \xrightarrow{\quad \text{RG flow} \quad} \CT_{\rm NLSM}(t) \xrightarrow{\quad \text{RG flow} \quad} \CT_{\rm CFT}.
$$

\begin{figure}[H]
\centering
\includegraphics[scale=0.7]{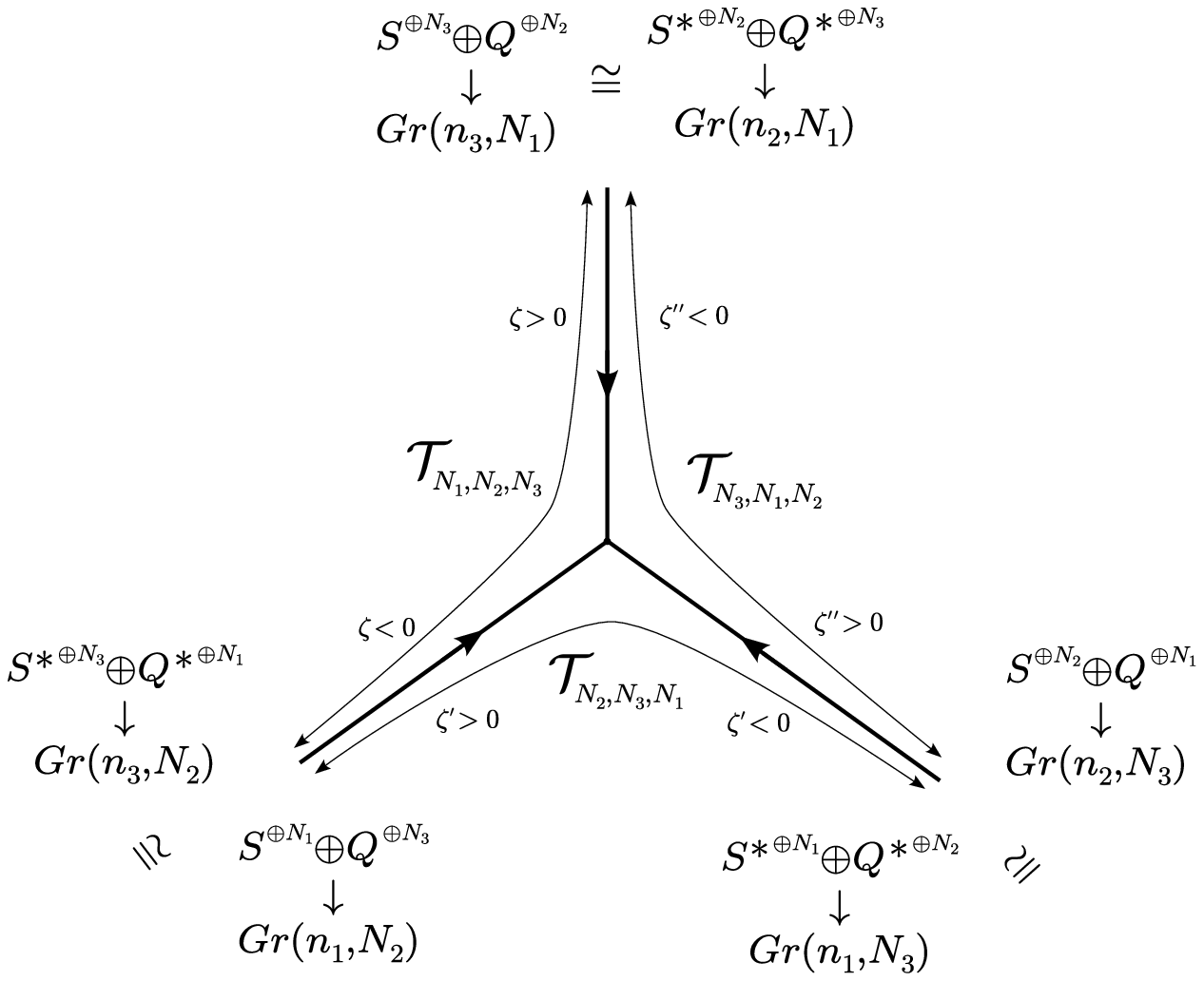}
\caption{Phases of gauge theories related by triality. Their parameter spaces can be glued to each other as shown. }\label{zetaspace}
\end{figure}

The target of a $(0,2)$ non-linear sigma-model is a holomorphic vector bundle $E\to M$ over a K\"ahler manifold $M$.
The anomaly cancellation requires  $ch_2(E)=ch_2(TM)$.  In our case, the target space is the vacuum moduli space of the gauge theory.
It is obtained by solving the $D$-term and ``$J$-term" constraints modulo the gauge symmetry action.
Let us analyze the sigma model description of the theory $\CT_{N_1,N_2,N_3}$ as a function of the FI parameter $t:=\zeta+i\theta$.
As the theory has an anomalous $\UU (1)_{(4)}$ symmetry, the $\theta$-angle is unphysical.
The $D$-term and $J$-term equations are
\bea
P_\alpha^a \bar P_{ a}^\beta - \Phi^\beta_s \bar \Phi_{ \alpha}^s - \zeta \delta^\beta_{\alpha} &=& 0\\
P_\alpha^a \Phi^\alpha_s & = & 0
\eea
They imply $\Phi =0$ (resp. $P$=0) for $\zeta >0$ (resp. $\zeta<0$). Dividing by the $\UU (n_3)$ gauge group, we get the space $Gr(n_3,N_1)$.
The Fermi fields engineer fibers of the holomorphic vector bundle.
As the field $\Psi$ transforms in the fundamental representation of the gauge group,
it forms a fiber of the universal subbundle (tautological bundle) $S$.
The field $\Gamma$ is neutral but it satisfies the $J$-term relation
\be
\Gamma_a^s P^a_\alpha=0.
\ee
Therefore, $\Gamma$ furnish a fiber of the universal quotient bundle (orthogonal bundle) $Q$, which is defined through the short exact sequence:
\be
0\longrightarrow S \longrightarrow {\cal O}^{N_1} \longrightarrow Q \longrightarrow0.
\ee
 All in all, for $\zeta >0$, the theory $\CT_{N_1,N_2,N_3}$ flows to the nonlinear sigma model with the target space,
\be
S^{\oplus N_3} \oplus Q^{\oplus N_2} \longrightarrow Gr(n_3,N_1).
\ee
For $\zeta <0$, the $D$-term equation gives vev only to $\Phi$.
Similar arguments lead to the target space\footnote{We have also conjugated the Fermi multiplets $\Gamma$ using equivalence between $J$-term and $E$-term interactions.} $S^{*\oplus N_3} \oplus Q^{*\oplus N_1} \longrightarrow Gr(n_3,N_2)$.
This space is isomorphic to $S^{\oplus N_1} \oplus Q^{\oplus N_3} \longrightarrow Gr(n_1,N_2)$ thanks to the equivalence relations
\bea
S\to Gr(k,n) \,&\qquad \cong \qquad & Q^* \to Gr(n-k,n),\\
Q \to Gr(k,n) & \qquad \cong \qquad & \,S^*\to Gr(n-k,n).
\eea
By cyclically permuting $N_i$'s we know that the theory $\CT_{N_2,N_3,N_1}$ also flows to the same target space but for $\zeta'>0$, where $\zeta'$ is its own FI parameter. This suggests that the parameter spaces of the theories related by triality are glued to each other as shown in Figure~\ref{zetaspace}. The RG flow is such that they flow to the same fixed point, as expected.
In fact, the figure demonstrates a stronger version of triality that is valid even away from the fixed point.

The target space of SQCD and its behavior under triality was also discussed in the recent paper \cite{SharpeDualities}.

%%%%%%%%%%%%%%%%%%%%%%%%%%%%%%%%%%%%%%%%%%%%%%%%%%%%%%%%%%%%%%%%%%%%%%%%%%%%%%%%%%%%%%%%%%%%%

\subsection{``Gluing'' via level-rank duality}

\begin{figure}[ht]
\centering
\includegraphics[scale=0.8]{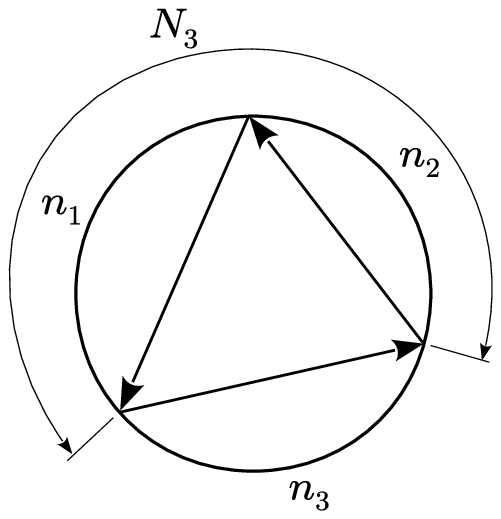}
\caption{The oriented triangle represents the $\CN=(0,2)$ SCFT labeled by $(N_1,N_2,N_3)$ modulo cyclic permutations.}
\label{triangleCFT}
\end{figure}

The triality implies that infra-red CFT is labeled by the triple $(N_1,N_2,N_3)$ modulo cyclic permutations. As pointed out earlier, the ${\mathbb Z}_2$ permutation of only two $N_i$'s is charge conjugation.
At the fixed point, the simple global symmetry $F$ of the theory is enhanced to the affine\footnote{\label{foot:affine}To avoid excessive notation, we use the same symbol for the affine algebra and its finite Lie subalgebra. The meaning in each case should be clear from the context.} symmetry $F_{2|k_F|}$. The chirality of the current algebra is determined by the sign of $k_F$. As all trace anomalies are negative, all the affine current algebras are left-moving. It is instructive to represent the IR fixed point as a directed triangle inscribed in a circle of circumference $N$, shown in Figure~\ref{triangleCFT}. Each side of the triangle represents a simple flavor symmetry factor. The arc length on the left is the rank and arc length on the right is the level. The supersymmetry preserving condition $n_i\geq 0$ is manifest in this picture.
Similar graphical structure has appeared in \cite{Gaiotto:2013gwa}, and for a good reason, see section \ref{CFT} and \cite{GGPjunctions}.

\begin{figure}[ht]
\centering
\includegraphics[scale=0.8]{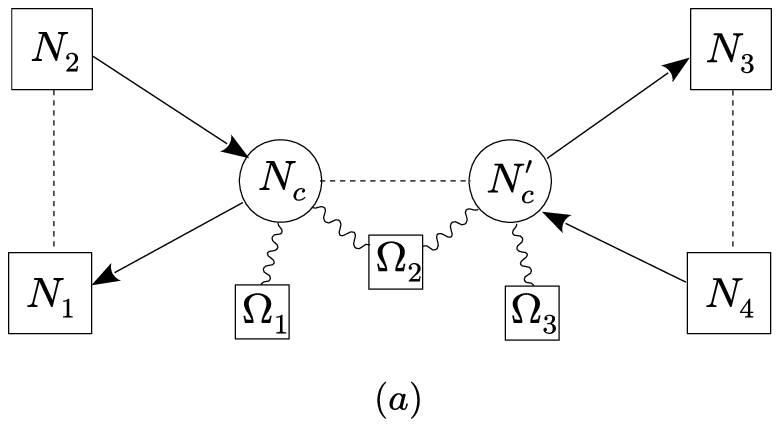}
\qquad
\includegraphics[scale=0.8]{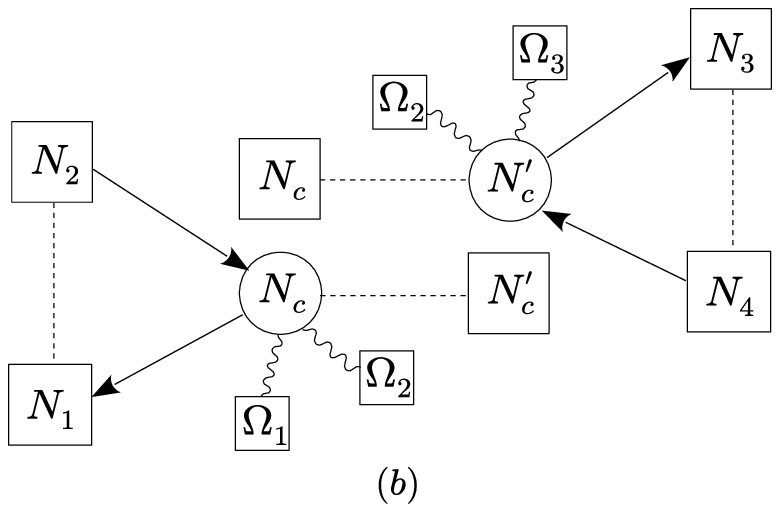}
\\
\includegraphics[scale=0.8]{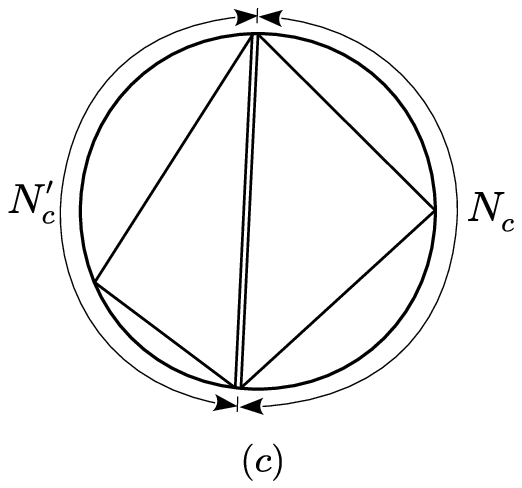}
\caption{A two-node quiver $(a)$ describes a $(0,2)$ theory obtained by ``gluing'' two copies of the elementary SQCD $(b)$. In the low-energy SCFT this gluing process is nicely represented by gluing inscribed triangles $(c)$.}
\label{2nodequiver}
\end{figure}

The picture generalizes to  quiver gauge theories as well. Consider the theory with two gauge nodes described in Figure~\ref{2nodequiver}. It can be thought of as ``gluing" two SQCD building blocks. This is schematically shown in the figure. The condition for anomaly cancellation requires that the gauge group rank of one theory is the flavor symmetry rank of the other. At low energies, this means that the two CFTs corresponding to component SQCDs can be glued to each other if they have level-rank dual affine symmetries.

\begin{figure}[ht]
\centering
\includegraphics[scale=0.9]{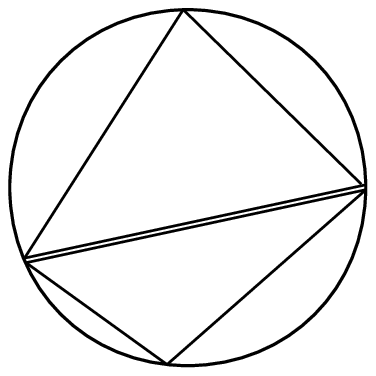}
\caption{The dual gauge theory that flows to the same fixed point as the one in the previous figure.}
\label{dual2node}
\end{figure}

The resulting CFT is elegantly understood as an inscribed quadrilateral obtained by gluing two triangles. Just like the SQCD $\CT_{N_1,N_2,N_3}$, the symmetries of the quiver theory are labeled by the edges of the quadrilateral. The rank is the arc length on the left and the level is the arc length on the right. A dual description of the same theory is shown in Figure~\ref{dual2node}. It corresponds to the other triangulation of the quadrilateral. The equivalence of different triangulations follows from~\cite{Gadde:2013lxa}.
This procedure can be repeated multiple times to obtain a  quiver with any number of gauge nodes. 
Although the gluing can be performed in any triality frame, in order to repeat the process outlined in Figure \ref{2nodequiver}, we need to perform a  triality operation on the ``boundary" gauge node so that the glued flavor symmetry is the one realized by $\Psi$ type Fermi fields.

%%%%%%%%%%%%%%%%%%%%%%%%%%%%%%%%%%%%%%%%%%%%%%%%%%%%%%%%%%%%%%%%%%%%%%%%%%%%%%%%%%%%%%%%%%%%%%%%%

\subsection{Symmetry enhancement in the Infra-Red}\label{enhance}

For certain special values of $N_i$'s, the theory $\CT_{N_1,N_2,N_3}$ has an enhanced global symmetry at low energy. This is a consequence of the triality and the following simple fact: the fundamental representation of $\UU (1)$ is the same as the determinant representation.

The most basic enhancement occurs for $N_2=N_3-N_1+2$, which implies $n_3=1$. The fundamental fermions $\Psi$ and determinant fermions $\Omega$ transform in the same way. The $\SU (2)_\Omega$ flavor symmetry acting on $\Omega$ and $\SU (N_3)$ flavor symmetry acting on $\Psi$ (and a $\UU (1)$) combine to $\SU (N_3+2)$ that rotates $\Psi$ and $\Omega$ into each other.

When we further specialize to $N_1=2$, we get $N_2=N_3$ and $n_2=1$. This means the gauge group in the frame $\CT_{N_3,N_1,N_2}$ is also $\UU (1)$. Similar to the above, in this frame $\SU (N_2)$ and $\SU (2)_\Omega$ combine to form $\SU (N_2+2)$.
Together, these two enhancements imply a much larger symmetry at the IR fixed point, namely $\SU (2) \times \SU (2N_3+2)$.

\begin{figure}[ht]
\centering
\includegraphics[scale=0.8]{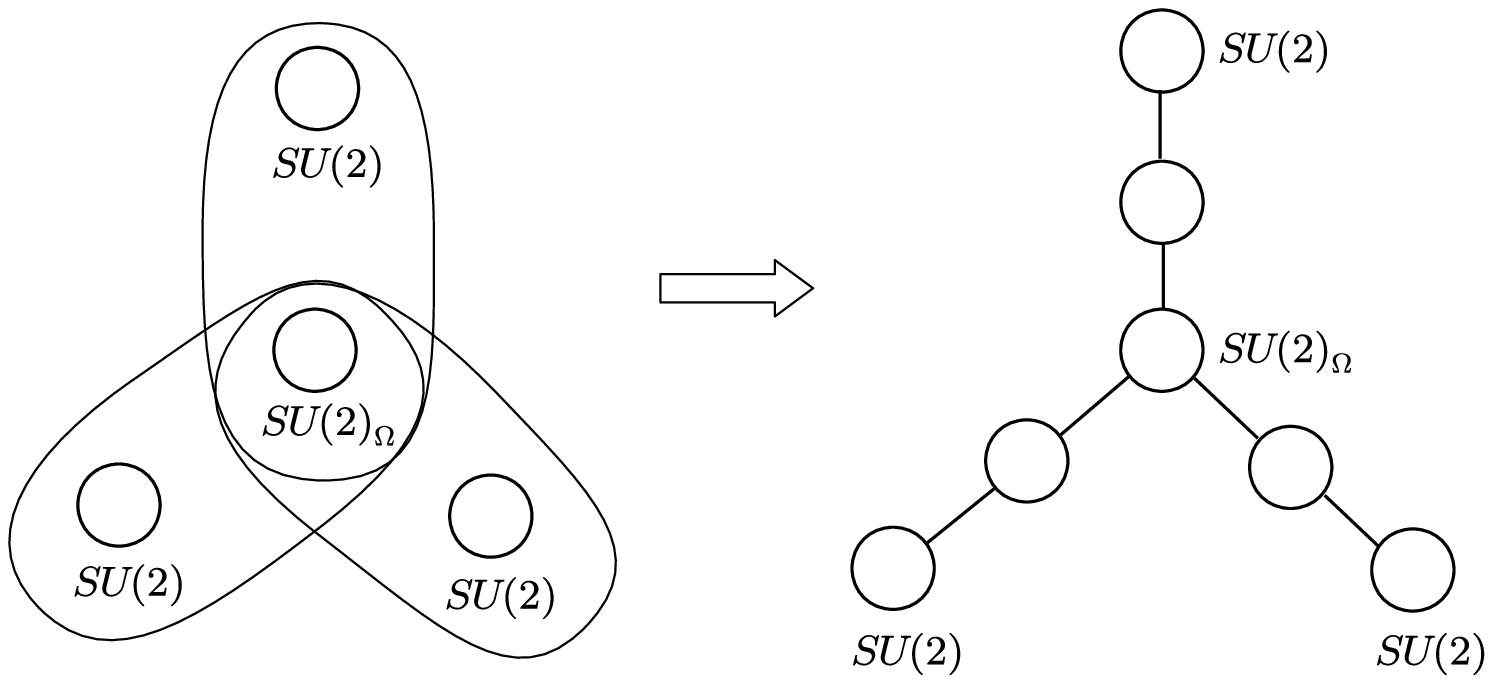}
\caption{Extended Dynkin diagram of $E_6$, viewed from three triality frames.}\label{E6enhancement}
\end{figure}

Now we see how to push this even further. We take $N_i=2$ for all $i=1,2,3$. The gauge group in all three triality frames is $\UU (1)$. Each of the three $\SU (2)$  factors combines with $\SU (2)_\Omega$ to form $\SU (4)$. This implies that the flavor symmetry is enhanced to $E_6$ in the infra-red. This is explained in Figure~\ref{E6enhancement}. It would be interesting to study symmetry enhancement for special cases of multi-node quiver theories.

%%%%%%%%%%%%%%%%%%%%%%%%%%%%%%%%%%%%%%%%%%%%%%%%%%%%%%%%%%%%%%%%%%%%%%%%%%%%%%%%%%%%%%%%%%%%%%

\section{Exact solution of the IR physics}
\label{CFT}

The SQCD that preserves supersymmetry in the infra-red generically flows to an interacting superconformal fixed point.  In this section we will study the SCFT at the fixed point  in great detail.  We identify the symmetry algebra in left-moving and right-moving sectors and use modular invariance of the partition function to pair their representations and derive the physical spectrum at the fixed point. The partition function can be used to identify the cohomology and to obtain the superconformal index (\emph{a.k.a.} flavored elliptic genus in the NS-NS sector).

From our analysis in section \ref{SQCD}, we know that the left-moving affine current algebra is
\be\label{basicaffine}
{\frH} :=\prod_{i=1}^3\, \SU (N_i)_{n_i}\times \UU (1)_{NN_i}
\ee
and the Sugawara central charge for $\frH$ is
\be\label{sugawarac}
c_{\frH} \; = \; \sum_{i=1}^3 \Big(\frac{n_i (N_i^2-1)}{n_i+N_i}+1\Big) \,.
\ee
Here we used the formula $c_{\frg}= k\, {\rm dim\,}\frg /(k+h^\vee_\frg)$ for an affine symmetry $\frg$ at level $k$,
with the dual Coxeter number $h^\vee_\frg$.
Remarkably, the central charge in \eqref{sugawarac} is exactly equal to the left-moving central charge $c_{\ttL}$ of the gauge theory, see eq. \eqref{gaugec}.  This implies that holomorphic stress tensor is  equal to the Sugawara stress tensor of the current algebra and hence the low-energy spectrum of $\CT_{N_1,N_2,N_3}$ consists of states of the type $|\psi\rangle_\ttL\otimes |\psi\rangle_\ttR$ where $|\psi\rangle_\ttL$ belongs to a module of the corresponding chiral WZW model. The spectrum simplifies immensely as there are only finitely many such modules labeled by the integrable representations $\lambda$ of the current algebra:
\be\label{hilbertfactor}
\CH= \bigoplus_{\lambda} \CH^{\lambda}_{\ttL\rm WZW} \otimes  \CH^\lambda_\ttR .
\ee
Here $\CH^\lambda_{\ttL\rm WZW}$ is the module of left-moving $\frH$ WZW model labeled by $\lambda$. In addition to constraining the left-moving spectrum, this decomposition also defines the right-moving subspace $\CH^{\lambda}_\ttR$ that forms a (not necessarily irreducible) representation of $\CN=2$ superconformal algebra.

\begin{figure}[ht]
\centering
\includegraphics[scale=1.0]{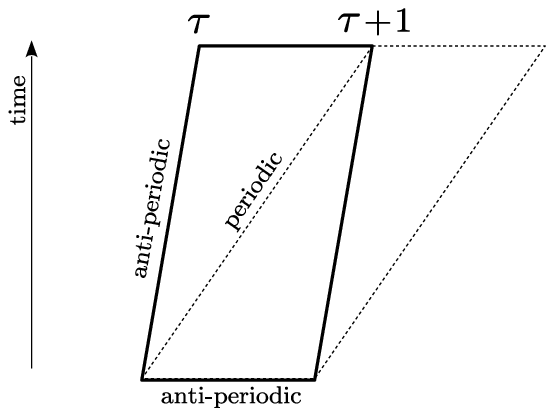}
\caption{The fermion boundary conditions used to define the partition function. They are invariant under the $S$-transformation.  However, the $T$-transformation changes the temporal boundary conditions and hence the partition function. It also destroys the $S$-invariance as the spatial and temporal boundary conditions no longer match.}
\label{ZBC}
\end{figure}

The partition function of the CFT in the NS-NS sector is defined as
$$
\qquad Z(\tau,\xi_i;\bar \tau, \bar \jacobi):={\Tr}_{\CH} \,  e^{2\pi i (\tau L_0+\sum_i \xi_i H_0^i-\bar \tau \bar L_0-\bar \jacobi\, \bar J_0)} \qquad q=e^{2\pi i \tau},\, y=e^{2\pi i \jacobi},\, z_i=e^{2\pi i \xi_i}.
$$
Here, $\tau$ is the complex structure of the torus, the chemical potential $\bar \jacobi$ couples to the R-symmetry in the right-moving sector and the chemical potentials $\xi_i$ couple to Cartan generators $H_0^i$ of the global symmetries $\frH$ in the left-moving sector. We will sometimes use the `exponentiated' variables $q,y$ and $z_i$  as defined above. Quantizing the theory in the NS-NS sector means we have to impose anti-periodic boundary conditions for the fermions along the spatial circle, and the absence of $(-1)^F$ insertion means that anti-periodic boundary conditions along the temporal circle are used.
This implies that the partition function is invariant only under the subgroup of the modular group, $\modsub\subset SL(2,{\mathbb Z})$,
generated by the elements $S$ and $T^2$, {\it cf.} Figure~\ref{ZBC}.
Although the classical argument suggests that the partition function should be strictly invariant under this modular subgroup, quantum mechanically this is not true. The theory has a modular anomaly which spoils the $S$-invariance. Instead we expect
\bea\label{zmod}
Z(-\frac{1}{\tau},\frac{\xi_i}{\tau};-\frac{1}{\bar \tau}, \frac{\bar \jacobi}{\tau})&=&\ph(\tau,\xi_i)\, \aph(\bar \tau,\bar \jacobi)\, Z(\tau,\xi_i;\bar \tau, \bar \jacobi)\nonumber\\
\ph(\tau,\xi_i)&=&\exp\Big(i\pi \frac{c_\ttL}{12}(\tau+\frac{1}{\tau})-i\pi \frac{\sum_{ij} 2k_{ij} \xi_i \xi_j}{\tau}\Big)\nonumber\\
\aph(\bar \tau,\bar \jacobi)&=&\exp\Big(-i\pi \frac{c_\ttR}{12}(\bar \tau+\frac{1}{\bar \tau})-i\pi \frac{c_\ttR}{3}\frac{\bar \jacobi^2}{\bar \tau}\Big).
\eea
where $k_{ij}$ is the mixed anomaly for symmetries $H_0^i$ and $H_0^j$. The modular anomaly factors $\phi$ and $\bar \phi$ come from holomorphic and anti-holomorphic sectors respectively.

From the structure of the Hilbert space \eqref{hilbertfactor}, the partition function of the gauge theory fixed point has the form
\be\label{ZCFT}
Z(\tau,\xi_i;\bar \tau, \bar \jacobi) \; = \; \sum_{\lambda} \chi_\lambda(\tau,\xi_i) K_\lambda(\bar \tau,\bar \jacobi),
\ee
where $\chi_\lambda(\tau,\xi_i)$ is the character of a module $\lambda$ of the affine algebra $ \frH$ and $K_\lambda(\bar \tau,\bar \jacobi)$ is defined abstractly as the $\CN=2$ character  over the right-moving module $\CH_{\ttR}^\lambda$. The affine characters are reviewed in appendix \ref{affinechar}. They transform under the modular $S$-transformation as,
\be
\chi_\lambda(-1/\tau,\xi_i/\tau) \; = \; \ph(\tau,\xi_i) \, S_{\lambda\, \mu} \, \chi_\mu(\tau,\xi_i)
\ee
where $\ph(\tau,\xi_i)$ is precisely the holomorphic modular anomaly defined in \eqref{zmod} and $S_{\lambda\,\mu}$ is a constant matrix known as the modular $S$-matrix. Because $(\tau,\xi_i)\to (\tau,-\xi_i)$ under $S^2$, it obeys $S^2=\CC$, where $\CC$ is the charge conjugation operator.
The charge conjugated $S$-matrix $\CC S$ is simply the complex conjugate $\bar S$, so we have $S\bar S=1$. The $S$-invariance of the partition function \eqref{ZCFT} implies that the right-moving characters should transform as
\be
K_\lambda(-1/\bar \tau,\bar \jacobi/\bar \tau) \; = \; \aph(\bar\tau,\bar\jacobi) \, \bar S_{\lambda\,\mu} K_\mu(\bar \tau,\bar \jacobi)
\ee
The presence of anti-holomorphic modular anomaly $\aph(\bar\tau,\bar\jacobi)$ simply means that $K_\lambda$ should be a character of the anti-holomorphic $\CN=2$ algebra with central charge $c_\ttR$. The $S$-matrix is identical to the one for the anti-holomorphic copy of  $\frH$. More importantly, it is also the one that transforms the level-rank dual characters. To be more explicit, let $\frH^t$ be the level-rank dual of $\frH$, or
\be
\frH^t \; = \; \prod_{i=1}^3\, \SU (n_i)_{N_i}\times \UU (1)_{Nn_i} \,.
\ee
Under level-rank duality, the equivalence classes of $\lambda$ modules of $\frH$ are mapped to equivalence classes of $\lambda^t$ modules of $\frH^t$. The map is nicely encoded in the level-rank duality matrix $L_{\lambda\,\lambda^t}$. This is reviewed in appendix \ref{affinechar}. Then, the $S$-matrix for dual characters obeys
\be\label{levelrankS}
\sum_{\mu\,  \mu^t}S_{\lambda\, \mu} L_{\mu\, \mu^t} S_{\mu^t \,\lambda^t}=L_{\lambda\, \lambda^t}.
\ee
To summarize, we have deduced the following properties of $K_\lambda(\bar \tau,\bar \jacobi)$ so far:
\begin{itemize}
\item It is a character of anti-holomorphic $\CN=2$  algebra with the central charge $c_\ttR$.
\item It transforms as a character of holomorphic $\frH^t$ symmetry under modular $S$-transformation.
\item It is a singlet under all affine symmetries (except, of course, the $\UU (1)$ R-symmetry of the $\CN=2$ algebra).
\end{itemize}
Searching for an object with these properties, a careful reader will notice
that characters of the supersymmetric Kazama-Suzuki coset $[\frG]/[\frH^t]$ fit the bill perfectly (we use square brackets to denote supersymmetry). Before committing to a specific numerator $[\frG]$ let us briefly review supersymmetric current algebras.

The supersymmetric extension $[\frg]_{k}$ of the current algebra $\frg$ at level $k$ is obtained by adding free adjoint fermions $\psi^a$ to the WZW model realizing $\frg$ at level $k_{\text{bos}}$. The current $J_a$ of the supersymmetric theory gets a contribution from the fermions too,
\be
J^a \; = \; J^a_{\text{bos}}-\frac{i}{k}\, f^a_{\,bc} \psi^b \psi^c \,.
\ee
Here $J_{\text{bos}}^a$ is the current of the bosonic WZW model.
Its level is determined from the OPE of free fermions, $k_{\text{bos}}=k-h_\frg^\vee$.
The Sugawara stress tensor is constructed out of the total currents $J^a$, and its central charge is
\be\label{superc}
c_{[\frg]_{k}} \; = \; \big(\frac{k_{\text{bos}}}{k}+\frac{1}{2}\big){\rm dim}\, \frg \,.
\ee
The first term here is the contribution from the bosonic WZW model and the second term is the fermion contribution.
The supersymmetric WZW model can be gauged in a supersymmetric way to produce the supercoset $[\frg]/[\frh]$.
For generic supercosets, the theory has $\CN=1$ supersymmetry, but it can be enhanced to $\CN=2$ supersymmetry
if the ordinary Lie coset $\frg/\frh$ is K\"ahler.
This is known as the Kazama-Suzuki construction \cite{KS1} of the $\CN=2$ supercoset (or, the KS supercoset for short).

In our theory, the denominator of the KS supercoset $[\frG]/[\frH^t]$ is the supersymmetric affine algebra,
\be
[\frH^t] \; = \; \prod_{i=1}^{3}\,[\UU(n_i)]_N \,.
\ee
The central charge is simply the difference $c_{[\frG]/[\frH^t]}=c_{[\frG]}-c_{[\frH^t]}$.
Comparing it to the right-moving central charge \eqref{gaugec} of the gauge theory, we get a remarkably simple result:
\be
c_{[\frG]} \; = \; N^2 \,.
\ee
When combined with the most obvious condition $\frH^t\subset \frG$, this suggests
\be
[\frG] = [\UU (N)]_N = [\UU (1)]_{N^2} \times [\SU (N)]_N \,.
\ee
Then, the ordinary Lie coset $\frG/\frH^t$ is K\"ahler and the supersymmetry is indeed enhanced to $\CN=2$, as desired.
The bosonic level of $ [\SU (N)]_N$ is $0$; it admits only the trivial representation.
So, effectively, the bosonic part of $[\frG]$ is simply $\UU (1)_{N^2}$. This Kazama-Suzuki coset also appeared in \cite{Gaiotto:2013gwa}.

The characters of the supercoset $C^{\Lambda, \upsilon}_{\lambda^t}$
are labeled by a representation $\Lambda$ of the (bosonic part of $[\frG]) \sim \UU (1)_{N^2}$,
a representation $\upsilon$ of the coset fermions under $\SO({\rm dim}\,\frG/\frH^t)$ at level 1,
and a representation $\lambda^t$ of the bosonic part of $[\frH]$.
For brevity, let us introduce $D:={\rm dim}\,\frG/\frH^t$.
The supercoset characters are defined by the branching rule \cite{KS1},
\be\label{cosetcharacter}
\chi_\Lambda^{\UU (1)_{N^2}} (\bar\tau,0) \, \chi_\upsilon^{\SO(D)_1}(\bar \tau, \bar \xi_l=\bar \jacobi)
\; = \; \sum_{\lambda^t} \, C^{\Lambda, \upsilon}_{\lambda^t}(\bar \tau, \bar \jacobi) \, \chi_{\lambda^t}^{\frH^t}
\left( \bar \tau, \bar \xi_k = \tfrac{\sum_{\alpha\in \Delta_+ }\alpha^k}{h_{\frG}^\vee}\bar \jacobi \right) \,.
\ee
Here, $\bar \xi_l$ and $\bar \xi_k$ are the chemical potentials for the Cartan generators of $\SO(D)$ and $\frH^t$, respectively,
while $\Delta_+$ is the set of positive roots in $\frG /\frH^t$.
Positive roots are the holomorphic directions in the coset.
The choice of  $\Delta_+$ is equivalent to a choice of complex structure, and the coset characters do depend on this choice.
In our case, there are only two choices of complex structure and they are related to each other by charge conjugation.
In section~\ref{quiver} we will see how to deal with multiple complex structures. For now, we can ignore this issue.

The label $\Lambda$ takes values in $\{ 1,\ldots, N^2 \}$ corresponding to $\UU (1)_{N^2}$ representations.
The group $\SO(D)_1$ admits four representations: singlet $({\bf 0})$, vector $({\bf v})$, spinor $({\bf s})$ and conjugate spinor $({\bf \bar s})$. It is easy to see that the characters of the representations
\be
\Lambda_0:=\bigoplus_{r=1}^{N} rN,\qquad \qquad \qquad \upsilon_0:={\bf 0}\oplus {\bf v}
\ee
are invariant under modular $S$-transformations modulo modular anomaly.
In fact, they are partition functions of free fermions with anti-periodic boundary conditions along spatial and temporal directions,
best expressed in terms of $q$-theta functions\footnote{defined as $\theta(a;q)=\prod_{i=0}^{\infty}(1-a q^i)(1-q^{i+1}/a)$.}
\be
\chi_{\Lambda_0}^{\UU (1)_{N^2}}(\bar\tau,0) \; = \; \theta(-\bar q^{\frac12};\bar q)\qquad \qquad  \chi_{\upsilon_0}^{SO(D)_1}(\bar \tau,\bar\xi_l =\bar \jacobi) \; = \; \theta(-\bar q^{\frac12}\bar y;\bar q)^D
\ee
Now we are ready to make a proposal for the character of the right-moving sector of the low-energy theory:
\be\label{CtoK}
K_\lambda(\bar \tau,\bar \jacobi) \; = \; \sum_{[\lambda^t]} L_{\lambda\,\lambda^t}C^{\Lambda_0, \upsilon_0}_{\lambda^t} (\bar \tau,\bar \jacobi)
\ee
Both the coset characters $C^{\Lambda_0, \upsilon_0}_{\lambda^t}$ and the level-rank duality matrix $L_{\lambda\, \lambda^t}$ are identical for  the representations $\lambda^t$ that belong to the same equivalence class $[\lambda^t]$, see appendix \ref{affinechar}. The sum has been performed only over the equivalence classes to avoid over-counting.
Thanks to the $S$-invariance of the left-hand side of \eqref{cosetcharacter}, this choice of $K_\lambda(\bar \tau,\bar \jacobi)$ has all the properties listed above. Substituting it into \eqref{ZCFT} gives us the partition function of the gauge theory fixed point\footnote{Instead of $\upsilon=\upsilon_0$, we could also choose another $S$-invariant combination $\upsilon={\bf s}\oplus {\bf \bar s}$. This choice corresponds to periodic boundary conditions for the right-moving fermions along the spatial circle and destroys $\modsub$ invariance of the partition function.}.
More generally, we propose\footnote{Instead of summing over the equivalence class $[\lambda^t]$ if we sum over all the representations $\lambda^t$, we simply get $n_1n_2 n_3$ copies of the same Hilbert space.}:
 \be\label{solution}
\boxed{ \CH= \bigoplus_{\lambda\,[\lambda^t]}\, L_{\lambda \,\lambda^t}\, \CH^{\lambda}_{\ttL\rm WZW} \otimes \CH^{\lambda^t}_{\ttR\rm KS}}
 \ee
where the right-moving module $\CH^{\lambda^t}_{\ttR\rm KS}$ is the module of the Kazama-Suzuki coset $[\frG]/[\frH^t]$ labeled by $\Lambda_0,\upsilon_0$ and $\lambda^t$ and the left-moving $\CH^\lambda_{\ttL {\rm WZW}}$ is the module of $\frH$ WZW model labeled by $\lambda$.
This is the complete Hilbert space of the low-energy sector of the $\CN=(0,2)$ SQCD in the NS-NS sector.

In the special case when all $N_i$'s are equal, the triality implies a ${\mathbb Z}_3$ symmetry that cyclically permutes flavor symmetry factors $\SU(N_i)_{n_i}\times \UU(1)_{NN_i}$. In this case, the theory also has a ${\mathbb Z}_2$ symmetry that is a combination of charge conjugation and the odd permutation. In section~\ref{casestudy} we demonstrate both of these symmetries in a concrete example.

%%%%%%%%%%%%%%%%%%%%%%%%%%%%%%%%%%%%%%%%%%%%%%%%%%%%%%%%%%%%%%%%%%%%%%%%%%%%%%%%%%%%%%%%%%%%%%%%%%%%%%%%

\subsection{$\bar\CQ$-cohomology and the index}\label{sec:index}

Among the two supercharges $\bar Q^+$ and $\bar Q^-$ of the $(0,2)$ gauge theory, we can pick either one to define cohomology. Here the superscripts $\pm$ stand for the R-symmetry charge.
Let us pick $\bar\CQ=\bar Q^+$.
By definition, the cohomology consists of states that are annihilated by $\bar\CQ$
modulo those of the form $\bar\CQ|\psi\rangle$ for some $|\psi\rangle$.
Important property of the $\bar\CQ$-cohomology is that it remains invariant under the RG-flow and,
therefore, can be computed using the low-energy solution of the gauge theory.
At low energies, we have $ \bar\CQ=\bar G_{-\frac12}^+$,
where $\bar G^+(\bar z)$ is one of the anti-holomorphic supercurrents in the NS sector.
It obeys the anti-commutation relation
\be
\{ \bar\CQ, \bar\CQ^\dagger\} \; = \; 2\bar L_0- \bar J_0
\ee
where hermitian conjugate operator $\bar\CQ^\dagger$ is simply the conformal supercharge $\bar G^-_{\frac12}$.
The harmonic representatives of the cohomology (\emph{i.e.} the states for which $\{\bar\CQ, \bar\CQ^\dagger\}=0$)
can only appear as primaries of the right-moving modules.
Denoting such modules by ${\hat \lambda}^t$ and their primaries by $|\psi\rangle_{\ttR}^{\hat \lambda^t}$,
the $\bar\CQ$-cohomology of the theory is given by
\be
H^* (\bar\CQ) \; = \; \sum_{\lambda\, [\hat \lambda^t]} L_{\lambda\, \hat \lambda^t}\, \CH_{\ttL\rm WZW}^{\lambda} \otimes |\psi\rangle_{\ttR}^{\hat \lambda^t}.
\ee
The ``cohomological partition function'' (= the Poincar\'e series of $H^* (\bar\CQ)$)
can be obtained from the full partition function \eqref{ZCFT} by setting $\bar y\to \bar y \,\bar q^{-\frac12}$
and taking the limit $\bar q \to 0$ (while keeping $q$ fixed).
Indeed, only the states with $2\bar L_0-\bar J_0=0$ contribute in this limit.
It would be interesting to compute the cohomology or the Poincar\'e polynomial
directly in the gauge theory, {\it i.e.} in the UV, and to compare with the result obtained from the IR SCFT.

The superconformal index in the NS-NS sector is defined as
\be
{\cal I} \; = \; {\rm Tr} (-1)^F q^{L_0} z_i^{H_0^i} e^{-\beta (\bar L_0-\frac12 \bar J_0)}.
\ee
The factor $(-1)^F$ ensures cancellation between bosonic and fermionic states with $\bar L_0-\frac12 \bar J_0\neq 0$
and makes the index independent of $\beta$.
This factor can be engineered in the partition function $Z(q, z_i; \bar q, \bar y)$ by the modular $T$-transformation $\tau \to \tau+1$:
the multiplicative shift in $q$ gives the extra factor $(-1)^{2(L_0-\bar L_0)}$ which is same as $(-1)^F$.
Alternatively, we can also understand this factor as a result of changing the temporal boundary condition from anti-periodic to periodic via a $T$-transformation. This is illustrated in Figure~\ref{ZBC}.

Finally, in order to obtain the index from the partition function we need to set $\bar y=\bar q^{-\frac12}$.
This limit gets rid of the fugacity $\bar y$ that couples to a non $\bar \CQ$-commuting charge. Moreover, $\bar q$ now couples to $\bar L_0-\frac12 \bar J_0$, just like $e^{-\beta}$. As expected, this limit is independent of $\bar q$. It should be noted that the partition function ceases to have any modular properties after taking this limit. This is expected because the superconformal index in the NS-NS sector does not have any modular properties. It is defined with anti-periodic boundary condition for fermions along the spatial circle and periodic boundary conditions along temporal circle.

In the next subsection we will study the case of theory $\CT_{222}$. We will compute its spectrum, partition function, and the superconformal index. We will also match the index with the one computed in the UV gauge theory.

%%%%%%%%%%%%%%%%%%%%%%%%%%%%%%%%%%%%%%%%%%%%%%%%%%%%%%%%%%%%%%%%%%%%%%%%%%%%%%%%%%%%

\subsection{A case study: $\CT_{222}$}
\label{casestudy}

The theory $\CT_{222}$ is the gauge theory in Figure~\ref{sqcd} with all $N_i=2$. The left-moving affine symmetry $\frH$ and its level-rank dual $\frH^t$ are
\be
\frH=\Big(\SU(2)_1 \times \UU(1)_6\Big)^3\, \qquad \qquad \frH^t=\Big(\UU(1)_3\Big)^3 \,.
\ee
It is convenient to split their representation labels $\lambda$ and $\lambda^t$ into the triples $(\lambda_1,\lambda_2,\lambda_3)$ and $(\lambda_1^t,\lambda_2^t,\lambda_3^t)$ respectively. The label $\lambda_i$ denotes representation with respect to the $i$-th copy of $\SU(2)_1\times \UU(1)_6$ and takes the values in $\lambda_i\in \{\cdot, \square\} \otimes \{-2,-1,0,1,2,3\}$.
Similarly, the label $\lambda_i^t$ denotes a representation of the $i$-th copy of $\UU(1)_3$ and takes the values in $\lambda_i^t\in \{-1,0,1\}$. The level-rank duality matrix $L_{\lambda\, \lambda_t}$ is given in Table~\ref{levelrankpair}.
\begin{table}[ht]
\begin{center}
\begin{tabular}{|c||c|c|c|c|c|c|}
\hline
 & $(\cdot,-2)$ & $(\cdot, 0)$ & $(\cdot, 2)$ & $(\square,1)$ & $(\square,3)$ & $(\square,-1)$ \\
\hline
\hline
-1& $1$ & & & $1$& & \\
\hline
0 & & $1$ & & & $1$ &\\
\hline
1 & & & $1$ &  & & $1$\\
\hline
\end{tabular}
\end{center}
\caption{Paring of level-rank dual modules for $\frH=\SU(2)_1\times \UU(1)_6$ and $\frH^t=\UU(1)_3$. The entries not shown are zero.}
\label{levelrankpair}
\end{table}

Earlier in this section, we described how to reduce the problem of finding the spectrum
of this gauge theory to the problem of finding the spectrum of the Kazama-Suzuki coset $[\frG]/[\frH^t]=[\UU(3)]_3/[\UU(1)_3]^3$. The central charge of this coset is $1$, which should ring a bell. Indeed, the coset at hand is the familiar $\CN=2$ minimal model with $c=1$. It has three primaries labeled by $(h,Q)\in \{(0,0),(\frac16,\frac13),(\frac16,-\frac13)\}$ where $h$ and $Q$ are eigenvalues of $\bar L_0$ and $\bar J_0$ respectively. The character of $\CN=2$ algebra at $c=1$ can be written explicitly\footnote{The Pochhammer symbol $(x,q)_\infty$ is defined as follows, $(x,q)_\infty=\prod_{i=1}^\infty(1-x q^i)$.}:
\be
\chi^{\mathcal{N}=2}_{(h,Q)}(\bar{q},\bar y) \; = \; (\bar q;\bar q)_\infty^{-1}\sum_{n\in\mathbb{Z}}{\bar q}^{\frac{3}{2}(n+Q)^2}\bar y^{n+Q} \,.
\ee
It takes the form of an affine $\UU(1)_3$ character. This is because the $\UU(1)$ R-symmetry is an affine symmetry at level $3$ and the stress tensor is just the Sugawara stress tensor for this symmetry.

The right-moving characters $K_\lambda(\bar \tau,\bar\jacobi)$  are found using \eqref{cosetcharacter} and \eqref{CtoK}.
Specializing to the present case,
\be
\theta(-\bar q^{1/2})\theta(-\bar y\bar q^{1/2})^{3}=
\sum_{\lambda^t}C^{\Lambda_0,\upsilon_0}_{\lambda_1^t,\lambda_2^t,\lambda_3^t}(\bar{q},\bar y)
\chi^{\UU(1)_{3}}_{\lambda_1^t}(\bar{q},\bar y^{\frac23})
\chi^{\UU(1)_{3}}_{\lambda_2^t}(\bar{q},1)
\chi^{\UU(1)_{3}}_{\lambda_3^t}(\bar{q},\bar y^{-\frac23}).
\ee
Solving this equation we get,
\begin{center}
 \begin{tabular}{|r|r|r|c}
    $\lambda_1^t$ &$\lambda_2^t$ &$\lambda_3^t$ & $C_{\lambda_1^t,\lambda_2^t,\lambda_3^t}^{\Lambda_0,\upsilon_0}$\\
   \hline
   0 & 0 & 0 &\multirow{3}{*}{$\chi^{\CN=2}_{(0,0)}$} \\
   1 & 1 & 1 &\\
  -1 & -1 & -1 &\\
  \hline
     1 & 0 & -1 & \multirow{3}{*}{$\chi^{\CN=2}_{(\frac16,\frac13)}$}\\
   -1 & 1 & 0 &\\
   0 & -1 & -1 &\\
  \hline
  -1 & 0 & 1 &  \multirow{3}{*}{$\chi^{\CN=2}_{(\frac16,-\frac13)}$} \\
   1 & -1 & 0 &\\
   0 & 1 & -1 &\\
  \hline
\end{tabular}
\end{center}
All other $C^{\Lambda_0,\upsilon_0}_{\lambda^t}$ are zero. The characters $K_\lambda$ are determined using level-rank duality matrix.
This gives the complete low-energy spectrum as a specific pairing of modules of the left-moving algebra $\frH$ and modules of the right-moving $\CN=2$ algebra. The partition function is computed using \eqref{ZCFT}:
\bea\label{222Z}
Z_{\CT_{222}}&=&
\chi^{\CN=2}_{(0,0)}(\bar \tau,\bar \jacobi)
\Big(
\Xi_{0,0,0}(\tau)
+\Xi_{1,1,1}(\tau)
+\Xi_{-1,-1,-1}(\tau)
\Big)\\
&+&
\chi^{\CN=2}_{(\frac16,\frac13)}(\bar \tau,\bar \jacobi)
\Big(
\Xi_{1,0,-1}(\tau)
+\Xi_{-1,1,0}(\tau)
+\Xi_{0,-1,1}(\tau)
\Big)\nonumber\\
&+&
\chi^{\CN=2}_{(\frac16,-\frac13)}(\bar \tau,\bar \jacobi)
\Big(
\Xi_{-1,0,1}(\tau)
+\Xi_{1,-1,0}(\tau)
+\Xi_{0,1,-1}(\tau)
\Big)\nonumber
\eea
The shorthand notation $\Xi_{a,b,c}(\tau)$ stands for $\Xi_{a,b,c}(\tau,\xi_1,\xi_2,\xi_3)$, defined as
\bea
\Xi_{a,b,c}(\tau, \xi_1, \xi_2,\xi_3)&:=&\Xi_a(\tau,\xi_1)\Xi_b(\tau,\xi_2)\Xi_c(\tau,\xi_3)\\
\Xi_{-1}(\tau,\xi)&:=&
\chi^{\SU(2)_1\times \UU(1)_6}_{(\square,-1)}(\tau,\xi)
+\chi^{\SU(2)_1\times \UU(1)_6}_{(\cdot,2)}(\tau,\xi)\nonumber\\
\Xi_0(\tau,\xi)&:=&
\chi^{\SU(2)_1\times \UU(1)_6}_{(\cdot,0)}(\tau,\xi)
+\chi^{\SU(2)_1\times \UU(1)_6}_{(\square,3)}(\tau,\xi)\nonumber\\
\Xi_{1}(\tau,\xi)&:=&
\chi^{\SU(2)_1\times \UU(1)_6}_{(\square,1)}(\tau,\xi)
+\chi^{\SU(2)_1\times \UU(1)_6}_{(\cdot,-2)}(\tau,\xi)\nonumber.
\eea
The left-moving characters in \eqref{222Z} are obviously invariant under the ${\mathbb Z}_3$ symmetry,
\emph{i.e.} the cyclic permutation of the three $\SU(2)_1\times \UU(1)_6$ factors.
This is consistent with the triality of the UV gauge theory description.
They also have manifest ${\mathbb Z}_2$ symmetry which is a combination of charge conjugation and odd permutation.

Remarkably, the left-moving  characters combine to form $\EE_6$ characters at level $1$. The $(\EE_6)_1$ admits only three modules,  the vacuum module $\bullet$, the  fundamental module ${\square}$ and the anti-fundamental module $\bar \square$. In terms of their characters, the partition function takes a much more compact form,
\be
Z_{\CT_{222}}=\chi^{\CN=2}_{(0,0)}(\bar \tau,\bar \jacobi)\chi^{(\EE_6)_1}_{\bullet}(\tau,\xi_i)+
\chi^{\CN=2}_{(\frac16,\frac13)}(\bar \tau,\bar \jacobi)\chi^{(\EE_6)_1}_{\square}(\tau,\xi_i)+
\chi^{\CN=2}_{(\frac16,-\frac13)}(\bar \tau,\bar \jacobi)\chi^{(\EE_6)_1}_{\bar \square}(\tau,\xi_i),
\ee
where the variables $\xi_i$ stand for collective ${\EE_6}$ fugacities.
Correspondingly, the three holomorphic modules of $(\EE_6)_1$ elegantly pair up with three anti-holomorphic modules of $\CN=2$  $c=1$ algebra to form the complete low-energy spectrum of the $\CT_{222}$ gauge theory.
In section \ref{enhance}, we advocated the enhancement of the global symmetry to $(\EE_6)_1$ using triality,
and the partition function provides us with a concrete evidence of this fact.
The first few terms in the expansion of $Z_{\CT_{222}}$ are
$$
Z_{\CT_{222}} \; = \; 1+({\bf 27}\,\bar y^\frac13 + {\bar {\bf 27}}\, \bar y^{-\frac13})q^\frac23\bar q^\frac16+\bar q+{\bf 78}\,q+ ({\bf 27}\,\bar y^\frac23 + {\bar {\bf 27}}\, \bar y^{-\frac23})q^\frac23\bar q^\frac23+{\bf 78}\, q\bar q + \ldots \,.
$$
They are contributions of the light states in the spectrum, \emph{i.e.} states with $L_0,\bar L_0\leq 1$.

The $\CN=2$ primaries $(0,0)$ and $(\frac16, \frac13)$ obey the BPS condition $\bar L_0-\frac12 \bar J_0=0$.
They form the $\bar \CQ$-cohomology of the theory,
\be
H^* (\bar \CQ) \; = \; \CH_{\ttL, (\EE_6)_1}^{\bullet}\otimes |\psi_{\CN=2}^{(0,0)} \rangle_\ttR \, \oplus\,
 \CH_{\ttL,(\EE_6)_1}^{\square} \otimes |\psi_{\CN=2}^{(\frac16,\frac13)} \rangle_\ttR \,,
\ee
where $|\psi_{\CN=2}^{(h,Q)} \rangle $ is the primary of the $(h,Q)$ module of the $\CN=2$ algebra. The superconformal index is computed from the partition function using a $T$-transformation and then setting $\bar y=\bar q^{-\frac12}$. Only the BPS modules $(0,0)$ and $(\frac16,\frac13)$ contribute in this limit. Their characters reduce to $+1$ and $-1$, respectively, so that
\be
\CI_{\CT_{222}} \; = \; \chi^{(\EE_6)_1}_{\bullet}-\chi^{(\EE_6)_1}_{\square} \,.
\ee
If our proposal for the low-energy physics of $\CN=(0,2)$ SQCD is true, this index has to agree with
the gauge theory computation in \cite{Gadde:2013lxa}, which indeed is the case to first ten
orders in the $q$-expansion.

The modular invariant pairing between the characters of holomorphic $(\EE_6)_1$ Affine algebra and antiholomorphic $\mathcal{N}=2$ superconformal algebra is less mysterious if we think of the $\CN=2$ superconformal algebra at $c=1$ as $\UU(1)_3$ affine algebra. It has a canonical pairing with $\SU(3)_1$ which follows from the conformal embedding
\be
\UU(3)_1 \supset \UU(1)_3 \times \SU(3)_1.
\ee
In turn, the $\SU(3)_1$ characters can be canonically paired with $(\EE_6)_1$ into $(\EE_8)_1$ characters,
\be
(\EE_8)_1\supset\SU(3)_1\times (\EE_6)_1.
\ee 
As this is a maximal rank embedding at level $1$, this is also a conformal embedding. The affine group $(\EE_8)_1$ admits only one integrable representation, naturally, its character is invariant under modular $S$-transformation.

%%%%%%%%%%%%%%%%%%%%%%%%%%%%%%%%%%%%%%%%%%%%%%%%%%%%%%%%%%%%%%%%%%%%%%%%%%%%%%%%%%%%%%%

\subsection{Generalization to quiver theories}
\label{quiver}

As explained in section \ref{SQCD}, the infra-red fixed points of a multi-node quiver theory can be described by a polygon inscribed in a circle. A quiver theory associated to an $m$-gon has the flavor symmetry $\prod_{i=1}^m\SU(N_i) \times \UU(1)_{(i)}$. From the anomalies we see that at low energies this symmetry is promoted to the left-moving affine symmetry
\be\label{quiverH}
\frH \; = \; \prod_{i=1}^m \, \SU(N_i)_{n_i} \times \UU(1)_{NN_i} \,,
\ee
where $N=\sum_i N_i/(m-1)$. Following our conventions so far, let us also define $n_i=N-N_i$. The Sugawara central charge of this affine symmetry agrees with the left-moving central charge $c_\ttL$ of the quiver theory. In this sense, $\frH$ is a natural generalization of the affine symmetry of the ``elementary" theory $\CT_{N_1,N_2,N_3}$. The left-moving states are modules of $\frH$. Motivated by the analysis of the  $\CT_{N_1,N_2,N_3}$, we guess that the right-moving sector is the Kazama-Suzuki coset $[\frG]/[\frH^t]$ where $[\frG]=[\UU(N)]_N$ and $\frH^t$ is the level-rank dual of $\frH$,
\be
\frH^t \; = \; \prod_{i=1}^m \, \SU(n_i)_{N_i} \times \UU(1)_{Nn_i} \,.
\ee
Happily, the central charge of this coset matches the right-moving central charge $c_\ttR$ of the quiver theory.
As before, the left-moving WZW modules and the right-moving KS coset modules have the same canonical paring
which results in $\modsub$-invariant partition function modulo modular anomaly.
We propose that the formula \eqref{solution} describes the spectrum of
the low-energy quiver gauge theory as well but with $\frH$ given in \eqref{quiverH}.

\begin{figure}[ht]
\centering
\includegraphics[scale=1.0]{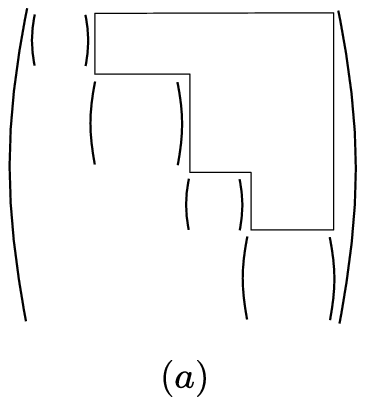}
\qquad
\includegraphics[scale=1.0]{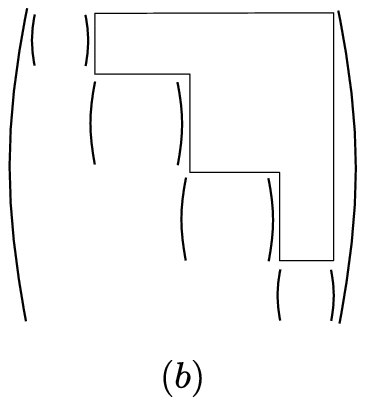}
\caption{The embedding of $\frH^t$ into $\frG$ determines the set of positive roots in the coset.
The positive roots are the holomorphic directions.
Two such embeddings shown here lead to different sets of positive roots and, therefore, to different complex structures.}
\label{complexKS}
\end{figure}

Generalized triality implies that the IR fixed point is labeled by the ordered $m$-tuple $\{ N_i \}_{i = 1, \ldots, m}$ modulo cyclic permutations.
In other words, for a given set $\{N_i\}$, the IR fixed point is labeled by elements of $S^m/{\mathbb Z}_m$.
This choice is in one-to-one correspondence with the choice of complex structure on the coset $\frG/\frH^t$.
This is described in Figure~\ref{complexKS}.

As a quick check of our proposal, consider a special case when all $N_i$'s are equal.
We expect that the fixed point has ${\mathbb Z}_m$ symmetry that cyclically permutes flavor nodes.
The proposal for the fixed point indeed has this property.
Moreover, we also get a ${\mathbb Z}_2$ symmetry that is a combination of charge conjugation and reflection among the flavor nodes.
Together they generate the dihedral group ${\mathbb D}_{2m}$.

As further evidence, the superconformal index computed from the SCFT description
can be checked against the UV computation. For simplicity, consider $m=4$.
In this case, the coset $[\frG]/[\frH^t]$ can be schematically written as
\bea
\frac{[\UU(N)]_N}{\prod_{i=1}^{4}[\UU(n_i)]_N}&=&
\frac{[\UU(N)]_N}{\prod_{i=1}^{2}[\UU(n_i)]_N \times [\UU(n_3+n_4)]_N}
\times
\Big(\frac{[\UU(N)]_N}{[\UU(n_3+n_4)]_N\times [\UU(n_1+n_2)]_N}\Big)^{-1} \nonumber\\
&\times&
\frac{[\UU(N)]_N}{ [\UU(n_1+n_2)]_N\prod_{i=3}^{4}[\UU(n_i)]_N}
\eea
We have used the shorthand notation $[\UU(n_i)]_{N}$ for $[\SU(n_i)]_N\times U(1)_{N n_i}$. From this equation we can write the right-moving character of the $m=4$ theory in terms of the right-moving characters of two elementary ``component" theories:
\be\label{IR-gluing}
K_{\lambda_1,\lambda_2,\lambda_3,\lambda_4}(\bar\tau,\bar\eta) \; = \; \sum_{\lambda\, \lambda^t}K_{\lambda_1,\lambda_2,\lambda}(\bar\tau,\bar\eta) \tilde L_{\lambda\,\lambda^t} K_{\lambda^t,\lambda_3,\lambda_4}(\bar\tau,\bar\eta)
\ee
where $\lambda$ is a module of $\SU(n_3+n_4)_{(n_1+n_2)}\times \UU(1)_{N(n_3+n_4)}$ and $\lambda^t$ its level-rank dual. The matrix $\tilde L$ is the generalized inverse of the level-rank duality matrix $L$, see appendix \ref{affinechar}.
As outlined in section \ref{sec:index}, after performing a $T$-transformation and taking the limit $\bar y=\bar q^{-\frac12}$, the right-moving characters become the ``structure constants" for the superconformal index expanded in terms of left-moving affine characters. One can verify that the same gluing equation is obeyed by the structure constants computed in the UV $\CN=(0,2)$ gauge theory. We demonstrate this explicitly in appendix~\ref{indexglue}. This implies that the agreement of the superconformal index for the quiver theories follows from that for the SQCD.

%%%%%%%%%%%%%%%%%%%%%%%%%%%%%%%%%%%%%%%%%%%%%%%%%%%%%%%%%%%%%%%%%%%%%%%%%%%%%%%%%%%%%%%%%%%%%%%%%%%%%%

\section{Meson spectroscopy}
\label{sec:massive}

It is believed that large $N_c$ limit of four-dimensional quantum chromodynamics (QCD$_4$)
is a weakly coupled theory of neutral massive particles, the mesons and ``glueballs''.
If we denote by $\sigma \sim \frac{\sqrt{N_c}}{g^2} \bar \psi \Gamma \psi$ and $S \sim \frac{1}{g^2} \Tr F_{\mu \nu}^2$
the corresponding irreducible gauge invariant operators\footnote{that have probability of order 1 to create meson and glueball states},
then mesons and glueball interactions to leading order are given by the tree graphs of an effective Lagrangian
\be
\CL_{\text{eff}} (\sigma, S)
\label{massiveLeff}
\ee
where all interaction terms scale as positive powers of $1/N_c$.
As a result, the $n$-point function of meson fields at large $N_c$ behaves as
\be
\langle \, T \, \underbrace{ \sigma \; \cdots \; \sigma }_{n} \, \rangle_{\text{conn}} \; \sim \; N_c^{1 - \frac{n}{2}}
\ee
The glueballs interact more weakly than mesons:
\be
\langle \, T \, \underbrace{ S \; \cdots \; S }_{n} \, \rangle_{\text{conn}} \; \sim \; N_c^{2 - n}
\ee
and the meson-glueball mixing is suppressed (because it requires quark/antiquark in meson to annihilate into gluons),
\be
\langle \, T \, \underbrace{ \sigma \; \cdots \; \sigma }_{n} \, \underbrace{ S \; \cdots \; S }_{m} \, \rangle_{\text{conn}}
\; \sim \; \frac{1}{N_c^{m+\frac{n}{2}-1}}
\ee
Also, because the interactions of mesons are too weak to cause bound states, mesons with ``exotic'' quantum numbers
(like $\bar \psi \psi \bar \psi \psi$) do not occur in the leading $1/N_c$ expansion.\\

\begin{figure}[htb] \centering
\includegraphics[width=4.0in]{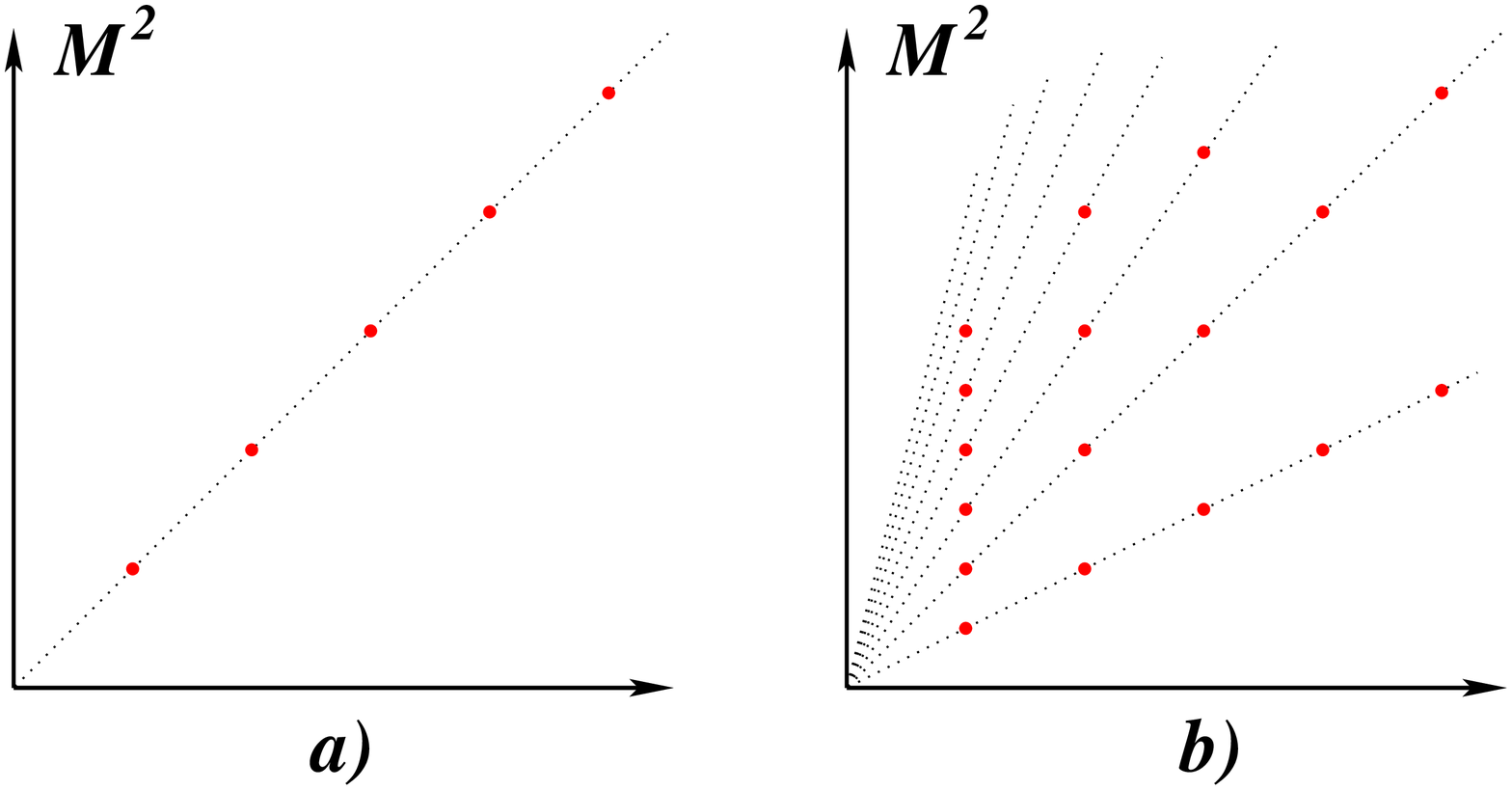}
\caption{\label{fig:SQCD2}
In the large $N_c$ limit the spectrum of non-supersymmetric ($\CN=0$) QCD$_2$ with $N_f=1$
has one asymptotically linear Regge trajectory $(a)$,
and infinitely many Regge trajectories (with integer slopes) for $N_f = N_c$ or one massive matter multiplet
in the adjoint representation $(b)$.}
\end{figure}

Motivated by these phenomenological facts, here wish to study the spectrum of mesons and glueballs
in two-dimensional QCD with $\CN=(0,2)$ supersymmetry at large as well as finite $N_c$.
Since confinement in 2d is generic, we are going to find that the effective $\CN=(0,2)$ theory
is described by a Lagrangian of the form \eqref{massiveLeff} with multiple copies of the $(0,2)$ chiral superfields $\sigma$ and $S$
that describe colorless states. (Even though 2d gauge field has no physical degrees of freedom, its superpartner $\lambda_-$ does.
For this reason we shall refer to the gaugino bilinear $\Tr \lambda_{-\alpha}^{\beta} \lambda^{\alpha}_{-\beta}$
as the glueball $(0,2)$ chiral superfield $S$.)

As usual, the large $N_c$ limit simplifies the dynamics by removing the interaction between confined states.
In this limit the closed stings are free since
\be
g_s \sim \frac{1}{N_c}
\ee
while the 't Hooft coupling\footnote{We do not use the standard notation $\lambda$ to avoid
confusion with the gluino fields.} $g^2 N_c$ sets the ``string tension''
\be
\frac{1}{\alpha'} = g^2 N_c
\label{tHooftcoupling}
\ee
However, the number of colorless states also grows rapidly with $N_c$. The reason is clear from the relations \eqref{trialityframe}
which, among other things, imply that one can not take the limit $N_c \to \infty$ without letting the ranks of at least {\it two}
flavor symmetries scale with $N_c$. This is where our definition \eqref{nudef} of the ratios $\nu_i$ becomes very handy,
which can be kept fixed along with the 't Hooft coupling \eqref{tHooftcoupling}
and the parameter $m^2 N_c$ (whose role will become clear momentarily).
In other words, when taking $N_c \to \infty$ we shall consider the Veneziano limit {\it a la} \cite{Veneziano:1976wm}:
\be
N_c \to \infty \,, \quad
g^2 N_c = \text{fixed} \,, \quad
m^2 N_c = \text{fixed} \,, \quad
\nu_i = \text{fixed}
\label{Venezianolim}
\ee

We also use the light-cone quantization, which makes all unphysical degrees of freedom manifestly non-dynamical.
In particular, there is no gluon self-interaction in light-cone gauge in 1+1 dimensions:
\be
A_- = A^+ = 0
\label{lcgauge-usual}
\ee
where the theory reduces to quantum mechanics with $x^+$ as the ``time'' direction\footnote{Other
light-cone conventions include:
$$
g^{+-} = g^{-+} = 1
\qquad
\partial_{\pm} = \frac{\partial}{\partial x^{\pm}}
\qquad \gamma^0 = \sigma_2 \qquad \gamma^1 = i \sigma_1
$$
},
\be
x^{\pm} = \frac{1}{\sqrt{2}} (x^0 \pm x^1)
\ee
Similarly, we can choose $x^-$ to be the time variable with the gauge condition $A_+ = 0$.
With either of these two choices, summarized in Table~\ref{tab:aboptions},
there are no dynamical gluons and, therefore, no need to introduce Fadeev-Popov ghosts.
This simplifies the analysis dramatically.
Quantization on constant $x^+$ surfaces gives the momentum operators $P^+ = T^{++}$ and $P^- = T^{+-}$.
The main goal then is to solve the eigenvalue problem
\be
M^2 \vert \varphi \rangle = 2 P^+ P^- \vert \varphi \rangle
\label{MPP}
\ee
in the basis of colorless states.
We are going to find masses of states in the form
\be
M^2 \; = \; g^2 N_c \, \CF \left( \nu_i , \tfrac{m^2}{g^2} \right)
\ee
It would be interesting to perform a more systematic study of the dependence on the dimensionless
parameters $\frac{m^2}{g^2}$ and $\nu_i$, and, in particular, to see if there are any phase transitions
similar to the Berezinskii-Kosterlitz-Thouless (BKT) type conformal phase transition in QCD$_4$
at a critical value of $\frac{N_f}{N_c} \approx 4$ found via holographic dual \cite{Kaplan:2009kr,Jarvinen:2011qe}.
In fact, it would be interesting to approach our $(0,2)$ SQCD via gauge/gravity duality as well.

In general, the light-cone quantization describes the Hilbert space seen by an observer moving with the speed of light to the right,
which can see only massive particles and right-moving massless particles, but misses left-moving massless particles \cite{Kutasov:1994xq}.
(One does not miss, though, any massive bound states of these massless constituents.)
This is not a problem for us since, first of all, here we are interested in massive states and, moreover,
because we already gave a detailed account of all massless states in section~\ref{CFT}.

\begin{table}[htb]
\centering
\renewcommand{\arraystretch}{1.3}
\begin{tabular}{|@{\quad}c@{\quad}|@{\quad}c@{\quad}| }
\hline  {\bf Option a)}: \quad $x^+ = $ ``time'' & {\bf Option b)}: \quad $x^- =$ ``time'' \\
\hline
\hline integrate out $A_+$, $\psi_+$, $\rho_+$ & integrate out $A_-$, $\psi_-$, $\lambda_-$, $\gamma_-$ \\
\hline adjoint $\lambda_-$ &  \\
(anti-)fundamental: & (anti-)fundamental: \\
$\phi$, $p$, $\psi_-$ & $\phi$, $\psi_+$ $p$, $\rho_+$ \\
neutral $\gamma_-$ &  \\
\hline
\end{tabular}
\caption{$\SU(N_c)$ representations of SQCD partons with different choices of the light-cone time.}
\label{tab:aboptions}
\end{table}

In our $(0,2)$ SQCD the spectrum of left-moving and right-moving fields is rather different,
{\it cf.} Figure~\ref{fig:qcd-su}.
Therefore, the theory will look differently depending on the choice of the light-cone time
and it is rather non-trivial that both choices must lead to the {\it same} massive spectrum:

\begin{itemize}

\item[$a)$] In one option, we see the modes of $\phi$, $p$, $\lambda_-$, $\psi_-$, $\gamma_-$ (and integrate out $A_+$, $\psi_+$, $\rho_+$).
All fields except $\lambda_-$ are in the fundamental representation of $\SU(N_c)$, so we get closed strings from $\lambda_-$ bits
and other partons in bifundamental representations $({\bf \bar N_c} , {\bf N_i})$ or $({\bf N_i}, {\bf \bar N_j})$ for which $N_{i,j} \sim N_c$.
The remaining partons whose color-flavor content does not scale as $N_c^2$ become open string bits.
Note, $\gamma_-$ does not couple directly to the gauge field $A$, while $\psi_-$ couples to the gauge field {\it only}.

\item[$b)$] In the other option, we see the modes of $\phi$, $\psi_+$, $p$, $\rho_+$
(and integrate out $A_-$, $\psi_-$, $\lambda_-$, $\gamma_-$).
All of these fields are in the fundamental representation of $\SU(N_c)$, so we get lots of mesons,
which become closed string states if the color-flavor content scales as $N_c^2$ and open string states otherwise.

\end{itemize}

\begin{figure}[htb] \centering
\includegraphics[width=5.0in]{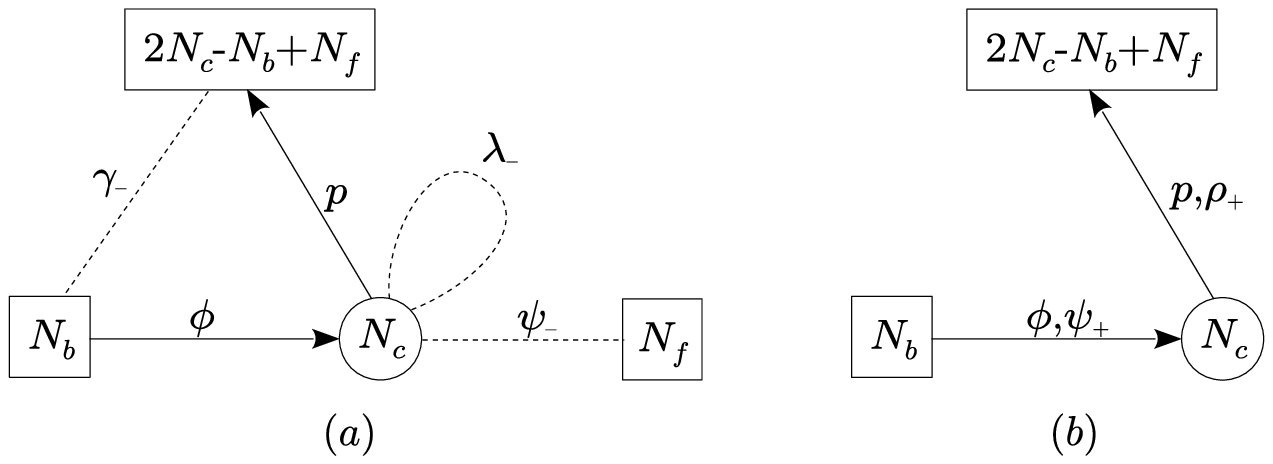}
\caption{\label{fig:qcd-lc}
Propagating fields in the light-cone approach $(a)$ with $x^+$ as ``time'' and $(b)$ with $x^-$ as ``time''.}
\end{figure}

\noindent
We summarize these two choices in Table~\ref{tab:aboptions} and also in a graphical form in Figure~\ref{fig:qcd-lc};
the latter is obtained from Figure~\ref{fig:qcd-su} by omitting non-propagating degrees of freedom and writing
$\CN = (0,2)$ supermultiplets in components.

In order to carry this out in practice, we write the Lagrangian of $\CN=(0,2)$ SQCD.
After integrating out the auxiliary fields, it takes the form
\be
\CL = \CL_{\text{kin}} + A_+ J^+ + A_- J^- + \CL_{\text{int}}
\label{Ltot}
\ee
where $\CL_{\text{kin}}$ contains the standard kinetic terms of all the component fields,
\be
J^{+\alpha}_{\beta} =
\frac{1}{g^2} \bar \lambda_{- \gamma}^\alpha \lambda_{- \beta}^{\gamma}
+ \frac{1}{g^2} \lambda_{- \gamma}^\alpha \bar \lambda_{- \beta}^{\gamma}
+ \bar \psi_{- i}^\alpha \psi_{- \beta}^i
- i \bar \phi^{s}_\beta \partial_- \phi^{\alpha}_s
+ i \bar p_{ a}^\alpha \partial_- p_{\beta}^a
\label{Jplus}
\ee
\be
J_{ \beta}^{-\alpha} = - \bar \psi_{+\beta}^s  \psi_{+ s}^\alpha
+ \bar \rho_{+a}^\alpha \rho_{+ \beta}^a
- i \bar \phi^{s}_\beta \partial_+ \phi^{\alpha}_s
+ i \bar p_{a}^\alpha \partial_+ p_{\beta}^a
\label{Jminus}
\ee
are the left-moving and right-moving $\SU(N_c)$ gauge currents, and
\bea
\CL_{\text{int}} & = &
\frac{g^2}{2} \left( \phi^{\dagger} \phi - p p^{\dagger} \right)^2 +  m^2 |\phi p|^2 \nonumber \\
& - & \sqrt{2}i \bar \phi \lambda_- \psi_+ + \sqrt{2}i \bar p \lambda_- \rho_+
+ m \gamma_{-} \phi \rho_{+} + m \gamma_{-} \psi_{+} p \label{Lint} \\
& + & \sqrt{2} i \phi \bar \psi_+ \bar \lambda_- - \sqrt{2} i p \bar \rho_+ \bar \lambda_-
+ m \bar \rho_+ \bar \phi \bar \gamma_- + m \bar \psi_+ \bar \gamma_- \bar p
\nonumber
\eea
contains the remaining interaction terms (with color and flavor indices suppressed).

Starting with the Lagrangian \eqref{Ltot}--\eqref{Lint}
and integrating out $A_+$, $\psi_+$ and $\rho_+$ --- as in the scenario $a)$ --- we get:
\bea
\CL_{\text{int}}^{(a)} = \frac{g^2}{2} \left( \phi^{\dagger} \phi - p p^{\dagger} \right)^2 +  m^2 |\phi p|^2
& + & i \left( \sqrt{2} \bar \phi \lambda_- + i m p \gamma_- \right) \frac{1}{\partial_-} \left( \sqrt{2} \bar \lambda_- \phi - i m \bar \gamma_- \bar p \right) \label{Linta} \\
+ g^2 J^+ \frac{1}{\partial_-^2} J^+ & + & i \left( \sqrt{2} i \bar p \lambda_- + m \gamma_- \phi \right) \frac{1}{\partial_-} \left( - \sqrt{2} i \bar \lambda_- p + m \bar \phi \bar \gamma_- \right)
\nonumber
\eea
On the other hand, integrating out $A_-$, $\lambda_-$, $\gamma_-$, and $\psi_-$ --- as in the scenario $b)$ --- we would find:
\bea
\CL_{\text{int}}^{(b)} = \frac{g^2}{2} \left( \phi^{\dagger} \phi - p p^{\dagger} \right)^2 + m^2 |\phi p|^2
& + & 2 i g^2 \left( \bar \phi \psi_+ - \bar p \rho_+ \right) \frac{1}{\partial_+} \left( \phi \bar \psi_+ - p \bar \rho_+ \right) \label{Lintb} \\
+ g^2 J^- \frac{1}{\partial_+^2} J^-
& + & m^2 \left( \phi \rho_+ + \psi_+ p \right) \frac{1}{\partial_+} \left( \bar \rho_+ \bar \phi + \bar p \bar \psi_+ \right)
\nonumber
\eea
In what follows we make a more traditional choice of the light-cone gauge \eqref{lcgauge-usual},
{\it i.e.} option $a)$ in Table~\ref{tab:aboptions}.
Since only left-moving fermions remain after integrating out $\psi_+$ and $\rho_+$,
in the rest of our discussion we shall omit the label ``$-$'' to avoid clutter.
It would be interesting to repeat similar analysis with the other choice, {\it i.e.} option $b)$ in Table~\ref{tab:aboptions},
that, of course, should lead to the same massive spectrum, but not necessarily the same massless spectrum.

Note, if we choose a gauge fixing condition we need to shift the supercharge ${\bar{Q}}$
by a certain gauge transformation $\delta_{\Lambda}$ with the generator $\Lambda$:
\begin{equation}
{\bar{\CQ}} = {\bar{Q}}+\delta_{\Lambda}
\end{equation}
\begin{equation}
 \delta_{\Lambda} A= d \Lambda -i[\Lambda,A]
\end{equation}
In particular, for the condition \eqref{lcgauge-usual} we have
\begin{equation}
\Lambda = \frac{2ig}{\d_-}\lambda
\end{equation}
One can show that the corresponding Noether charge is given by
\begin{equation}
{\bar{\CQ}} = 2g\int {J^+}^\beta_\alpha \frac{1}{\d_-}\bar{\lambda}^\alpha_\beta+\sqrt{2}\int \bar{\gamma}^a_s\bar{J}^s_a (\bar \phi, \bar p)
\label{QJJ}
\end{equation}
where $J^+_{\alpha \beta}$ is the longitudinal momentum current \eqref{Jplus}
and $J^{a}_s (\Phi,P)$ is the $\CN = (0,2)$ superpotential \eqref{JPPhi}.

Next, we need to consider quantization of scalar and spinor fields at fixed light-cone time $x^+ = 0$.
Thus, for a complex scalar (such as $\phi$ or $p$) we have:
\bea
\phi & = & \frac{1}{\sqrt{2\pi}} \int_0^{+ \infty} \frac{d k^+}{\sqrt{2 k^+}}
\left( \phi (k^+) e^{- ik^+ x^-} + \bar \phi^{\dagger} (k^+) e^{i k^+ x^-} \right) \\
\bar \phi & = & \frac{1}{\sqrt{2\pi}} \int_0^{+ \infty} \frac{d k^+}{\sqrt{2 k^+}}
\left( \bar \phi (k^+) e^{- i k^+ x^-} + \phi^{\dagger} (k^+) e^{ik^+ x^-} \right)
\nonumber
\eea
where the creation and annihilation operators obey the standard commutation relations:
\be
[ \phi (k^+) , \phi^{\dagger} (\tilde k^+)] \; = \; \delta (k^+ - \tilde k^+) \; = \; [ \bar \phi (k^+) , \bar \phi^{\dagger} (\tilde k^+)]
\ee
Note, the modes $\phi (k^+)$ and $\bar \phi^{\dagger} (k^+)$ transform in the same representation
of the gauge and global symmetry groups, whereas $\phi^{\dagger} (k^+)$ and $\bar \phi (k^+)$
transform in the conjugate representation.
For instance, for the field $\phi$ in our $\CN = (0,2)$ SQCD it means that the modes
$\phi (k^+)$ and $\bar \phi^{\dagger} (k^+)$ transform in the bifundamental representation $({\bf N_c}, {\bf \bar N_b})$
of the symmetry group $\SU(N_c) \times \SU(N_b)$, {\it cf.} \eqref{reps},
whereas $\phi^{\dagger} (k^+)$ and $\bar \phi (k^+)$ transform as $({\bf \bar N_c}, {\bf N_b})$.
Paying attention to such facts will be important in constructing color (and flavor) singlets in what follows.

Similarly, for a complex spinor field (such as $\psi$, $\gamma$, or $\lambda$) we have:
\bea
\psi & = & \frac{1}{2\sqrt{\pi}} \int_0^{+ \infty} d k^+
\left( \psi (k^+) e^{-ik^+ x^-} + \bar \psi^{\dagger} (k^+) e^{i k^+ x^-} \right) \label{psimodes} \\
\bar \psi & = & \frac{1}{2\sqrt{\pi}} \int_0^{+ \infty} d k^+
\left( \bar \psi (k^+) e^{-i k^+ x^-} + \psi^{\dagger} (k^+) e^{ik^+ x^-} \right)
\nonumber
\eea
where $\psi (k^+)$ and $\bar \psi^{\dagger} (k^+)$ transform in the same representation,
while $\psi^{\dagger} (k^+)$ and $\bar \psi (k^+)$ transform in the conjugate representation, and obey
\be
\{ \psi (k^+) , \psi^{\dagger} (\tilde k^+) \} \; = \; \delta (k^+ - \tilde k^+) \; = \; \{ \bar \psi (k^+) , \bar \psi^{\dagger} (\tilde k^+) \}
\ee
Let us remind that when writing explicitly the quantum operator ${\bar{\CQ}}$ in terms of these modes one needs to do normal ordering.

Once we introduced the mode expansion of all the fields, the physical states can be constructed as
$\SU(N_c)$ singlets of the form
\be
\vert \varphi \rangle \; \sim \; \frac{1}{N_c^{r/2}\sqrt{s}} \CO (k_1^+) \ldots \CO (k_r^+) \vert 0 \rangle
\label{lcphysstates}
\ee
where $\vert 0 \rangle$ is the Fock vacuum and each ``string bit'' $\CO (k_i^+)$
stands for creation operator of a boson or fermion carrying longitudinal momentum $k_i^+$.
In addition to the standard normalization of the $r$-parton state,
we have a symmetry factor $1/\sqrt{s}$, where $s$ is the number of cyclic permutations that give the same state.
Some states vanish due to fermionic statistics, {\it e.g.}
\be
\Tr [\lambda^{\dagger} (k^+) \lambda^{\dagger} (k^+) ] \vert 0 \rangle = 0
\ee

Since all physical states \eqref{lcphysstates} are already eigenstates of the operator $P^+$:
\be
P^+ = \sum_{i=1}^r k_i^+
\label{lcPplustot}
\ee
the problem of computing the mass spectrum \eqref{MPP} boils down to diagonalizing the operator $P^-$,
which is our next and final step.
Namely, the standard practice in analyzing the mass spectrum of 2d gauge theories is to discretize
the values of $k^+$ by compactifying the ``space'' direction $x^-$.
Indeed, with the periodic boundary conditions for both bosons and fermions
\be
\phi (x^-) = \phi \left(x^- + 2 \pi \frac{K}{P^+}\right) \,, \qquad
\psi (x^-) = \psi \left(x^- + 2 \pi \frac{K}{P^+}\right)
\ee
the partons in \eqref{lcphysstates} carry integer quanta of the longitudinal momentum
\be
k_i^+ = \frac{n_i P^+}{K} \qquad , \qquad n_i =1, \ldots, K
\ee
so that \eqref{lcPplustot} becomes
\be
\sum_{i=1}^r n_i = K
\label{lcKplustot}
\ee
The positive integer $K$ is called the {\it harmonic resolution},
and taking $K \to \infty$ while keeping $P^+$ fixed corresponds to the continuum limit.
For finite value of the harmonic resolution $K$, the physical states are labeled
by partition of $K$ into integers $1 \le n_i \le K$ and all integrals $\int d k^+$
are replaced by the corresponding sums $\sum_n$.

As a useful warm-up and to illustrate how this works,
let us consider a free fermion $\psi$ in a bifundamental representation of $\SU(N_c) \times \SU(N_f)$,
which is basically one of our ingredients in Figure~\ref{fig:qcd-lc}$a$.
As we explained around \eqref{psimodes}, quantization of $\psi$ leads to creation and annihilation operators
$\psi (n)$, $\bar \psi^{\dagger} (n)$, $\psi^{\dagger} (n)$ and $\bar \psi (n)$
that in the present DLCQ approach are labeled by an integer $1 \le n \le K$.
Moreover, $\psi (n)$ and $\bar \psi^{\dagger} (n)$ transform as $({\bf \bar N_c}, {\bf N_f})$,
whereas $\psi^{\dagger} (n)$ and $\bar \psi (n)$ transform as $({\bf N_c}, {\bf \bar N_f})$.
Relevant to the construction of physical states \eqref{lcphysstates} are the creation operators
$\bar \psi^{\dagger} (n)$ and $\psi^{\dagger} (n)$ that we can summarize in a quiver diagram
\begin{equation}
 \xymatrix{
      N_c ~\bullet \ar@/^/[r]^{\bar \psi^{\dagger}}
         &
         \bullet~ N_f \ar@/^/[l]^{\psi^{\dagger}}}
  \label{freefermquiv}
\end{equation}
Acting with these creation operations on the Fock vacuum $\vert 0 \rangle$ gives a basis of physical states \eqref{lcphysstates}
labeled by partitions of $K$.
For example, for $K=1$ we have only two states,
$(\psi^\dagger)_i^{\alpha}(1) \, |0\rangle$ and $(\bar \psi^\dagger)^i_{\alpha} (1) \, |0\rangle$, {\it etc.}

To make our exercise a little more interesting and to anticipate what is going to come next,
let us consider a subset of states that are complete singlets under the symmetry group $\SU(N_c) \times \SU(N_f)$.
Clearly, there are no such states for $K=1$ and only one state for $K=2$:
\be
(\psi^\dagger)_i^{\alpha}(1) \; (\bar \psi^\dagger)^i_{\alpha} (1) \; |0\rangle
\ee
In general, such states correspond to closed loops in the quiver diagram, in the present case \eqref{freefermquiv}.
Indeed, for $K=3$ we find two singlets labeled by partitions $(n_1,n_2) = (1,2)$ and $(2,1)$,
\be
(\psi^\dagger)_i^{\alpha}(1) \; (\bar \psi^\dagger)^i_{\alpha} (2) \; |0\rangle
\qquad , \qquad
(\psi^\dagger)_i^{\alpha}(2) \; (\bar \psi^\dagger)^i_{\alpha} (1) \; |0\rangle
\ee
whereas for $K=4$ there are four possible ways to make complete $\SU(N_c) \times \SU(N_f)$ singlets:
\be
\begin{array}{c}
(\psi^\dagger)_i^{\alpha}(1) \; (\bar \psi^\dagger)^i_{\alpha} (3) \; |0\rangle \\
(\psi^\dagger)_i^{\alpha}(2) \; (\bar \psi^\dagger)^i_{\alpha} (2) \; |0\rangle \\
(\psi^\dagger)_i^{\alpha}(3) \; (\bar \psi^\dagger)^i_{\alpha} (1) \; |0\rangle \\
(\psi^\dagger)_i^{\alpha}(1) \; (\bar \psi^\dagger)^i_{\beta} (1) \; (\psi^\dagger)_j^{\beta}(1) \; (\bar \psi^\dagger)^j_{\alpha} (1) \; |0\rangle \\
\end{array}
\ee
Note, not included here is
$(\psi^\dagger)_i^{\alpha}(1) \; (\bar \psi^\dagger)^i_{\alpha} (1) \; (\psi^\dagger)_j^{\beta}(1) \; (\bar \psi^\dagger)^j_{\beta} (1) \; |0\rangle$
since it is not a single-trace state.
Continuing in this fashion we find a total of six singlet single-trace states at $K=5$ and so on:
\be
 \begin{tabular}{|c|c|}
 \hline
  $K$ & \# of singlet states \\
  \hline
  \hline
  1 & 0 \\
  2 & 1 \\
  3 & 2 \\
  4 & 4 \\
  5 & 6 \\
  6 & 12 \\
  7 & 18 \\
  8 & 34 \\
   \hline
 \end{tabular}
 \label{KKKfermion}
\ee

Now we have all the necessary tools to study the spectrum of massive states in $\CN=(0,2)$ SQCD
for various values of the harmonic resolution $K$ and in the continuum limit $K \to \infty$.
For each value of $K$, the problem is to diagonalize the (mass)$^2$ operator \eqref{MPP} or, equivalently, $P^-$
on the states \eqref{lcphysstates} labeled by partitions of $K$.
In the past, this was done for 2d gauge theories with $\CN=(1,1)$ supersymmetry
in \cite{Matsumura:1995kw,Antonuccio:1998kz,Antonuccio:1998jg,Armoni:1998kv},
for $\CN=(2,2)$ supersymmetry in \cite{Antonuccio:1998mq},
and even for $\CN=(8,8)$ supersymmetry in \cite{Antonuccio:1998tm},
but never in enough details for models with $\CN=(0,2)$ supersymmetry.

\begin{figure}[ht]
\centering
 \includegraphics[scale=0.7]{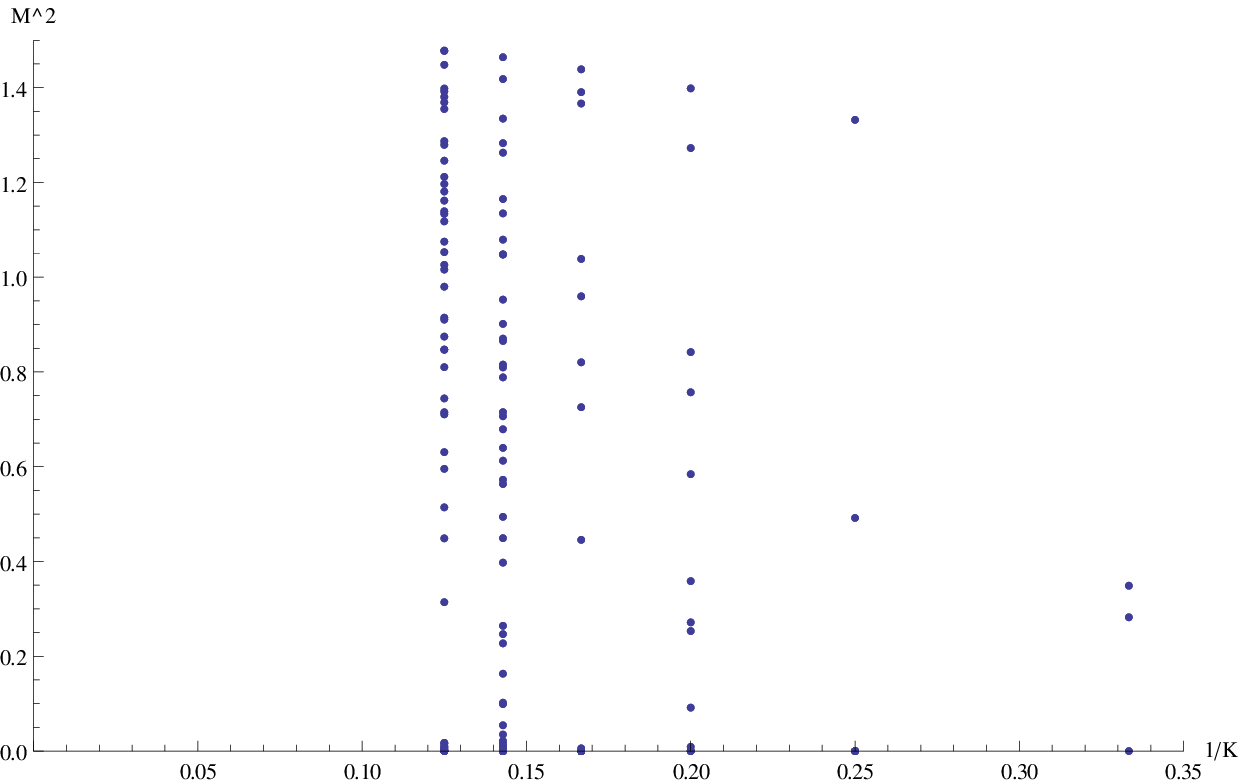}
\caption{DLCQ spectrum of light ($M^2 < 1.5$ in units of $2g^2N_c$)
flavor-singlet mesons and glueballs in $\CN=(0,2)$ SQCD with $N_1=N_2=N_3$ and $g=m$.}
\label{fig:dlcq-tp-low-masses}
\end{figure}

In $\CN=(0,2)$ SQCD, the (mass)$^2$ operator has the following general structure:
\be
2 P^+ P^- \; = \; \frac{K}{\alpha'} \left( F_{\text{current-current}} +  \frac{m^2}{g^2} F_{J-\text{interaction}} \right)
\label{PPFF}
\ee
Moreover, it is easy to see from \eqref{Linta} that, besides a diagonal quadratic
term in $F_{\text{current-current}}$,
all terms in $F_{\text{current-current}}$ and $F_{J-\text{interaction}}$ are quartic in the oscillator modes.
The mass spectrum is expected to converge for increasing values of $K$.
The first appearance of a state at a given resolution $K$ is called the {\it trail head} \cite{Gross:1997mx}.
The procedure of finding trail heads is usually easy if one plots the eigenvalues of $M^2$ (or $P^-$)
as a function of $1/K$, {\it e.g.} one can spot a few trail heads in Figure~\ref{fig:dlcq-tp-low-masses}.

Equivalently, since\footnote{The supercharges $\bar Q$ and $\bar Q^\dagger$ are the two supercharges of the $\CN=(0,2)$ super-Poincar\'e algebra. In particular $\bar Q^\dagger$ does \emph{not} stand for the conformal supercharge.
To avoid the confusion with the $\pm$ notation used here, we do not use supercharge notation of section \ref{sec:index}. }
\be
 2P^- \; = \; \{ {\bar{\CQ}}, {\bar{\CQ}}^\dagger \}
 \label{PQQnaive}
\ee
one can study the action of the supersymmetry generator $\CQ$ on the states \eqref{lcphysstates}.
In particular, the massless spectrum can be computed as:
\be
\ker P^- \; = \; \ker {\bar{\CQ}} \cap \ker {\bar{\CQ}}^\dagger
\ee
However for discrete light-cone quantization one needs to be more careful with the use of (\ref{PQQnaive}).
Indeed, as pointed out in \cite{Antonuccio:1998mq}, for finite values of $K$ one can not preserve all supersymmetry commutation relations.
As a result, there are several candidates for the light-cone Hamiltonian, related by different choice of normal ordering
and converging to the same operator $P^-$ in the continuum limit, {\it i.e.} in the limit $K \to \infty$:
\be
\frac{1}{2} ( {\bar{\CQ}} + {\bar{\CQ}}^\dagger )^2 \,, \qquad
\frac{1}{2} \{ {\bar{\CQ}}, {\bar{\CQ}}^\dagger \} \,, \qquad
-\frac{1}{2} ( {\bar{\CQ}} -  {\bar{\CQ}}^\dagger )^2
\label{Pchoices}
\ee
Although for finite $K$ details may be slightly different,
generally this choice does not affect qualitative features of the massive spectrum,
as we illustrate below by following \cite{Antonuccio:1998mq} and choosing the first expression in \eqref{Pchoices}
for our analysis in the rest of this section.
Then, for balance we will choose the second expression in \eqref{Pchoices} for the analysis in appendix~\ref{sec:K2}.
(Another important distinction between the results of this section and appendix~\ref{sec:K2} is that
here we deal with $\SU(N_c)$ gauge theory, while there we consider $\UU(N_c)$ gauge group.)
The first and the last choice \eqref{Pchoices} lead to the same massive spectrum.

Also note that supercharge does not preserve the number of partons because all of the terms in \eqref{QJJ}
contain three creation or annihilation operators. One of the implications is that all of 5 singlets
listed in \eqref{flavsinglet1} turn out to be massless for $K=2$. Indeed, since
the single-trace flavor singlet sector of $\SU(N_c)$ SQCD with $K=2$ does not have
1-parton states\footnote{something that we saw earlier in \eqref{KKKfermion}}
the action of ${\bar{\CQ}}$ is automatically trivial.

The explicit computation of the DLCQ spectrum for $K=2$ and $K=3$ is summarized in appendix~\ref{sec:K2}.
Namely, we go through the entire process is great detail, first by listing the physical states \eqref{lcphysstates}
and then analyzing the action of ${\bar{\CQ}}$ and $P^-$. Aside from the calculation of the mass spectrum, it also gives
us valuable information about mixing of different states that transform in the same representation of the flavor symmetry.

One such sector, namely the states in the trivial (singlet) representation of the flavor symmetry
plays a very important role in our 2d theory here and in the real QCD$_4$ \cite{Gusken:1999te,Beneke:2002jn}.
Indeed, these are the states that dominate in the Veneziano limit \eqref{Venezianolim} which,
as we explained earlier, is the only sensible way to take $N_c \to \infty$
(since at least {\it two} of the $N_i$, $i=1,2,3$ must become large in this limit in order to avoid dynamical SUSY breaking).
Therefore, in the rest of this section we present detailed results for the flavor-singlet states in
$\CN=(0,2)$ SQCD with $\SU(N_c)$ gauge group and the light-cone Hamiltonian given by the first expression in \eqref{Pchoices}.

Moreover in the limit \eqref{Venezianolim} of large $N_c$, $N_f$ and $N_b$, we only need to focus on
singlet ``single-trace'' states with all gauge and flavor indices contracted in a way that corresponds
to a single closed path in the quiver diagram in Figure~\ref{fig:qcd-lc}$a$.
The reason for this is exactly the same as in the standard 't Hooft limit of $\SU(N_c)$ gauge theory \cite{'tHooft:1973jz},
where single-trace operators correspond to closed string states and provide a good description of the physics as $N_c \to \infty$.

Similarly, the limit \eqref{Venezianolim} of large $N_c$, $N_f$ and $N_b$ in our model is described by closed string states
that are ``single-trace'' in the generalized sense of \cite{Gadde:2009dj}
where a similar limit of the 4d supersymmetric gauge theory was studied.
Our first task is to do the taxonomy of such single-trace states that are complete singlets under gauge and flavor symmetries.
Here the experience with a free fermion \eqref{freefermquiv} comes in handy and the result is, {\it cf.} \eqref{KKKfermion}:
\begin{equation}
 \begin{tabular}{|c|c|c|}
 \hline
  $K$ & \# of flavor singlets & \# of massless singlets \\
  \hline
  \hline
  1 & 0 & 0 \\
  2 & 5 & 5 \\
  3 & 24 & 2 \\
  4 & 78 & 14 \\
  5 & 266 & 6 \\
  6 & 947 & 47 \\
  7 &  3374 & 16 \\
  8 & 12476 & 152 \\
  \hline
 \end{tabular}
\label{flavsingKKK}
\end{equation}
For example, at $K=2$ we find a total of five flavor-singlet mesons and glueballs:
\be\label{flavsinglet1}
\begin{array}{c}
(\phi^\dagger)^s_{\alpha} (1) \; (\bar \phi^\dagger)_s^{\alpha}(1) \; |0\rangle \\
(p^\dagger)_a^{\beta} (1) \; (\bar p^\dagger)^a_{\beta}(1) \; |0\rangle \\
(\psi^\dagger)_i^{\alpha} (1) \; (\bar \psi^\dagger)^i_{\alpha}(1) \; |0\rangle \\
(\gamma^\dagger)_s^a (1) \; (\bar \gamma^\dagger)^s_a (1) \; |0\rangle \\
(\lambda^\dagger)_{\beta}^{\alpha}(1) \; (\bar \lambda^\dagger)^{\beta}_{\alpha} (1) \; |0\rangle \\
\end{array}
\ee
all of which turn out to be massless for reasons explained earlier.
In a similar way, one can explicitly write down physical states for other values of the harmonic resolution $K = 3,4, \ldots$,
in fact, not only in the singlet sector of the theory (as demonstrated in appendix \ref{sec:K2}).

\begin{figure}[ht]
\centering
 \begin{tabular}{cc}
\includegraphics[scale=0.6]{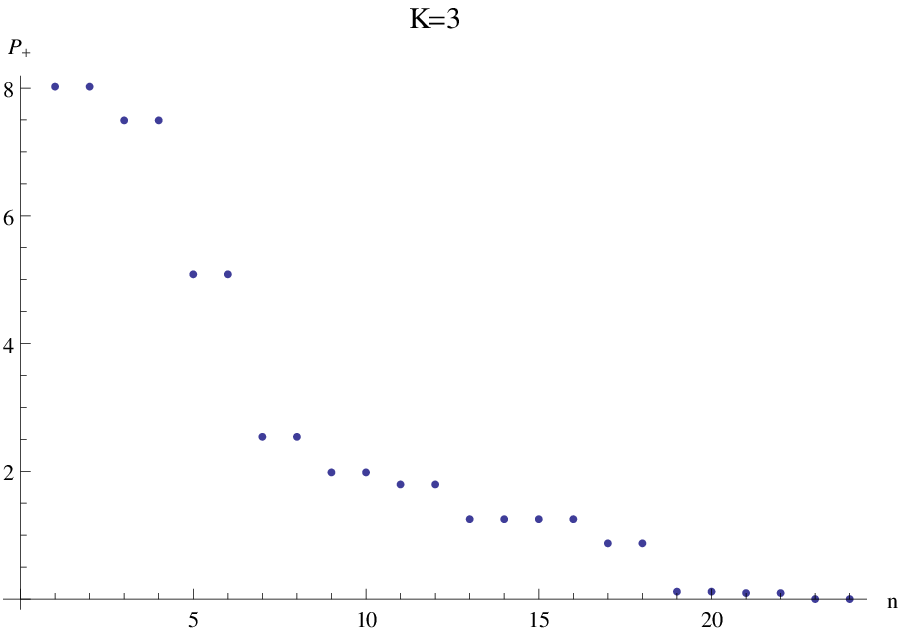} & \includegraphics[scale=0.6]{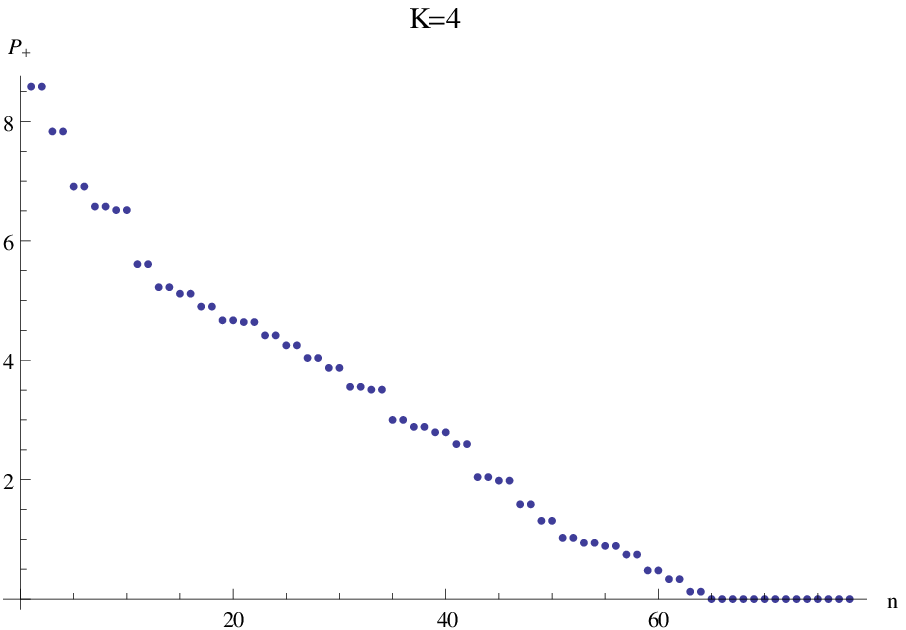} \\
\includegraphics[scale=0.6]{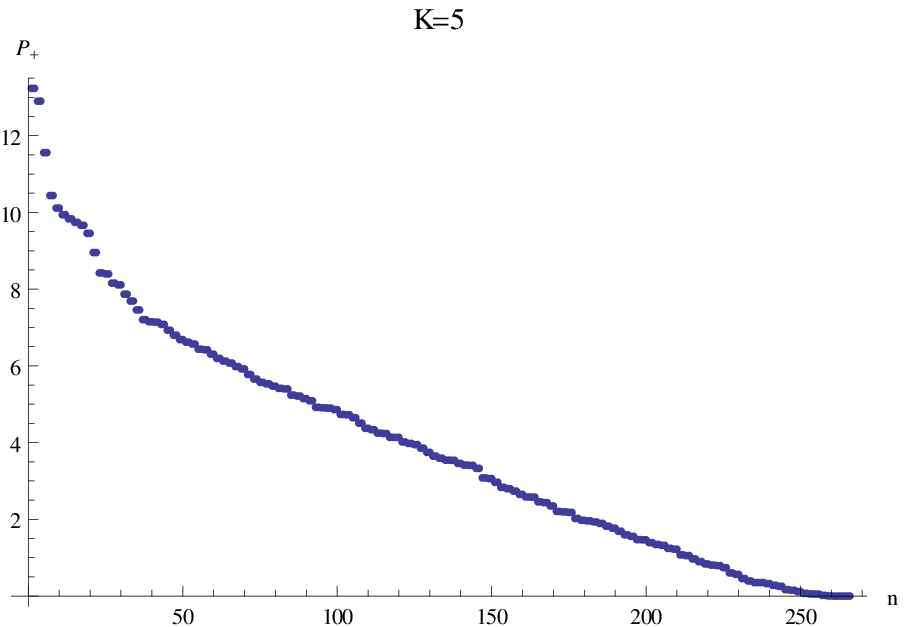} & \includegraphics[scale=0.6]{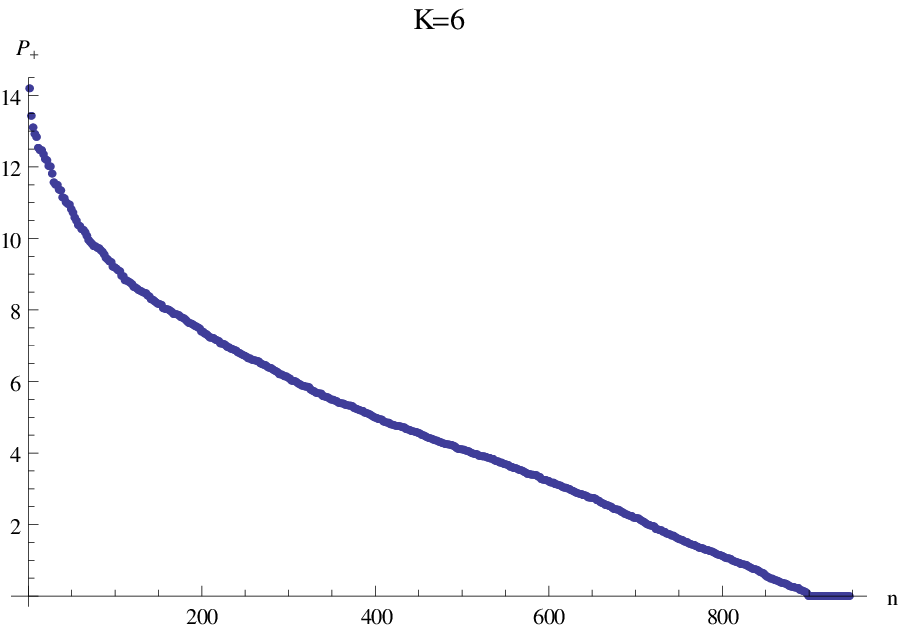} \\
\includegraphics[scale=0.6]{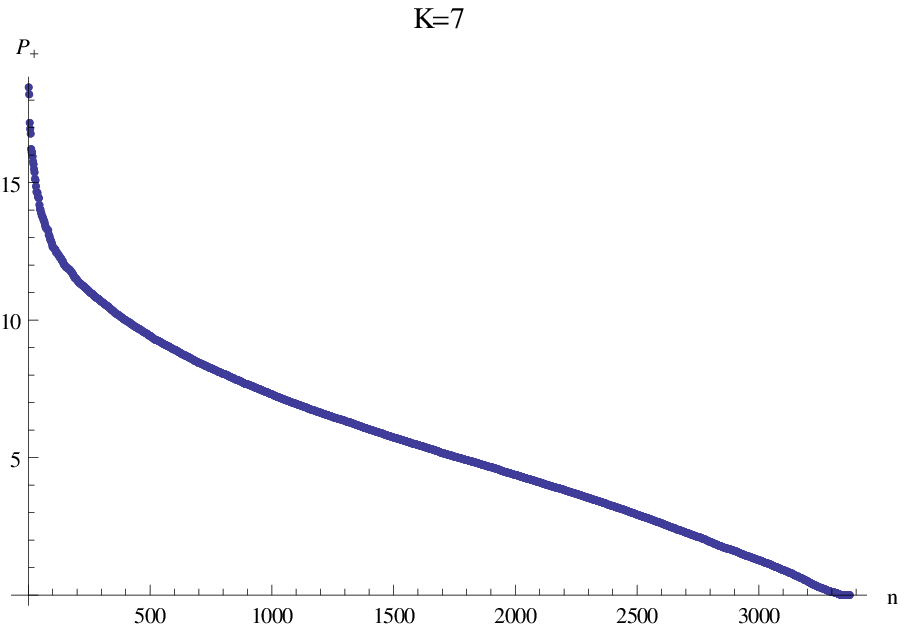} & \includegraphics[scale=0.6]{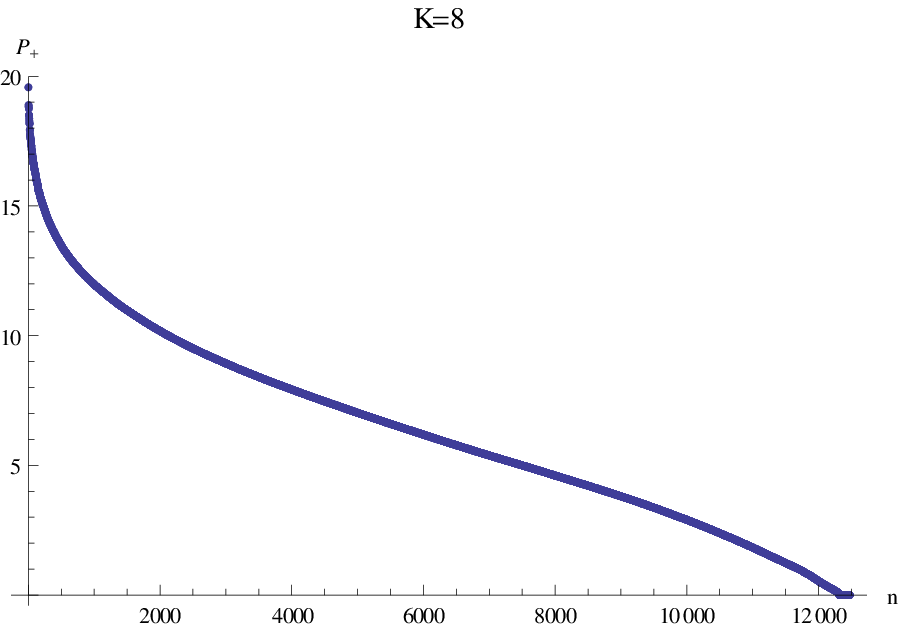}
 \end{tabular}
\caption{DLCQ spectrum of $\CN=(0,2)$ SQCD with $N_1=N_2=N_3$ and $g=m$. Each plot shows the ordered eigenvalues of $P_+$ in units of $2 g^2 N_c$. (The energy density of states $\rho (E)$ is minus the inverse derivative).}
\label{fig:dlcq-tp}
\end{figure}

Next, we study the action of ${\bar{\CQ}}$ and $P^-$ on these states which, in turn, determines the mass spectrum and the number of
massless states for each value of $K$. For generic values of the parameters $g^2 N_c$, $m^2 N_c$ and $\nu_i$
the results are summarized in \eqref{flavsingKKK} and in Figure~\ref{fig:dlcq-tp}.
(See also Figure~\ref{fig:dlcq-tp-low-masses} for a different presentation of light states.)

\begin{figure}[ht]
\centering
 \includegraphics[scale=0.6]{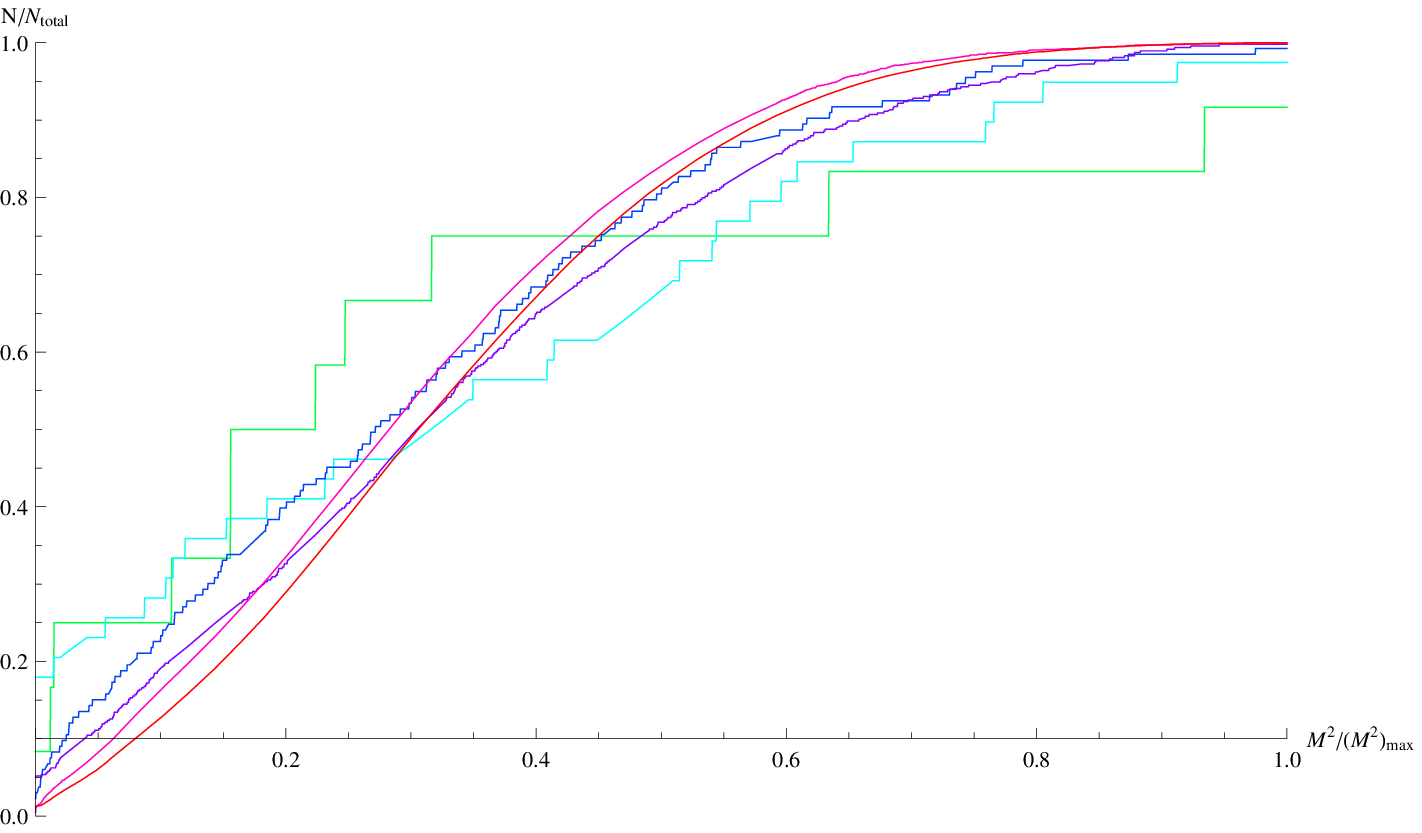}
\caption{Normalized distribution $F\left(\frac{M^2}{M^2_\text{max}}\right)=\frac{\text{number of states with mass less than $M$}}{\text{total number of states}}$ for $N_1=N_2=N_3$, $g=m$, and the harmonic resolution $K=3,4,5, \ldots, 8$. A similar distribution was obtained in \cite{Lunin:2000im}.
}
\label{fig:dlcq-norm-distr}
\end{figure}

All the plots in Figures~\ref{fig:dlcq-tp-low-masses} and \ref{fig:dlcq-tp} show a clear convergence with increasing values of $K$.
Moreover, it is easy to see --- especially from the normalized plot in Figure~\ref{fig:dlcq-norm-distr} ---
that eigenvalues of the (mass)$^2$ operator \eqref{PPFF} at finite values of $K$
often give a very good approximation to masses of states in the continuum limit ($K \to \infty$).

Moreover, as we can see from Figures~\ref{fig:dlcq-tp} and \ref{fig:dlcq-norm-distr},
the density of states approximately remains constant
in a wide range of energies that extends all the way from $E=0$ to the upper limit of the discrete light-cone approximation:
\be
\rho (E) \; \simeq \; \text{const}
\ee
This behavior is typical for theories with finitely many Regge trajectories\footnote{There is almost no distinction
between Regge trajectories for bosonic and fermionic fundamental flavors \cite{Aoki:1995dh}.} (as in Figure~\ref{fig:SQCD2}$a$)
and has to be contrasted with large $N_c$ limit of non-supersymmetric ($\CN=0$) QCD with one adjoint matter multiplet
or $N_f = N_c$ massive quarks in the fundamental representation.
The latter theory has infinitely many asymptotically linear Regge trajectories illustrated in Figure~\ref{fig:SQCD2}$b$
and an exponentially growing density of single particle states.
In the limit when $N_f = N_c$ quarks become massless or when the mass of the adjoint matter multiplet is turned off,
the non-supersymmetric QCD$_2$ exhibits a transition from confinement to screening \cite{Gross:1997mx,Armoni:1997bu}.
In particular, QCD string made out of the adjoint bits dissociates in this limit into stable constituent ``particles''
which become free in the massless limit and form a single Regge trajectory.
Therefore, we conclude that, even though our $\CN=(0,2)$ SQCD has superpotential \eqref{JPPhi} with a mass parameter $m$,
it nevertheless is much closer to the screening phase of QCD$_2$ with {\it massless} adjoint multiplet or $N_f = N_c$
quarks.\footnote{A somewhat similar behavior is also found in the $N_f \gg N_c$ limit of non-supersymmetric ($\CN=0$) QCD$_2$,
where infinitely many Regge trajectories collapse to a few massive mesons \cite{Armoni:1998ny}
via a non-abelian version of the Schwinger mechanism, {\it cf.} \cite{Dalley:1997df}.}
This, perhaps is not too surprising since after integrating out massive multiplets (in the limit $\frac{m^2}{g^2} \gg 1$)
we end up with $\CN=(0,2)$ gauge theory coupled to massless matter.
In particular, this explains why our $\CN=(0,2)$ SQCD does not have an exponentially growing density of states.

\begin{figure}[ht]
\centering
 \includegraphics[scale=1]{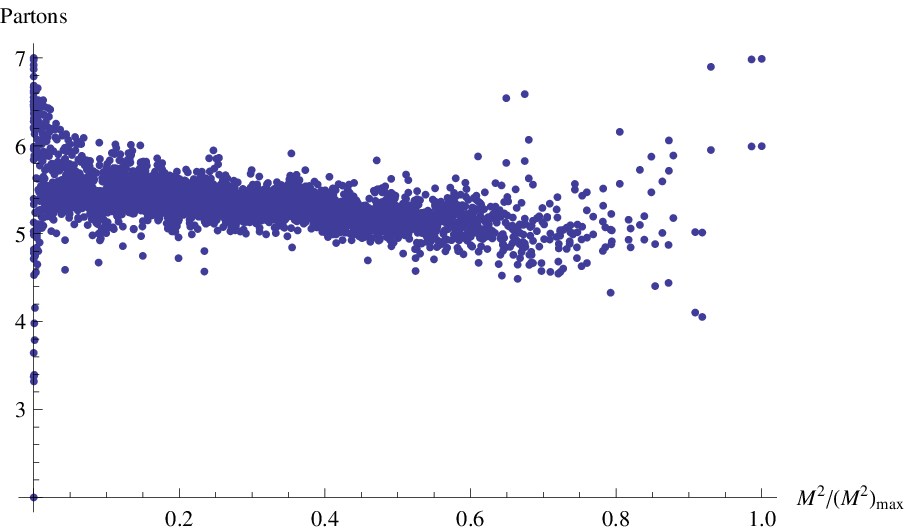}
\caption{Parton number vs. mass for $K=7$.}
\label{fig:dlcq-partons}
\end{figure}

Note, nothing prevents mixing of states within the sector of color and flavor singlets.
And such states do indeed mix, {\it cf.} Figure~\ref{fig:dlcq-partons}.
This graph also shows that spectrum is dominated by string-like states made of many partons.
It would be interesting to identify the closed string which describes the Veneziano limit \eqref{Venezianolim} of 2d $\CN=(0,2)$ SQCD.

%%%%%%%%%%%%%%%%%%%%%%%%%%%%%%%%%%%%%%%%%%%%%%%%%%%%%%%%%%%%%%%%%%%%%%%%%%%%%%%%%%%%%%%%%%%%%%

\acknowledgments{We would like to thank D.~Gaiotto, B.~Jia, I.~Melnikov, V.~Schomerus, and E.~Sharpe for useful discussions. A.G. would also like to thank the Kavli Institute for Theoretical Physics for providing a hospitable and enjoyable work environment during the late stages of this project.
The work of A.G. is supported in part by the John A. McCone fellowship and by DOE Grant DE-FG02-92-ER40701.
The work of S.G. is supported in part by DOE Grant DE-FG03-92-ER40701FG-02.
The work of P.P. is supported in part by the Sherman Fairchild scholarship and by NSF Grant PHY-1050729.
Opinions and conclusions expressed here are those of the authors and do not necessarily reflect the views of funding agencies.
}

%%%%%%%%%%%%%%%%%%%%%%%%%%%%%%%%%%%%%%%%%%%%%%%%%%%%%%%%%%%%%%%%%%%%%%%%%%%%%%%%%%%%%%%%%%%%%%

\appendix

\section{Characters of affine Lie algebras}
\label{affinechar}

In our paper we define a character of level-$k$ integrable representation $V_\lambda$ of affine Lie algebra $\mathfrak{A}$
\begin{equation}
 \chi_{\lambda}^{\mathfrak{A}_k}(\tau,\xi)=\Tr_{V_\lambda}e^{2\pi i \tau L_0+\sum_i \xi_i H_0^i}
\label{char-def}
\end{equation}
where $H_0^i$ are Cartan elements of the ordinary Lie algebra inside $\mathfrak{A}$. Alternatively one can use $\mathbb{C}^*$ variables $q=e^{2\pi i\tau}$ and $x_i=e^{2\pi i\xi_i}$. Note that we do not include the usual $q^{c/24}$ factor, that the $q$-expansion of the character starts with $q^h$ where $h$ is the conformal dimension of the primary\footnote{This quantity is unambiguously determined if we consider $L_0$ as an element of the universal envelopping algebra via Sugawara construction}. For simplicity let us restrict ourselves to simply-laced Lie algebras ${\mathfrak{A}}$ of ADE type. The characters can be explicitly expressed through level-$k$ theta functions of the root lattice $M=\oplus_{i} \mathbb{Z}\alpha_i$ of the ordinary Lie algebra which are defined as follows
\begin{equation}
 \Theta^{{\mathfrak{A}}}_{\lambda,k}(\tau,\xi):=
 \sum_{\mu\in M}e^{\pi i k\tau \left(\mu+\lambda/k\right)^2+\sum_{i}2\pi i(k\mu+\lambda,\xi)}
\end{equation}
where $\lambda$ is an element of the weight lattice and $\xi=\sum_i\xi_i\alpha_i$. Then \cite{KacPeterson}
\begin{equation}
  \chi_{\lambda}^{\mathfrak{A}_k}(\tau,\xi)=
  \frac{\sum_{w\in W}\epsilon(w)\Theta^{{\mathfrak{A}}}_{w(\lambda+\rho),k+g}(\tau,\xi)}{\sum_{w\in W}\epsilon(w)\Theta^{{\mathfrak{A}}}_{w(\lambda),g}(\tau,\xi)}
\end{equation}
where $W$ is the Weyl group of the ordinary Lie algebra, $\epsilon(w)$ denotes parity of its element $w$, $g$ is the dual coxeter number and $\rho$ is the sum of all fundamental weights. Weight $\lambda$ effectively takes values in $M^*/kM$. This is known as the Weyl-Kac character formula.

Under $T$ and $S$ transformation characters transform as follows:
\begin{equation}
 \chi_{\lambda}^{\mathfrak{A}_k}\left(\tau+1,\xi\right)=e^{2\pi ih_\lambda}\chi_{\lambda}^{\mathfrak{A}_k}\left(\tau,\xi\right),
\end{equation}
\begin{equation}
 \chi_{\lambda}^{\mathfrak{A}_k}\left(-\frac{1}{\tau},\frac{\xi}{\tau}\right)=
 e^{-\pi i\frac{c}{12}\left(\tau+1/\tau\right)+\pi i(\xi,\xi)/\tau}\sum_{\rho}S_{\lambda,\nu}
\chi_{\rho}^{\mathfrak{A}_k}\left({\tau},{\xi}\right)
\end{equation}
where $h_\lambda$ is the conformal dimension of the primary and $S$ is a constant matrix given by
\begin{equation}
 S_{\lambda,\nu}=\frac{i^{|\Delta_+|}}{|M^*/M|^{1/2}(k+g)^{{\mathrm{rank}\,{\mathfrak{A}}}/2}}
\sum_{w\in W}\epsilon(w)e^{-2\pi i(w(\lambda+\rho),\nu+\rho)/(k+g)}
\end{equation}
where $|\Delta_+|$ is the number of positive roots of the ordinary Lie algebra ${\mathfrak{A}}$. Matrix $S$ satisfies the following properties:
\begin{equation}
 S^2=\mathcal{C},\qquad S^T=S,\qquad S\mathcal{C}=\mathcal{C}S=\bar{S}
\end{equation}
where $\mathcal{C}$ is the charge conjugation matrix.

For level $1$ one can use simpler formulas:
\begin{equation}
  \chi_{\lambda}^{\mathfrak{A}_1}(\tau,\xi)=\frac{\Theta^{{\mathfrak{A}}}_{\lambda,1}(\tau,\xi)}{(q;q)_\infty^{\mathrm{rank}\,{\mathfrak{A}}}},
\end{equation}
 \begin{equation}
 S_{\lambda,\nu}=\frac{1}{|M^*/M|^{1/2}}e^{-2\pi i(\lambda,\nu)}.
\end{equation}

Affine $\UU(1)$ characters have the following explicit expressions:
\begin{equation}
 \chi^{\UU(1)_k}_{r}(\tau,\xi)=(q;q)_\infty^{-1}\sum_{n\in \mathbb{Z}}e^{\pi i \tau k\left(n+r/k\right)^2+2\pi i(kn+r)\xi}.
\end{equation}
where $r\in \mathbb{Z}/k\mathbb{Z}$. In the simplest case
\begin{equation}
 \chi^{\UU(1)_1}_{0}(\tau,\xi)=\frac{\theta_{3}(\xi;\tau)}{(q;q)_\infty}=\theta(-xq^{1/2};q)
\end{equation}

The modular transformation properties are given by
\begin{equation}
 \chi^{\UU(1)_k}_{r}(\tau+1,\xi)=e^{2\pi ih_r}\tilde{\chi}^{\UU(1)_k}_{r}(\tau,\xi),
\label{twisted-chars}
\end{equation}
\begin{equation}
 \chi^{\UU(1)_k}_{r}\left(-\frac{1}{\tau},\frac{\xi}{\tau}\right)=
e^{-\pi i\frac{1}{12}\left(\tau+1/\tau\right)+\pi ik\xi^2/\tau}\sum_{s}S_{r,s}{\chi}^{\UU(1)_k}_{s}(\tau,\xi)
\end{equation}
where $h_r$ denotes the conformal dimension of the primary and $\tilde{\chi}^{\UU(1)_k}_{r}$ denote ``twisted" affine $\UU(1)$ characters \textit{defined} by the formula (\ref{twisted-chars})\footnote{The ``twist" can be interpreted as insertion of $(-1)^{F}$ into the trace (\ref{char-def}). The characters for simply-laced algebras as well as $\UU(1)_k$ characters for even level $k$ are not affected by this twist.}.

Level rank duality between integrable representations of $\SU(n)_k\times \UU(1)_{(n+k)n}$ and $\SU(k)_n\times \UU(1)_{(n+k)k}$ that we use in this paper can be defined by the following conformal embedding\footnote{Which differs from the common one by extra $\UU(1)$ factors.} \cite{LRFrenkel,LRJM}:
\begin{equation}
 \UU(1)_{1}\times \UU(nk)_{1}\supset
\left(\SU(n)_k\times \UU(1)_{(n+k)n}\right)\times \left(\SU(k)_n\times \UU(1)_{(n+k)k}\right)
\end{equation}
Given explicitly by the following character decomposition formula\cite{LRHasegawa}
\begin{multline}
 \theta\left(-\prod_{i=1}^n{x_i}\prod_{j=1}^k{y_j}q^{1/2}\right)
 \prod_{i=1}^n\prod_{j=1}^k\theta\left(-{x_i}/{y_j}q^{1/2}\right)
=\\
\sum_{\lambda,r;\lambda^t,r^t}L_{(\lambda,r),(\lambda^t,r^t)}
\chi^{\SU(n)_k}_\lambda(q,\tilde{x})
\chi^{\UU(1)_{(n+k)n}}_{r}\left(q,X\right)
\chi^{\SU(k)_n}_{\lambda^t}(q,\tilde{y})
\chi^{\UU(1)_{(n+k)k}}_{r^t}\left(q,Y\right)
\label{LR-matrix}
\end{multline}
where $L_{(\lambda,r),(\lambda^t,r^t)}$ is the level-rank matrix \textit{defined} by this formula. Rectangular matrix $L_{(\lambda,r),(\lambda^t,r^t)}$ is a constant with elements equal to $1$ for level-rank dual pairs $(\lambda,r)$ and $(\lambda^t,r^t)$. It defines one-to-one map between equivalence classes of representations $\SU(n)_k\times \UU(1)_{(n+k)n}$ and $\SU(k)_n\times \UU(1)_{(n+k)k}$ appearing in the decomposition (\ref{LR-matrix}). The equivalence between representations of $\SU(n)_k\times \UU(1)_{(n+k)n}$ is given by the action of $\mathbb{Z}_n$ group. It acts as the outer automorphism group on the representations of $\SU(n)_k$ and shifts $\UU(1)_{(n+k)n}$ representations by a multiple of $(n+k)$. Similarly, the equivalence between representations of $\SU(k)_n\times \UU(1)_{(n+k)k}$ is given by the action of the analagous $\mathbb{Z}_k$ group. It follows that the generalized inverse $\tilde{L}$ of $L$ is proporsional to the transposed of $L$. Namely,
\begin{equation}
 \tilde{L}=\frac{1}{nk}L^T
\label{LR-inverse}
\end{equation}
\begin{equation}
 L\tilde{L}L=L
\end{equation}
\begin{equation}
 \tilde{L}{L}\tilde{L}=\tilde{L}
\end{equation}
Since the l.h.s. of (\ref{LR-matrix}) is modular invariant (up to a modular anomaly) it follows that
\begin{equation}
 SLS^t=L
\end{equation}
where $S$ is the $S$-matrix for representations of $\SU(n)_k\times \UU(1)_{(n+k)n}$  and $S^t$ is the $S$-matrix for its level-rank dual $\SU(k)_n\times \UU(1)_{(n+k)k}$.

%%%%%%%%%%%%%%%%%%%%%%%%%%%%%%%%%%%%%%%%%%%%%%%%%%%%%%%%%%%%%%%%%%%%%%%%%%%%%%%%%%%%%%%%%%%%%%%%%%%%%%%%%%%%%%%

\section{Gluing index in the UV}
\label{indexglue}

Consider basic theory $\CT_{N_1N_2N_c'}$ associated to the left triangle in Fig. \ref{2nodequiver} inscribed into the circle of circumference $N$. The flavor symmetries of the theory are $\SU (N_1)\times \UU(1)_{(1)}$, $\SU (N_2)\times \UU(1)_{(2)}$, $\SU (N_c')\times \UU(1)_{(3)}$. Let us denote the fugacities corresponding to $\UU(N_1)$, $\UU(N_2)$, $\UU(N_c')$ flavor and $\UU(N_c)$ gauge group by $x$, $y$, $z$ and $\xi$ respectively. They can be decomposed into $\SU\times \UU(1)$ fugacities as $x=(\tilde{x},X=(\prod_{a}x_a)^{1/N_1})$, $y=(\tilde{y},Y=(\prod_{r}y_r)^{1/N_2})$, etc. The index of $\CT_{N_1N_2N_c'}$ is given by
\begin{equation}
 \CI_{\CT_{N_1N_2N_c'}}(x,y,z)=
\int \prod_{\alpha}\frac{d\xi_\alpha}{\xi_\alpha} \tilde{\CI}_{\CT_{N_1N_2N_c'}}(x,y,\xi) K(\xi,z)
 \label{indexA}
\end{equation}
where
\begin{equation}
 \tilde{\CI}_{\CT_{N_1N_2N_c'}}(x,y,\xi)
=\frac{\prod\limits_{\alpha\neq\beta}\theta(\xi_\alpha/\xi_\beta)\prod\limits_{a,r}\theta(q^\frac{1+R_3}{2}y_r/x_a)\theta(q^\frac{1}{2}\Xi^{N_c}X^{-N_1}Y^{-N_2})}{\prod\limits_{a,\alpha}\theta(q^\frac{R_1}{2}x_a/\xi_\alpha)\prod\limits_{r,\alpha}\theta(q^\frac{R_2}{2}\xi_\alpha/y_{r})}
\end{equation}
and
\begin{equation}
 K(\xi,z)=\prod_{\alpha,i} \theta(q^\frac{1}{2}z_i/\xi_\alpha)\theta(q^\frac{1}{2}\Xi^{N_c'}Z^{N_c}).
 \label{kernel}
\end{equation}
The index have the following decmposition into characters:
\begin{multline}
 \CI_{\CT_{N_1N_2N_c'}}(x,y,z)=\sum_{\lambda,\mu,\nu,\alpha,\beta,\gamma}C_{\CT_{N_1N_2N_c'}}^{(\lambda,\alpha),(\mu,\beta),(\nu,\gamma)}
 \chi^{\SU(N_1)_{n_1}}_\lambda(x)\chi^{\SU(N_2)_{n_2}}_\mu(y)\chi^{\SU(N_c')_{N_c}}_\lambda(z)\times\\
 \tilde{\chi}^{\UU(1)_{NN_1}}_\alpha(X) \tilde{\chi}^{\UU(1)_{NN_2}}_\beta(Y) \tilde{\chi}^{\UU(1)_{NN_c'}}_\gamma(Z)
 \label{chardecomp}
\end{multline}
where the coefficients $C_{\CT_{N_1N_2N_c'}}^{(\lambda,\alpha),(\mu,\alpha),(\nu,\gamma)}$ are \textit{integer numbers} and as usual $n_i=N-N_i$.

The function (\ref{kernel}) has decomposition into characters of the following form:
\begin{equation}
 K(\xi,z)=\sum_{\rho,\nu,\delta,\gamma}K^{(\rho,\delta),(\nu,\gamma)}\chi^{\SU(N_c)_{N_c'}}_\rho(\xi)\chi^{\SU(N_c')_{N_c}}_\nu(z)
 \tilde{\chi}^{\UU(1)_{NN_c}}_\delta(\Xi)\tilde{\chi}^{\UU(1)_{NN_c'}}_\gamma(Z).
 \label{kernelchar}
\end{equation}
 From (\ref{LR-matrix}) it follows that the matrix $K$ is simply related to the level-rank duality matrix:
\begin{equation}
 K^{(\rho,\delta),(\nu,\gamma)}=(-1)^{2(h_\rho+h_\delta+h_\nu+h_\gamma)}
L_{(\rho,\delta),(\nu,\gamma)}
\end{equation}
where $2(h_\rho+h_\delta+h_\nu+h_\gamma)$ is an integer.

From (\ref{indexA}), (\ref{chardecomp}) and (\ref{kernelchar}) it follows that (using uniqueness of character decomposition):
\begin{multline}
 \sum_{\lambda,\mu,\alpha,\beta}C_{\CT_{N_1N_2N_c'}}^{(\lambda,\alpha),(\mu,\beta),(\nu,\gamma)}
 \chi^{\SU(N_1)_{n_1}}_\lambda(x)\chi^{\SU(N_2)_{n_2}}_\mu(y)
 \tilde{\chi}^{\UU(1)_{NN_1}}_\alpha(X) \tilde{\chi}^{\UU(1)_{NN_2}}_\beta(Y) =\\
 \int \prod_{\alpha}\frac{d\xi_\alpha}{\xi_\alpha} \tilde{\CI}_{\CT_{N_1N_2N_c'}}(x,y,\xi)
 \sum_{\rho,\delta}K^{(\rho,\delta),(\nu,\gamma)}\chi^{\SU(N_c)_{N_c'}}_\rho(\xi)
 \tilde{\chi}^{\UU(1)_{NN_c}}_\delta(\Xi)
\label{formulaA}
\end{multline}

Now consider theory $\CT_{N_3N_4N_c}$ associated to the right triangle in Fig. \ref{2nodequiver}. Let us denote fugacities corresponding to its flavor symmetries $\UU (N_3)$, $\UU (N_4)$ and $\UU (N_c)$ by $u$, $v$ and $\xi$ respectively. Consider the frame where the gauge group is $\UU (N_c')$ and denote the corresponding fugacities by $z$. Similarly to (\ref{formulaA}) one can write:
\begin{multline}
 \sum_{\lambda,\mu,\alpha,\beta}C_{\CT_{N_3N_4N_c}}^{(\rho,\delta),(\lambda',\alpha'),(\mu',\beta')}
 \chi^{\SU(N_3)_{n_3}}_{\lambda'}(u)\chi^{\SU(N_4)_{n_4}}_{\mu'}(v)
 \tilde{\chi}^{\UU(1)_{NN_3}}_{\alpha'}(U) \tilde{\chi}^{\UU(1)_{NN_4}}_{\beta'}(V) =\\
 \int \prod_{i}\frac{dz_i}{z_i} \tilde{\CI}_{\CT_{N_3N_4N_c}}(u,v,z)
 \sum_{\nu,\gamma}K^{(\rho,\delta),(\nu,\gamma)}\chi^{\SU(N_c')_{N_c}}_\nu(z)
 \tilde{\chi}^{\UU(1)_{NN_c'}}_\delta(Z)
\label{formulaB}
\end{multline}

Now let us consider the quiver theory $\CT_{N_1N_2N_3N_4}$ associated to the quadrilateral in Fig. \ref{2nodequiver} obtained by gluing $\CT_{N_1N_2N_c'}$ and $\CT_{N_3N_4N_c}$ along the common edge. Its index is given by:
\begin{equation}
 I_{\CT_{N_1N_2N_3N_4}}(x,y,u,v)=\int\prod_{\alpha} \frac{d\xi_\alpha}{\xi_\alpha}\prod_i\frac{d z_i}{z_i}\;\tilde{\CI}_{\CT_{N_1N_2N_c'}}(x,y,\xi) \tilde{\CI}_{\CT_{N_3N_4N_c}}(u,v,z) K(\xi,z)
 \label{formulaC}
\end{equation}
Let $\tilde{K}$ be generalized inverse of the matrix $K$. That is:
\begin{equation}
 \sum_{\nu,\rho,\gamma,\delta}K^{(\rho',\delta'),(\nu,\gamma)}\tilde{K}_{(\nu,\gamma),(\rho,\delta)}K^{(\rho,\delta),(\nu',\gamma')}=K^{(\rho',\delta'),(\nu',\gamma')}
\end{equation}
Similarly to (\ref{LR-inverse}) $\tilde{K}=K^T/(N_cN_c')$.

From (\ref{kernelchar}), (\ref{formulaA}), (\ref{formulaB}) and (\ref{formulaC}) it follows that
\begin{multline}
 I_{\CT_{N_1N_2N_3N_4}}(x,y,u,v)=\sum_{\lambda,\mu,\lambda',\mu',\alpha,\beta,\alpha',\beta'}C_{\CT_{N_1N_2N_3N_4}}^{(\lambda,\alpha),(\mu,\beta),(\lambda',\alpha'),(\mu',\beta')} \times \\
 \chi^{\SU(N_1)_{n_1}}_\lambda(x)\chi^{\SU(N_2)_{n_2}}_\mu(y)
 \chi^{\SU(N_3)_{n_3}}_{\lambda'}(u)\chi^{\SU(N_4)_{n_4}}_{\mu'}(v) \times\\
 \tilde{\chi}^{\UU(1)_{NN_1}}_\alpha(X) \tilde{\chi}^{\UU(1)_{NN_2}}_\beta(Y)
  \tilde{\chi}^{\UU(1)_{NN_3}}_{\alpha'}(U) \tilde{\chi}^{\UU(1)_{NN_4}}_{\beta'}(V)
\end{multline}
 where
\begin{equation}
 C_{\CT_{N_1N_2N_3N_4}}^{(\lambda,\alpha),(\mu,\beta),(\lambda',\alpha'),(\mu',\beta')}=\sum_{\nu,\rho,\gamma,\delta}C_{\CT_{N_1N_2N_c'}}^{(\lambda,\alpha),(\mu,\beta),(\nu,\gamma)}\tilde{K}_{(\nu,\gamma),(\rho,\delta)}C_{\CT_{N_3N_4N_c}}^{(\rho,\delta),(\lambda',\alpha'),(\mu',\beta')}.
\end{equation}
This is perfectly consistent with the gluing prescription (\ref{IR-gluing}) that follows from the proposed IR solution.

Decomposition in Fig. \ref{dual2node} of the same quadrilateral gives a different expression:
\begin{equation}
 C_{\CT_{N_1N_2N_3N_4}}^{(\lambda,\alpha),(\mu,\beta),(\lambda',\alpha'),(\mu',\beta')}=\sum_{\nu',\rho',\gamma',\delta'}C_{\CT_{N_2N_3N_c''}}^{(\mu',\beta'),(\lambda,\alpha),(\nu',\gamma')}\tilde{K}_{(\nu',\gamma'),(\rho',\delta')}C_{\CT_{N_4N_1N_c'''}}^{(\rho',\delta'),(\mu,\beta),(\lambda',\alpha')}.
\end{equation}
Similarly one can obtain coefficients of decomposition into characters for more general quiver theories associated to inscribed polygons using ``three-point functions'' $C$ and ``propagators'' $\tilde{K}$ satisfying crossing symmetry.

%%%%%%%%%%%%%%%%%%%%%%%%%%%%%%%%%%%%%%%%%%%%%%%%%%%%%%%%%%%%%%%%%%%%%%%%%%%%%%%%%%%%%%%%%%%%%%

\section{DLCQ spectrum of $\UU(N_c)$ SQCD at finite $N_c$}
\label{sec:K2}

In this appendix we explicitly compute the DLCQ meson spectrum of $\UU(N_c)$ SQCD at finite $N_c$.

\subsection*{DLCQ spectrum for $K=2$}

In total, we have 19 different types of meson states\footnote{including the superpartners and meson-like modes of the fermions $\gamma$ and $\bar \gamma$ that are gauge singlets}
\begin{equation}
\small{
\begin{array}{c}
 (\bar{\gamma}^\dagger)^r_a(2)|0\rangle \\
 (\gamma^\dagger)^a_r(2)|0\rangle \\
(\gamma^\dagger)_r^a(1)(\gamma^\dagger)_s^b(1)|0\rangle \\
(\gamma^\dagger)_r^a(1)(\bar{\gamma}^\dagger)_b^s(1)|0\rangle \\
(\bar{\gamma}^\dagger)^r_a(1)(\bar{\gamma}^\dagger)^s_b(1)|0\rangle \\
(\lambda^\dagger)^\alpha_\alpha(2)|0\rangle \\
 \end{array}\;\;
\begin{array}{c}
 (\bar{\lambda}^\dagger)^\alpha_\alpha(2)|0\rangle \\
(\lambda^\dagger)^\beta_\alpha(1)(\bar{\lambda}^\dagger)^\alpha_\beta(1)|0\rangle \\
(\lambda^\dagger)^\alpha_\alpha(1)(\bar{\lambda}^\dagger)^\beta_\beta(1)|0\rangle \\
 (\bar{\phi }^\dagger)_r^\alpha(1)(\bar{p}^\dagger)_\alpha^a(1)|0\rangle\\
({\phi }^\dagger)^r_\alpha(1)({p}^\dagger)^\alpha_a(1)|0\rangle\\
(\bar{p}^\dagger)_\alpha^a(1)({\psi }^\dagger)_i^\alpha(1)|0\rangle\\
\end{array}\;\;
\begin{array}{c}
({p}^\dagger)^\alpha_a(1)({{\bar{\psi}} }^\dagger)^i_\alpha(1)|0\rangle\\
(\bar{\phi }^\dagger)_r^\alpha(1)(\phi^\dagger)_\alpha^s(1)|0\rangle \\
 (\bar{p }^\dagger)^b_\alpha(1)(p^\dagger)^\alpha_a(1)|0\rangle \\
 ({\psi }^\dagger)_i^\alpha(1)({\bar{\psi}}^\dagger)_\alpha^j(1)|0\rangle \\
 (\bar{\omega }^\dagger)_A^\alpha(1)(\omega^\dagger)_\alpha^B(1)|0\rangle \\
(\phi^\dagger)_\alpha^r(1)({\psi }^\dagger)_i^\alpha(1)|0\rangle\\
 (\bar{\phi}^\dagger)^\alpha_r(1)({{\bar{\psi}} }^\dagger)^i_\alpha(1)|0\rangle\\
\end{array}
}\end{equation}
Where $(\omega^\dagger)^{1,2}_\alpha$ and $(\bar{\omega}^\dagger)^{1,2}_\alpha$ are creation operators for $\Omega_{1,2}$ fermions. Now let us consider subspaces corresponding to different representations of flavor groups.

Subspace of singlets:
\begin{equation}
\small{
\begin{array}{l}
 \text{$|$1$\rangle $}=\left(\bar{\gamma }^{\dagger }\right)_c^s(1) \left(\gamma ^{\dagger }\right)_s^c(1) \text{$|$0$\rangle $} \\
 \text{$|$2$\rangle $}=\left(\lambda ^{\dagger }\right)_{\alpha }^{\alpha }(2) \text{$|$0$\rangle $} \\
 \text{$|$3$\rangle $}=\left(\bar{\lambda }^{\dagger }\right)_{\alpha }^{\alpha }(2) \text{$|$0$\rangle $} \\
 \text{$|$4$\rangle $}=\left(\bar{\lambda }^{\dagger }\right)_{\beta }^{\alpha }(1) \left(\lambda ^{\dagger }\right)_{\alpha }^{\beta
   }(1) \text{$|$0$\rangle $} \\
\end{array}\qquad
\begin{array}{l}
 \text{$|$5$\rangle $}=\left(\bar{\lambda }^{\dagger }\right)_{\beta }^{\beta }(1) \left(\lambda ^{\dagger }\right)_{\alpha }^{\alpha
   }(1) \text{$|$0$\rangle $} \\
 \text{$|$6$\rangle $}=\left(\bar{\phi }^{\dagger }\right)_t^{\alpha }(1) \left(\phi ^{\dagger }\right)_{\alpha }^t(1) \text{$|$0$\rangle $} \\
 \text{$|$7$\rangle $}=\left(\bar{p}^{\dagger }\right)_{\beta }^b(1) \left(p^{\dagger }\right)_b^{\beta }(1) \text{$|$0$\rangle $} \\
 \text{$|$8$\rangle $}=\left({\psi }^{\dagger }\right)_k^{\alpha }(1) \left({\bar{\psi}} ^{\dagger }\right)_{\alpha }^k(1) \text{$|$0$\rangle $} \\
 \text{$|$9$\rangle $}=\left(\bar{\omega }^{\dagger }\right)_F^0(1) \left(\omega ^{\dagger }\right)_0^F(1) \text{$|$0$\rangle $} \\
\end{array}}
\end{equation}
Reminder: $\bar{\phi}$, $p$, ${\psi}$ transform as $\mathbf{N_2}$, $\mathbf{N_1}$ and $\mathbf{N_3}$ respectively.
The action of ${\bar{\CQ}}$ on this subspace is given by the following matrix:
\begin{equation}\footnotesize{
{\bar{\CQ}}|_\text{singlets}=\left(
\begin{array}{ccccccccc}
 0 & 0 & 0 & 0 & 0 & 0 & 0 & 0 & 0 \\
 0 & 0 & 0 & 0 & 0 & -\frac{i g {N_2}}{\sqrt{2}} & \frac{i g {N_1}}{\sqrt{2}} & -\frac{i g {N_3}}{\sqrt{2}} & -i \sqrt{2} g \\
 0 & 0 & 0 & 0 & 0 & 0 & 0 & 0 & 0 \\
 0 & 0 & 0 & 0 & 0 & 0 & 0 & 0 & 0 \\
 0 & 0 & 0 & 0 & 0 & 0 & 0 & 0 & 0 \\
 0 & 0 & -\frac{i g}{\sqrt{2}} & 0 & 0 & 0 & 0 & 0 & 0 \\
 0 & 0 & \frac{i g}{\sqrt{2}} & 0 & 0 & 0 & 0 & 0 & 0 \\
 0 & 0 & \frac{i g}{\sqrt{2}} & 0 & 0 & 0 & 0 & 0 & 0 \\
 0 & 0 & \frac{i g {N_c}}{\sqrt{2}} & 0 & 0 & 0 & 0 & 0 & 0 \\
\end{array}
\right)}
\end{equation}
For ${{\bar{\CQ}}}^\dagger$ we have:
\begin{equation}\footnotesize{
{{\bar{\CQ}}}^\dagger|_\text{singlets}=\left(
\begin{array}{ccccccccc}
 0 & 0 & 0 & 0 & 0 & 0 & 0 & 0 & 0 \\
 0 & 0 & 0 & 0 & 0 & 0 & 0 & 0 & 0 \\
 0 & 0 & 0 & 0 & 0 & \frac{i g {N_2}}{\sqrt{2}} & -\frac{i g {N_1}}{\sqrt{2}} & -\frac{i g {N_3}}{\sqrt{2}} & -i \sqrt{2} g \\
 0 & 0 & 0 & 0 & 0 & 0 & 0 & 0 & 0 \\
 0 & 0 & 0 & 0 & 0 & 0 & 0 & 0 & 0 \\
 0 & \frac{i g}{\sqrt{2}} & 0 & 0 & 0 & 0 & 0 & 0 & 0 \\
 0 & -\frac{i g}{\sqrt{2}} & 0 & 0 & 0 & 0 & 0 & 0 & 0 \\
 0 & \frac{i g}{\sqrt{2}} & 0 & 0 & 0 & 0 & 0 & 0 & 0 \\
 0 & \frac{i g {N_c}}{\sqrt{2}} & 0 & 0 & 0 & 0 & 0 & 0 & 0 \\
\end{array}
\right)}
\end{equation}
In this appendix we use the following relation between supercharges and the light-cone hamiltonian:
\be
2P_+=\{{\bar{\CQ}},{\bar{\CQ}}^\dagger\}.
\ee
Therefore,
$$
{\tiny
 P_+|_\text{singlets}=
 \left(
\begin{array}{ccccccccc}
 0 & 0 & 0 & 0 & 0 & 0 & 0 & 0 & 0 \\
 0 & \frac{1}{2} g^2 ({N_2}+{N_1}) & 0 & 0 & 0 & 0 & 0 & 0 & 0 \\
 0 & 0 & \frac{1}{2} g^2 ({N_2}+{N_1}) & 0 & 0 & 0 & 0 & 0 & 0 \\
 0 & 0 & 0 & 0 & 0 & 0 & 0 & 0 & 0 \\
 0 & 0 & 0 & 0 & 0 & 0 & 0 & 0 & 0 \\
 0 & 0 & 0 & 0 & 0 & \frac{g^2 {N_2}}{2} & -\frac{1}{2} \left(g^2 {N_1}\right) & 0 & 0 \\
 0 & 0 & 0 & 0 & 0 & -\frac{1}{2} \left(g^2 {N_2}\right) & \frac{g^2 {N_1}}{2} & 0 & 0 \\
 0 & 0 & 0 & 0 & 0 & 0 & 0 & \frac{g^2 {N_3}}{2} & g^2 \\
 0 & 0 & 0 & 0 & 0 & 0 & 0 & \frac{1}{4} g^2 ({N_2}+{N_1}-{N_3}) {N_3} & \frac{1}{2} g^2 ({N_2}+{N_1}-{N_3}) \\
\end{array}
\right)}
$$
The eigenvectors of $P_+$ are
\begin{equation}\small
 \left(
\begin{array}{ccccccccc}
 0 & 0 & 0 & 0 & 1 & 0 & 0 & 0 & 0 \\
 0 & 0 & 0 & 0 & 0 & 0 & 0 & 0 & 1 \\
 0 & 0 & 0 & 0 & 0 & 0 & 0 & 1 & 0 \\
 0 & 0 & 0 & 1 & 0 & 0 & 0 & 0 & 0 \\
 0 & 0 & 1 & 0 & 0 & 0 & 0 & 0 & 0 \\
 0 & \frac{{N_1}}{{N_2}} & 0 & 0 & 0 & 0 & -1 & 0 & 0 \\
 0 & 1 & 0 & 0 & 0 & 0 & 1 & 0 & 0 \\
 -\frac{2}{{N_3}} & 0 & 0 & 0 & 0 & \frac{2}{{N_2}+{N_1}-{N_3}} & 0 & 0 & 0 \\
 1 & 0 & 0 & 0 & 0 & 1 & 0 & 0 & 0 \\
\end{array}
\right)
\end{equation}
The corresponding eigenvalues are
\begin{equation}\small
0, 0, 0, 0, 0, \frac{g^2 (N_2 + N_1)}{2}, \frac{g^2 (N_2 + N_1)}{2}, \frac{g^2 (N_2 + N_1)}{2}, \frac{g^2 (N_2 + N_1)}{2}
\end{equation}
In particular, the are five massless states given by
\begin{equation}\small
\ker P_+|_\text{singlets}= \text{Span}\left\{
\begin{array}{c}
 \text{$|$9$\rangle $} {N_3}-2 \text{$|$8$\rangle $} \\
 \text{$|$7$\rangle $} {N_2}+\text{$|$6$\rangle $} {N_1} \\
 \text{$|$5$\rangle $} \\
 \text{$|$4$\rangle $} \\
 \text{$|$1$\rangle $} \\
\end{array}\right\}
\end{equation}

Now let us consider the subspace in the representation $\mathbf{N_2 \times \bar{N}_1}$:
\begin{equation}\small
\begin{array}{l}
 \text{$|$1$\rangle $}\text{}_a^r=\left(\bar{\lambda }^{\dagger }\right)_{\alpha }^{\alpha }(1) \left(\bar{\gamma }^{\dagger
   }\right)_a^r(1) \text{$|$0$\rangle $} \\
 \text{$|$2$\rangle $}\text{}_a^r=\left(\lambda ^{\dagger }\right)_{\alpha }^{\alpha }(1) \left(\bar{\gamma }^{\dagger
   }\right)_a^r(1) \text{$|$0$\rangle $} \\
 \text{$|$3$\rangle $}\text{}_a^r=\left(\bar{\gamma }^{\dagger }\right)_a^r(2) \text{$|$0$\rangle $} \\
 \text{$|$4$\rangle $}\text{}_a^r=\left(\phi ^{\dagger }\right)_{\alpha }^r(1) \left(p^{\dagger }\right)_a^{\alpha }(1) \text{$|$0$\rangle $} \\
\end{array}
\end{equation}
\begin{equation}\small
{\bar{\CQ}} |_\mathbf{N_2\times \bar{N}_1}=
\left(
\begin{array}{cccc}
 0 & 0 & 0 & 0 \\
 0 & 0 & 0 & 0 \\
 0 & 0 & 0 & 0 \\
 0 & 0 & 1 & 0 \\
\end{array}
\right)
,\qquad
{{\bar{\CQ}}}^\dagger|_\mathbf{N_2\times \bar{N}_1}=\left(
\begin{array}{cccc}
 0 & 0 & 0 & 0 \\
 0 & 0 & 0 & 0 \\
 0 & 0 & 0 & {N_1} \\
 0 & 0 & 0 & 0 \\
\end{array}
\right)
,\qquad
 P_+|_\mathbf{N_2\times \bar{N}_1}=
 \left(
\begin{array}{cccc}
 0 & 0 & 0 & 0 \\
 0 & 0 & 0 & 0 \\
 0 & 0 & \frac{{N_1}}{2} & 0 \\
 0 & 0 & 0 & \frac{{N_1}}{2} \\
\end{array}
\right)
\end{equation}
Expressions for the conjugate representation $\mathbf{\bar{N}_2\times N_1}$ can be obtained in a similar way.
For all other irreps the action of ${\bar{\CQ}}$ and ${{\bar{\CQ}}}^\dagger$ is trivial.
For example, in the $\mathbf{{N}_1\times \bar{N}_3}$ sector we have only one state:
\begin{equation}\small
 |1\rangle_i^a =(\bar{p}^\dagger)_\alpha^a(1)({\psi }^\dagger)_i^\alpha(1)|0\rangle
\end{equation}

\newpage

\subsection*{DLCQ spectrum for $K=3$}

Singlet states:
\begin{equation}\footnotesize
 \begin{array}{l}
 \text{$|$1$\rangle $}=\left(\lambda ^{\dagger }\right)_{\alpha }^{\alpha }(3)\text{$|$0$\rangle $} \\
 \text{$|$2$\rangle $}=\left(\bar{\lambda }^{\dagger }\right)_{\alpha }^{\alpha }(3)\text{$|$0$\rangle $} \\
 \text{$|$3$\rangle $}=\left(\bar{\lambda }^{\dagger }\right)_{\beta }^{\alpha }(2)\left(\lambda ^{\dagger }\right)_{\alpha }^{\beta
   }(1)\text{$|$0$\rangle $} \\
 \text{$|$4$\rangle $}=\left(\bar{\lambda }^{\dagger }\right)_{\beta }^{\alpha }(1)\left(\lambda ^{\dagger }\right)_{\alpha }^{\beta
   }(2)\text{$|$0$\rangle $} \\
 \text{$|$5$\rangle $}=\left(\bar{\lambda }^{\dagger }\right)_{\beta }^{\beta }(2)\left(\lambda ^{\dagger }\right)_{\alpha }^{\alpha
   }(1)\text{$|$0$\rangle $} \\
 \text{$|$6$\rangle $}=\left(\bar{\lambda }^{\dagger }\right)_{\beta }^{\beta }(1)\left(\lambda ^{\dagger }\right)_{\alpha }^{\alpha
   }(2)\text{$|$0$\rangle $} \\
 \text{$|$7$\rangle $}=\left(\lambda ^{\dagger }\right)_{\beta }^{\beta }(2)\left(\lambda ^{\dagger }\right)_{\alpha }^{\alpha
   }(1)\text{$|$0$\rangle $} \\
 \text{$|$8$\rangle $}=\left(\bar{\lambda }^{\dagger }\right)_{\beta }^{\beta }(2)\left(\bar{\lambda }^{\dagger }\right)_{\alpha }^{\alpha
   }(1)\text{$|$0$\rangle $} \\
 \text{$|$9$\rangle $}=\left(\lambda ^{\dagger }\right)_{\beta }^{\alpha }(2)\left(\lambda ^{\dagger }\right)_{\alpha }^{\beta
   }(1)\text{$|$0$\rangle $} \\
 \text{$|$10$\rangle $}=\left(\bar{\lambda }^{\dagger }\right)_{\beta }^{\alpha }(2)\left(\bar{\lambda }^{\dagger }\right)_{\alpha }^{\beta
   }(1)\text{$|$0$\rangle $} \\
 \text{$|$11$\rangle $}=\left(\bar{\lambda }^{\dagger }\right)_{\gamma }^{\beta }(1)\left(\lambda ^{\dagger }\right)_{\beta }^{\gamma
   }(1)\left(\lambda ^{\dagger }\right)_{\alpha }^{\alpha }(1)\text{$|$0$\rangle $} \\
 \text{$|$12$\rangle $}=\left(\bar{\lambda }^{\dagger }\right)_{\gamma }^{\alpha }(1)\left(\bar{\lambda }^{\dagger }\right)_{\beta }^{\beta
   }(1)\left(\lambda ^{\dagger }\right)_{\alpha }^{\gamma }(1)\text{$|$0$\rangle $} \\
 \text{$|$13$\rangle $}=\left(\bar{\lambda }^{\dagger }\right)_{\gamma }^{\alpha }(1)\left(\lambda ^{\dagger }\right)_{\beta }^{\gamma
   }(1)\left(\lambda ^{\dagger }\right)_{\alpha }^{\beta }(1)\text{$|$0$\rangle $} \\
 \text{$|$14$\rangle $}=\left(\bar{\lambda }^{\dagger }\right)_{\gamma }^{\beta }(1)\left(\bar{\lambda }^{\dagger }\right)_{\beta }^{\alpha
   }(1)\left(\lambda ^{\dagger }\right)_{\alpha }^{\gamma }(1)\text{$|$0$\rangle $} \\
 \text{$|$15$\rangle $}=\left(\lambda ^{\dagger }\right)_{\gamma }^{\beta }(1)\left(\lambda ^{\dagger }\right)_{\beta }^{\alpha }(1)\left(\lambda
   ^{\dagger }\right)_{\alpha }^{\gamma }(1)\text{$|$0$\rangle $} \\
 \text{$|$16$\rangle $}=\left(\bar{\lambda }^{\dagger }\right)_{\gamma }^{\beta }(1)\left(\bar{\lambda }^{\dagger }\right)_{\beta }^{\alpha
   }(1)\left(\bar{\lambda }^{\dagger }\right)_{\alpha }^{\gamma }(1)\text{$|$0$\rangle $} \\
 \text{$|$17$\rangle $}=\left(\bar{\gamma }^{\dagger }\right)_c^s(1)\left(\gamma ^{\dagger }\right)_s^c(2)\text{$|$0$\rangle $} \\
 \text{$|$18$\rangle $}=\left(\bar{\gamma }^{\dagger }\right)_c^s(2)\left(\gamma ^{\dagger }\right)_s^c(1)\text{$|$0$\rangle $} \\
 \text{$|$19$\rangle $}=\left(\bar{\phi }^{\dagger }\right)_t^{\alpha }(1)\left(\phi ^{\dagger }\right)_{\alpha }^t(2)\text{$|$0$\rangle $} \\
 \text{$|$20$\rangle $}=\left(\bar{\phi }^{\dagger }\right)_t^{\alpha }(2)\left(\phi ^{\dagger }\right)_{\alpha }^t(1)\text{$|$0$\rangle $} \\
 \text{$|$21$\rangle $}=\left(\bar{p}^{\dagger }\right)_{\beta }^b(1)\left(p^{\dagger }\right)_b^{\beta }(2)\text{$|$0$\rangle $} \\
 \text{$|$22$\rangle $}=\left(\bar{p}^{\dagger }\right)_{\beta }^b(2)\left(p^{\dagger }\right)_b^{\beta }(1)\text{$|$0$\rangle $} \\
\end{array}\qquad
\begin{array}{l}
 \text{$|$23$\rangle $}=\left({\psi }^{\dagger }\right)_k^{\alpha }(1)\left({\bar{\psi}} ^{\dagger }\right)_{\alpha }^k(2)\text{$|$0$\rangle $} \\
 \text{$|$24$\rangle $}=\left({\psi }^{\dagger }\right)_k^{\alpha }(2)\left({\bar{\psi}} ^{\dagger }\right)_{\alpha }^k(1)\text{$|$0$\rangle $} \\
 \text{$|$25$\rangle $}=\left(\bar{\omega }^{\dagger }\right)_F^0(1)\left(\omega ^{\dagger }\right)_0^F(2)\text{$|$0$\rangle $} \\
 \text{$|$26$\rangle $}=\left(\bar{\omega }^{\dagger }\right)_F^0(2)\left(\omega ^{\dagger }\right)_0^F(1)\text{$|$0$\rangle $} \\
 \text{$|$27$\rangle $}=\left(\gamma ^{\dagger }\right)_d^b(1)\left(\phi ^{\dagger }\right)_{\beta }^d(1)\left(p^{\dagger }\right)_b^{\beta
   }(1)\text{$|$0$\rangle $} \\
 \text{$|$28$\rangle $}=\left(\bar{\gamma }^{\dagger }\right)_l^t(1)\left(\bar{\phi }^{\dagger }\right)_t^{\alpha }(1)\left(\bar{p}^{\dagger
   }\right)_{\alpha }^l(1)\text{$|$0$\rangle $} \\
 \text{$|$29$\rangle $}=\left(\lambda ^{\dagger }\right)_{\gamma }^{\gamma }(1)\left(\bar{\gamma }^{\dagger }\right)_c^s(1)\left(\gamma ^{\dagger
   }\right)_s^c(1)\text{$|$0$\rangle $} \\
 \text{$|$30$\rangle $}=\left(\lambda ^{\dagger }\right)_{\gamma }^{\gamma }(1)\left(\bar{\phi }^{\dagger }\right)_t^{\alpha }(1)\left(\phi
   ^{\dagger }\right)_{\alpha }^t(1)\text{$|$0$\rangle $} \\
 \text{$|$31$\rangle $}=\left(\lambda ^{\dagger }\right)_{\gamma }^{\gamma }(1)\left(\bar{p}^{\dagger }\right)_{\beta }^b(1)\left(p^{\dagger
   }\right)_b^{\beta }(1)\text{$|$0$\rangle $} \\
 \text{$|$32$\rangle $}=\left({\psi }^{\dagger }\right)_u^{\beta }(1)\left({\bar{\psi}} ^{\dagger }\right)_{\beta }^u(1)\left(\lambda ^{\dagger
   }\right)_{\alpha }^{\alpha }(1)\text{$|$0$\rangle $} \\
 \text{$|$33$\rangle $}=\left(\bar{\omega }^{\dagger }\right)_G^0(1)\left(\omega ^{\dagger }\right)_0^G(1)\left(\lambda ^{\dagger }\right)_{\alpha
   }^{\alpha }(1)\text{$|$0$\rangle $} \\
 \text{$|$34$\rangle $}=\left(\bar{\lambda }^{\dagger }\right)_{\gamma }^{\gamma }(1)\left(\bar{\gamma }^{\dagger }\right)_c^s(1)\left(\gamma
   ^{\dagger }\right)_s^c(1)\text{$|$0$\rangle $} \\
 \text{$|$35$\rangle $}=\left(\bar{\lambda }^{\dagger }\right)_{\gamma }^{\gamma }(1)\left(\bar{\phi }^{\dagger }\right)_t^{\alpha }(1)\left(\phi
   ^{\dagger }\right)_{\alpha }^t(1)\text{$|$0$\rangle $} \\
 \text{$|$36$\rangle $}=\left(\bar{\lambda }^{\dagger }\right)_{\gamma }^{\gamma }(1)\left(\bar{p}^{\dagger }\right)_{\beta }^b(1)\left(p^{\dagger
   }\right)_b^{\beta }(1)\text{$|$0$\rangle $} \\
 \text{$|$37$\rangle $}=\left({\psi }^{\dagger }\right)_u^{\beta }(1)\left({\bar{\psi}} ^{\dagger }\right)_{\beta }^u(1)\left(\bar{\lambda }^{\dagger
   }\right)_{\alpha }^{\alpha }(1)\text{$|$0$\rangle $} \\
 \text{$|$38$\rangle $}=\left(\bar{\omega }^{\dagger }\right)_G^0(1)\left(\omega ^{\dagger }\right)_0^G(1)\left(\bar{\lambda }^{\dagger
   }\right)_{\alpha }^{\alpha }(1)\text{$|$0$\rangle $} \\
 \text{$|$39$\rangle $}=\left(\lambda ^{\dagger }\right)_{\gamma }^{\alpha }(1)\left(\bar{\phi }^{\dagger }\right)_t^{\gamma }(1)\left(\phi
   ^{\dagger }\right)_{\alpha }^t(1)\text{$|$0$\rangle $} \\
 \text{$|$40$\rangle $}=\left(\lambda ^{\dagger }\right)_{\gamma }^{\beta }(1)\left(\bar{p}^{\dagger }\right)_{\beta }^b(1)\left(p^{\dagger
   }\right)_b^{\gamma }(1)\text{$|$0$\rangle $} \\
 \text{$|$41$\rangle $}=\left({\psi }^{\dagger }\right)_u^{\alpha }(1)\left({\bar{\psi}} ^{\dagger }\right)_{\beta }^u(1)\left(\lambda ^{\dagger
   }\right)_{\alpha }^{\beta }(1)\text{$|$0$\rangle $} \\
 \text{$|$42$\rangle $}=\left(\bar{\lambda }^{\dagger }\right)_{\gamma }^{\alpha }(1)\left(\bar{\phi }^{\dagger }\right)_t^{\gamma }(1)\left(\phi
   ^{\dagger }\right)_{\alpha }^t(1)\text{$|$0$\rangle $} \\
 \text{$|$43$\rangle $}=\left(\bar{\lambda }^{\dagger }\right)_{\gamma }^{\beta }(1)\left(\bar{p}^{\dagger }\right)_{\beta }^b(1)\left(p^{\dagger
   }\right)_b^{\gamma }(1)\text{$|$0$\rangle $} \\
 \text{$|$44$\rangle $}=\left({\psi }^{\dagger }\right)_u^{\alpha }(1)\left({\bar{\psi}} ^{\dagger }\right)_{\beta }^u(1)\left(\bar{\lambda }^{\dagger
   }\right)_{\alpha }^{\beta }(1)\text{$|$0$\rangle $} \\
\end{array}
\end{equation}
The massless part is given by
\begin{equation}\small
\ker P_+|_\text{singlets}= \text{Span}\left\{
 \begin{array}{c}
\text{$|$43$\rangle $} {N_2}-\text{$|$36$\rangle $} {N_c} {N_2}+\text{$|$42$\rangle $} {N_1}-\text{$|$35$\rangle $} {N_1} {N_c} \\
 \text{$|$40$\rangle $} {N_2}-\text{$|$31$\rangle $} {N_c} {N_2}+\text{$|$39$\rangle $} {N_1}-\text{$|$30$\rangle $} {N_1} {N_c} \\
 \text{$|$34$\rangle $} \\
 \text{$|$29$\rangle $} \\
 -2 \text{$|$23$\rangle $}-2 \text{$|$24$\rangle $}+\text{$|$25$\rangle $} {N_3}+\text{$|$26$\rangle $} {N_3} \\
 \text{$|$12$\rangle $} \\
 \text{$|$11$\rangle $} \\
\end{array}\right\}
\end{equation}

The subspace in the representation $\mathbf{N_2\times \bar{N}_1}$:
\begin{equation}\footnotesize
\begin{array}{l}
 \text{$|$1$\rangle $}\text{}_a^r=\left(\bar{\gamma }^{\dagger }\right)_a^r(3) \text{$|$0$\rangle $} \\
 \text{$|$2$\rangle $}\text{}_a^r=\left(\phi ^{\dagger }\right)_{\alpha }^r(1) \left(p^{\dagger }\right)_a^{\alpha }(2) \text{$|$0$\rangle $} \\
 \text{$|$3$\rangle $}\text{}_a^r=\left(\phi ^{\dagger }\right)_{\alpha }^r(2) \left(p^{\dagger }\right)_a^{\alpha }(1) \text{$|$0$\rangle $} \\
 \text{$|$4$\rangle $}\text{}_a^r=\left(\bar{\gamma }^{\dagger }\right)_a^r(1) \left(\bar{\gamma }^{\dagger }\right)_c^s(1) \left(\gamma ^{\dagger
   }\right)_s^c(1) \text{$|$0$\rangle $} \\
 \text{$|$5$\rangle $}\text{}_a^r=\left(\lambda ^{\dagger }\right)_{\alpha }^{\alpha }(2) \left(\bar{\gamma }^{\dagger
   }\right)_a^r(1) \text{$|$0$\rangle $} \\
 \text{$|$6$\rangle $}\text{}_a^r=\left(\bar{\lambda }^{\dagger }\right)_{\alpha }^{\alpha }(2) \left(\bar{\gamma }^{\dagger
   }\right)_a^r(1) \text{$|$0$\rangle $} \\
 \text{$|$7$\rangle $}\text{}_a^r=\left(\bar{\lambda }^{\dagger }\right)_{\beta }^{\alpha }(1) \left(\lambda ^{\dagger }\right)_{\alpha }^{\beta
   }(1) \left(\bar{\gamma }^{\dagger }\right)_a^r(1) \text{$|$0$\rangle $} \\
 \text{$|$8$\rangle $}\text{}_a^r=\left(\bar{\lambda }^{\dagger }\right)_{\beta }^{\beta }(1) \left(\lambda ^{\dagger }\right)_{\alpha }^{\alpha
   }(1) \left(\bar{\gamma }^{\dagger }\right)_a^r(1) \text{$|$0$\rangle $} \\
 \text{$|$9$\rangle $}\text{}_a^r=\left(\bar{\gamma }^{\dagger }\right)_a^r(1) \left(\bar{\phi }^{\dagger }\right)_t^{\alpha }(1) \left(\phi
   ^{\dagger }\right)_{\alpha }^t(1) \text{$|$0$\rangle $} \\
 \text{$|$10$\rangle $}\text{}_a^r=\left(\bar{\gamma }^{\dagger }\right)_a^r(1) \left(\bar{p}^{\dagger }\right)_{\beta }^b(1) \left(p^{\dagger
   }\right)_b^{\beta }(1) \text{$|$0$\rangle $} \\
\end{array}
\qquad
 \begin{array}{l}
 \text{$|$11$\rangle $}\text{}_a^r=\left({\psi }^{\dagger }\right)_k^{\alpha }(1) \left({\bar{\psi}} ^{\dagger }\right)_{\alpha }^k(1) \left(\bar{\gamma
   }^{\dagger }\right)_a^r(1) \text{$|$0$\rangle $} \\
 \text{$|$12$\rangle $}\text{}_a^r=\left(\bar{\omega }^{\dagger }\right)_F^0(1) \left(\omega ^{\dagger }\right)_0^F(1) \left(\bar{\gamma }^{\dagger
   }\right)_a^r(1) \text{$|$0$\rangle $} \\
 \text{$|$13$\rangle $}\text{}_a^r=\left(\lambda ^{\dagger }\right)_{\alpha }^{\alpha }(1) \left(\bar{\gamma }^{\dagger
   }\right)_a^r(2) \text{$|$0$\rangle $} \\
 \text{$|$14$\rangle $}\text{}_a^r=\left(\lambda ^{\dagger }\right)_{\beta }^{\beta }(1) \left(\phi ^{\dagger }\right)_{\alpha
   }^r(1) \left(p^{\dagger }\right)_a^{\alpha }(1) \text{$|$0$\rangle $} \\
 \text{$|$15$\rangle $}\text{}_a^r=\left(\bar{\lambda }^{\dagger }\right)_{\alpha }^{\alpha }(1) \left(\bar{\gamma }^{\dagger
   }\right)_a^r(2) \text{$|$0$\rangle $} \\
 \text{$|$16$\rangle $}\text{}_a^r=\left(\bar{\lambda }^{\dagger }\right)_{\beta }^{\beta }(1) \left(\phi ^{\dagger }\right)_{\alpha
   }^r(1) \left(p^{\dagger }\right)_a^{\alpha }(1) \text{$|$0$\rangle $} \\
 \text{$|$17$\rangle $}\text{}_a^r=\left(\lambda ^{\dagger }\right)_{\beta }^{\alpha }(1) \left(\phi ^{\dagger }\right)_{\alpha
   }^r(1) \left(p^{\dagger }\right)_a^{\beta }(1) \text{$|$0$\rangle $} \\
 \text{$|$18$\rangle $}\text{}_a^r=\left(\bar{\lambda }^{\dagger }\right)_{\beta }^{\alpha }(1) \left(\phi ^{\dagger }\right)_{\alpha
   }^r(1) \left(p^{\dagger }\right)_a^{\beta }(1) \text{$|$0$\rangle $} \\
 \text{$|$19$\rangle $}\text{}_a^r=\left(\bar{\gamma }^{\dagger }\right)_a^t(1) \left(\bar{\phi }^{\dagger }\right)_t^{\alpha }(1) \left(\phi
   ^{\dagger }\right)_{\alpha }^r(1) \text{$|$0$\rangle $} \\
 \text{$|$20$\rangle $}\text{}_a^r=\left(\bar{\gamma }^{\dagger }\right)_c^r(1) \left(\bar{p}^{\dagger }\right)_{\alpha }^c(1) \left(p^{\dagger
   }\right)_a^{\alpha }(1) \text{$|$0$\rangle $} \\
\end{array}
\end{equation}
The massless part is given by
$$
\footnotesize
\ker P_+|_\mathbf{N_2\times \bar{N}_1}= \text{Span}\left\{
\begin{array}{c}
 \text{$|$20$\rangle $}-\frac{\text{$|$10$\rangle $} ({N_2} {N_1}-1)}{{N_2}-{N_1}}-\frac{\text{$|$1$\rangle $}
   \left(1-{N_1}^2\right)}{{N_2}-{N_1}}-\frac{\text{$|$9$\rangle $} \left({N_1}^2-1\right)}{{N_2}-{N_1}} \\
 \text{$|$19$\rangle $}-\frac{\text{$|$9$\rangle $} (1-{N_2} {N_1})}{{N_2}-{N_1}}-\frac{\text{$|$10$\rangle $}
   \left(1-{N_2}^2\right)}{{N_2}-{N_1}}-\frac{\text{$|$1$\rangle $} \left({N_2}^2-1\right)}{{N_2}-{N_1}} \\
 -\frac{\text{$|$16$\rangle $}}{{N_c}}+\text{$|$18$\rangle $}+\frac{4 i \text{$|$9$\rangle $} g {N_1} \left({N_c}^2-1\right)}{({N_2}-{N_1})
   {N_c}}+\frac{4 i \text{$|$10$\rangle $} g {N_2} \left({N_c}^2-1\right)}{({N_2}-{N_1}) {N_c}}-\frac{2 i \text{$|$1$\rangle $} \left(g
   {N_2} {N_c}^2+g {N_1} {N_c}^2-g {N_2}-g {N_1}\right)}{({N_2}-{N_1}) {N_c}} \\
 \text{$|$12$\rangle $}-\frac{2 \text{$|$11$\rangle $}}{{N_3}} \\
 \text{$|$8$\rangle $} \\
 \text{$|$7$\rangle $} \\
 \text{$|$4$\rangle $} \\
\end{array}\right\}
$$

The subspace in the representation $\mathbf{{N}_1\times \bar{N}_3}$ is
\begin{equation}\footnotesize
 \begin{array}{l}
 \text{$|$1$\rangle $}\text{}_i^a=\left({\psi }^{\dagger }\right)_i^{\alpha }(2) \left(\bar{p}^{\dagger }\right)_{\alpha
   }^a(1) \text{$|$0$\rangle $} \\
 \text{$|$2$\rangle $}\text{}_i^a=\left({\psi }^{\dagger }\right)_i^{\alpha }(1) \left(\bar{p}^{\dagger }\right)_{\alpha
   }^a(2) \text{$|$0$\rangle $} \\
 \text{$|$3$\rangle $}\text{}_i^a=\left({\psi }^{\dagger }\right)_i^{\alpha }(1) \left(\gamma ^{\dagger }\right)_t^a(1) \left(\phi ^{\dagger
   }\right)_{\alpha }^t(1) \text{$|$0$\rangle $} \\
 \text{$|$4$\rangle $}\text{}_i^a=\left({\psi }^{\dagger }\right)_i^{\alpha }(1) \left(\lambda ^{\dagger }\right)_{\beta }^{\beta
   }(1) \left(\bar{p}^{\dagger }\right)_{\alpha }^a(1) \text{$|$0$\rangle $} \\
\end{array}
\qquad
 \begin{array}{l}
 \text{$|$5$\rangle $}\text{}_i^a=\left({\psi }^{\dagger }\right)_i^{\alpha }(1) \left(\bar{\lambda }^{\dagger }\right)_{\beta }^{\beta
   }(1) \left(\bar{p}^{\dagger }\right)_{\alpha }^a(1) \text{$|$0$\rangle $} \\
 \text{$|$6$\rangle $}\text{}_i^a=\left({\psi }^{\dagger }\right)_i^{\beta }(1) \left(\lambda ^{\dagger }\right)_{\beta }^{\alpha
   }(1) \left(\bar{p}^{\dagger }\right)_{\alpha }^a(1) \text{$|$0$\rangle $} \\
 \text{$|$7$\rangle $}\text{}_i^a=\left({\psi }^{\dagger }\right)_i^{\beta }(1) \left(\bar{\lambda }^{\dagger }\right)_{\beta }^{\alpha
   }(1) \left(\bar{p}^{\dagger }\right)_{\alpha }^a(1) \text{$|$0$\rangle $} \\
\end{array}
\end{equation}
with two massless states
\begin{equation}\small
\ker P_+|_\mathbf{{N}_1\times \bar{N}_3}= \text{Span}\left\{
\begin{array}{c}
 \text{$|$7$\rangle $}-\text{$|$5$\rangle $} {N_c} \\
 \text{$|$6$\rangle $}-\text{$|$4$\rangle $} {N_c} \\
\end{array}\right\}
\end{equation}

%%%%%%%%%%%%%%%%%%%%%%%%%%%%%%%%%%%%%%%%%%%%%%%%%%%%%%%%%%%%%%%%%%%%%%%%%%%%%%%%%%%%%%%%%%%%%%

\newpage

\bibliographystyle{JHEP_TD}
\bibliography{classH}

\end{document}